\documentclass[epsf,useAMS,usenatbib,usegraphicx]{mn2e}
\usepackage[table]{xcolor}
\definecolor{lightgray}{gray}{0.9}
\usepackage{epsfig}
\usepackage{aas_macros} 
\usepackage{amsfonts}
\usepackage{amsmath}
\usepackage{amssymb}
\usepackage{appendix}
\usepackage{booktabs}
\usepackage{pdfpages}
\usepackage{upgreek}
\usepackage{url}
\newcommand{\blu}{\textcolor{black} } 
\newcommand{\red}{\textcolor{black} }

\definecolor{darkgreen}{rgb}{0.05000,0.650000,0.05000}
\definecolor{lilac}{rgb}{0.55000,0.350000,0.95000}

\newcolumntype{H}{>{\setbox0=\hbox\bgroup}c<{\egroup}@{}}
\newcommand{\ion}[2]{{#1}\,{\sc #2}}

\newcommand{\specline}[3]{{#1}\,{\sc #2}\:{#3}}

\title[The $\lambda$\,Boo stars]{An evaluation of the membership probability of 212 $\lambda$\,Boo stars: I. A Catalogue.}

\author[Simon J. Murphy et al.] 
{Simon~J.~Murphy$^{1,2,\dagger}$, 
Christopher~J.~Corbally$^{3}$, 
Richard~O.~Gray$^{4}$, \and 
Kwang-Ping~Cheng$^{5}$,
James~E.~Neff$^{6}$,
Chris~Koen$^{7}$,
Charles~A.~Kuehn$^{1,2}$,  \and
Ian~Newsome$^{4}$ and
Quinlin~Riggs$^{4}$\\
$^{1}$ Sydney Institute for Astronomy (SIfA), School of Physics, University of Sydney, NSW 2006, Australia\\
$^{2}$ Stellar Astrophysics Centre, Department of Physics and Astronomy, Aarhus University, 8000 Aarhus C, Denmark\\
$^{3}$ Vatican Observatory Research Group, Steward Observatory, Tucson, AZ 85721-0065, USA\\
$^{4}$ Department of Physics and Astronomy, Appalachian State University, Boone, NC 28608, USA\\
$^{5}$ Department of Physics, California State University, Fullerton, CA, USA\\
$^{6}$ Department of Physics and Astronomy, College of Charleston, Charleston, SC, USA\\
$^{7}$ Department of Statistics, University of the Western Cape, Private Bag X17, Bellville, 7535 Cape, South Africa\\
\\
$^{\dagger}$email: murphy@physics.usyd.edu.au
}
 
\begin{document} 

\maketitle 

\begin{abstract}
The literature on the $\lambda$\,Boo stars has grown to become somewhat heterogenous, as different authors have applied different criteria across the UV, optical and infrared regions to determine the membership status of $\lambda$\,Boo candidates. We aim to clear up the confusion by consulting the literature on 212 objects that have been considered as $\lambda$\,Boo candidates, and subsequently evaluating the evidence in favour of their admission to the $\lambda$\,Boo class. We obtained new spectra of $\sim$90 of these candidates and classified them on the MK system to aid in the membership evaluations. The re-evaluation of the 212 objects \blu{resulted in} \blu{64} members and \blu{103} non-members of the $\lambda$\,Boo class, with a further \blu{45} stars for which membership status is unclear. We suggest observations for each of the stars in the latter category that will allow them to be confidently included or rejected from the class. Our reclassification facilitates homogenous analysis on group members, and represents the largest collection of confirmed $\lambda$\,Boo stars known. 
\end{abstract}

\section{Introduction}
\label{sec:intro}

\subsection{The class of $\lambda$\,Boo stars}
\label{ssec:intro:LB}

The $\lambda$\,Boo stars are a rare class, making up about 2\:per\:cent of the population of stars of spectral type A. Their main characteristic is a surface depletion of refractory elements; typically, Fe-peak elements are underabundant by 1\,dex, and in extreme $\lambda$\,Boo stars by 2\,dex, while the volatile elements (C, N, O, S) have solar abundances. For many years this abundance pattern remained unexplained, but its origin is believed to lie in the accretion of material from either the interstellar medium \citep{kamp&paunzen2002} or a circumstellar disk \citep{venn&lambert1990, king1994}. In this environment, dust-gas separation occurs, with the refractory elements precipitating into dust grains. The strong radiation field of the A star ejects the dust whilst the star accretes the gas, leaving an enhancement of volatiles compared to refractories on the stellar surface.

Observationally, $\lambda$\,Boo stars lie on the main sequence or in the later pre-main sequence phases. Their calculated ages show a spread throughout the main-sequence phase \citep{iliev&barzova1995} and some pre-main sequence stars have $\lambda$\,Boo-like abundance patterns \citep{folsometal2012}. Late-A stars fall inside the classical instability strip, where pulsation is driven by \blu{the opacity mechanism operating on} helium. Hence, it is no surprise that so many $\lambda$\,Boo stars with late-A hydrogen-line types are found to pulsate (e.g.\ \citealt{weissetal1994, paunzenetal2002a}). Asteroseismology of these stars facilitates distinction between a star whose metal deficiencies are limited to the surface, i.e.\ a $\lambda$\,Boo star, and a star that is metal-weak throughout \citep[e.g.][]{murphyetal2013a}. Asteroseismology may also constrain their ages \citep{moyaetal2010a,sodoretal2014}.

The spectra of $\lambda$\,Boo stars show \ion{Ca}{ii} K line and metal-line types that are much earlier than the hydrogen line type, which is indicative of the metal weakness. The metal weakness is also apparent in the UV, where the reduced line-blanketing leads to a UV excess. UV characteristics of the group were discussed by \citet{faraggianaetal1990}, and we do not go into detail here. The infrared properties of the $\lambda$\,Boo stars have also been investigated \citep{andrillatetal1995}, and some $\lambda$\,Boo stars are known to have an infrared excess \citep{holweger&rentzsch-holm1995} that is attributable to the dusty environment in which they reside. The residual dust may play a role in planet formation if the dust is able to coalesce within the strong radiation field. There is some evidence that it can, exemplified by the dusty $\lambda$\,Boo star HR\,8799 (HD\,218396) which has at least four planetary companions (e.g.\ \citealt{soummeretal2011}). But what is not known is whether the fraction of $\lambda$\,Boo stars having an infrared excess is higher than that of normal stars, and an inhibiting factor in ascertaining that ratio is the heterogeneity of the $\lambda$\,Boo class. This forms the basis of the motivation of this work, which is discussed in \S\,\ref{ssec:motivation}.

Unlike common chemical peculiarities seen in A stars, namely those of the Am and Ap stars, the $\lambda$\,Boo phenomenon is not associated with slow rotation; on the contrary, the mean $v\sin i$ of the $\lambda$\,Boo stars is in accord with the modal value for normal A stars, that being $\sim$165\,km\,s$^{-1}$ \citep{abt&morrell1995}.

\subsection{Motivation for membership evaluations}
\label{ssec:motivation}

Progress in understanding the $\lambda$\,Boo stars has been hindered by a somewhat heterogeneous literature. The last catalogue of $\lambda$\,Boo stars was assembled by \citet{paunzen2001}, in the form of a list of new and confirmed $\lambda$\,Boo stars, though \blu{in print} that catalogue did not include some stars that were considered `classic' $\lambda$\,Boo stars at that time (e.g.\ HD\,111786). Heterogeneity continued to proliferate because one survey would demand rejection of a star while another would consider it a firm member of the $\lambda$\,Boo group. Without a consolidated evaluation, future studies were left unsure how to treat certain `members'. The extent of the heterogeneity of the $\lambda$\,Boo group was articulated by (and actually exacerbated by) \citet{gerbaldietal2003}, after which interest in the $\lambda$\,Boo stars dropped. That work has been heavily criticised \citep{stutz&paunzen2006, griffinetal2012} and we discuss it in more detail in \S\,\ref{sssec:supplementary}.

Therefore we aim here to provide a consistent, homogenous membership evaluation for every star that has been classified as a $\lambda$\,Boo star in the literature, or considered for membership in that group. One must note that surveys for $\lambda$\,Boo stars (e.g.\ \citealt{paunzenetal2001}) will have considered the membership of hundreds of stars, so we only re-evaluate stars that have been called `$\lambda$\,Boo', or stars that an author has  felt the need to stress are `definitely not $\lambda$\,Boo', presuming at least some evidence to the contrary has existed. We summarise the evidence for and against $\lambda$\,Boo membership for each of the 212 stars considered as $\lambda$\,Boo stars in the literature. We have avoided the inclusion of unpublished $\lambda$\,Boo candidates in order to uphold authenticity; a key exception is a dynamic list of $\lambda$\,Boo stars kept on Gray's website\footnote{\url{http://www1.appstate.edu/dept/physics/spectrum/lamboo.txt}}, as it was when our candidate list was frozen for analysis at the start of 2014.

We thus provide a snapshot of the state of the field at the start of 2014, which will act as a firm footing on which to continue investigations into $\lambda$\,Boo stars. In order to maintain as accurate a compilation as possible, we solicit notifications regarding evidence we may have overlooked, in case of future editions.

\subsection{Criteria for membership evaluations}

We considered the following forms of evidence when evaluating the membership status of $\lambda$\,Boo candidates.

First and foremost, we look for a spectral classification of `$\lambda$\,Boo', particularly if performed by Gray or Gray \& Corbally, who have been authorities on the classification of $\lambda$\,Boo stars for the last 25 years \citep{gray1988,gray&corbally1993} and equally for A stars as a whole \citep{gray&garrison1987,  gray&garrison1989, gray&garrison1989b, gray&corbally2009}. We take a conservative approach to $\lambda$\,Boo classifications made by Abt \citep{abt1984b, abt1985, abt&morrell1995}, who was a little lenient when admitting stars to the $\lambda$\,Boo group, having based some inclusions on a weak \specline{Mg}{ii}{4481} line only, when in fact that star could have belonged to any of a wider set of groups with that characteristic, such as classical shell stars, field horizontal branch stars, or generally metal-weak stars (see \citealt{griffinetal2012}).

We consider photometric measurements, especially Str\"omgren and $\Delta a$ \citep{maitzen&pavlovski1989, maitzen&pavlovski1989b}, that might support an optical assessment. \blu{A review of the utility of $\Delta a$ photometry in detecting chemically peculiar stars, including $\lambda$\,Boo stars, was written by \citet{paunzenetal2005}, who concluded that this photometric system is highly efficient in selecting $\lambda$\,Boo stars. The use of Geneva photometry for the same purpose was discussed therein, but we deemed there was insufficient added value in this photometric system when compared with Str\"omgren and $\Delta a$ for it to warrant application here.}

Local thermodynamic equilibrium (LTE) abundance analyses of refractory elements, especially iron-peak elements plus silicon and magnesium, are weighted strongly in our evaluations. Accompanying non-LTE (NLTE) abundance analyses for the volatile elements help to establish that the star exhibits the $\lambda$\,Boo phenomenon, rather than just generic metal weakness, hence NLTE analyses are highly sought after. An abundance analysis that supports the $\lambda$\,Boo spectral classification is the definitive assessment of the $\lambda$\,Boo phenomenon, but for a rapidly rotating star the difficulty in performing an abundance analysis restricts the availability of the former somewhat.

The UV character of $\lambda$\,Boo candidates provides valuable evidence for or against their inclusion in the class. We mostly use classifications from \citet{bascheketal1984} and from \citet{faraggianaetal1990}, where the latter established their own criteria for examining the UV character of $\lambda$\,Boo stars. \citet{solano&paunzen1998, solano&paunzen1999} added to those criteria, but did not re-evaluate $\lambda$\,Boo candidates from the literature. In this work we have also inspected the UV spectra of some stars that were not included in the aforementioned samples, but no detailed analysis of those was attempted. We will refine the UV criteria and conduct a uniform analysis of the UV spectra in a later paper.

We consider an infrared excess or circumstellar absorption in the optical as supporting evidence for membership, but with a low weighting, to avoid contaminating observational results with the expectation that $\lambda$\,Boo stars should be dusty if they accrete circumstellar material. Also as a low priority, we note inclusion, or equally, lack of \textit{ex}clusion, from earlier $\lambda$\,Boo catalogues.

We use parallax and proper motion measurements from the SIMBAD data base\footnote{\url{http://SIMBAD.u-strasbg.fr/SIMBAD/}} to calculate a transverse velocity, \blu{by converting angular motion at a known distance into a space velocity, so that Population II stars might be identified and rejected. This is more useful than the sole use of proper motions, since nearby objects can have large proper motions without correspondingly large space velocities. We do note, however, that} \textit{intermediate} Pop.\,II stars cannot be distinguished from $\lambda$\,Boo stars based on space velocities alone \citep{paunzenetal2014}. 

A summary of the evidence for each of the 212 stars is provided in \S\,\ref{sec:evaluations}. A membership recommendation was then made based on this evidence, as consistently as possible, into one of four classes: member, probable member, uncertain member, or non-member. Notice we have included two categories of `grey area'. These are used when evidence is in disagreement, inconclusive, or insufficient to make a firm decision. \blu{The need for two classes of grey area is founded. For example, some stars were suggested as $\lambda$\,Boo stars on very little evidence, and are awaiting verification or nullification. These would fall in the `uncertain member' group. On the other hand, a star that meets all but one of the criteria for a $\lambda$\,Boo star that we have adopted might be considered a `probable member', if that unsatisfied criterion were a normal UV spectrum, for instance. It is possible that the reader will disagree with some of our recommendations; the categorisation is unavoidably subjective at some level. Our decision-making process is kept transparent with the provision of all of the information involved in that decision. Hence the reader may also inform his or her own opinion and come to his or her own decision on the borderline cases.}

Elemental abundances form a continuous scale between normal and $\lambda$\,Boo, hence mild members of the class exist. Where we are persuaded to accept a star as a mild or marginal $\lambda$\,Boo, we indicate thusly, so as to distinguish between a definite mild $\lambda$\,Boo star and an uncertain member of the class, whose uncertainty arises from conflicting evidence (e.g. a $\lambda$\,Boo character in the UV, yet an apparently normal optical spectrum).

Particularly for stars in the `uncertain' and `probable member' categories, we have suggested observations that would help to arbitrate the membership of those stars in the $\lambda$\,Boo class. \blu{While high-resolution spectra and abundances for every object would be ideal, they are not necessary in all cases, and cannot always be obtained (e.g. where $v\sin i$ is too high). Thus we recommend full abundance analyses only when the existing evidence is insufficient. In any case, the first step in spectroscopic analyses should be an MK classification \citep{gray2014}, and so where practical and necessary} we obtained at least one spectrum from one or more of the following observatories: Dark Sky Observatory (DSO), the Vatican Advanced Technology Telescope (VATT), Siding Spring Observatory (SSO) or the South African Astronomical Observatory (SAAO). 

\subsubsection{Spectroscopic Observations and MK classification}

The VATT spectra were obtained over four runs with the VATTspec spectrograph on the Vatican Advanced Technology Telescope (VATT; 1.8-m, located on Mount Graham, Arizona). For these observations, the VATTspec is used with a 600\,g\,mm$^{-1}$ grating, which gives a resolution of \mbox{1.5\,\AA/2-pixels} in the vicinity of the \ion{Ca}{ii} H and K lines, with a spectral range of 3700--5540\,\AA. The spectra are recorded on a low-noise STA0520A CCD with 2688x512 pixels (University of Arizona Imaging Technology serial number 8228). Three pencil-style lamps, Hg, Ar, and Ne, were observed simultaneously for wavelength calibrations, and the spectroscopic data were reduced with IRAF using standard techniques.

Many of our spectra were obtained at DSO on the 0.8-m telescope as part of a regular observing programme. The observations were made with the GM spectrograph\footnote{\url{http://www.appstate.edu/~grayro/spectrum/GM/GM.html}} using the 1200\,g\,mm$^{-1}$ grating in the first order. The spectra have \mbox{1.8\AA/2-pixel} resolution, and cover the spectral range 3800--4600\,\AA. A hollow-cathode Fe-Ar comparison lamp was used for wavelength calibrations. Raw spectra were reduced using standard IRAF procedures.

Our SAAO spectra were obtained with the Grating Spectrograph on the 1.9-m telescope. The spectra cover the wavelength range 3800--5400\,\AA, and wavelength calibration is performed using a Cu/Ar lamp. The spectral resolution is \mbox{2\,\AA/2-pixel}.

Spectra from SSO were obtained during a single run in 2014\,Apr. We used the WiFeS spectrograph \citep{dopitaetal2007} on the ANU 2.3-m telescope at Siding Spring Observatory. Our spectra in the blue-violet region were obtained in B3000 mode, and have a resolution of about \mbox{2.5\,\AA/2-pixels}. The WiFeS data were reduced with the PyWiFeS software package \citep{childressetal2014}. Due to difficulty in rectifying the spectra over the Balmer jump, we trimmed the spectra to the range 3865--4960\,\AA. The spectra thus cover the region between the blue wing of H\,8 and the red wing of H\,$\beta$. 

We classified these spectra on the MK system. This enabled many of the initial `probable' and `uncertain' members to be moved to the `member' or `non-member' categories. Each spectrum was classified independently by all three of SJM, CJC and ROG, and after comparison of the initial classifications, the final spectral type assigned was agreed upon by iterative reclassification.

\subsection{Supplementary notes}
\label{sssec:supplementary}

\subsubsection{Spectral range}
\label{sssec:spectral-range}
\blu{At early spectral types (at A0 or B9) it is very difficult to be sure of a $\lambda$\,Boo classification, because metal lines are nearly absent anyway. Rapid rotation exacerbates the difficulty. The most notable metal lines at these spectral types are the \ion{Ca}{ii} K line and \specline{Mg}{ii}{4481}. Yet a weak K line or weak $\lambda$4481 line alone cannot confirm a $\lambda$\,Boo classification, since Am (or hot Am) stars can also have weak K lines, and shell stars have some features in common with $\lambda$\,Boo stars. Good reference lines of iron, e.g. \specline{Fe}{i}{4383}, are required for comparison with $\lambda$4481, and these are weak in late B stars. On the other hand, at the hot boundary, \ion{He}{i} lines provide additional constraints on the temperature type (and to a lesser extent, the luminosity type), allowing the strengths of the metal lines to be evaluated from a firmer footing. In practice, we find that a $\lambda$\,Boo classification can be considered at spectral types as early as A0.}

\blu{At later spectral types, i.e. for the F stars, different complications obfuscate the classification. One example is the population of Field Horizontal Branch (FHB) stars. The metal weakness seen in FHB stars cannot be adequately distinguished from those of $\lambda$\,Boo stars, unless abundances for the volatile elements C, N and O can be measured. While volatile elements are roughly solar in abundance for $\lambda$\,Boo stars, they share underabundances similar to the refractory elements in FHB stars \citep[see, e.g.,][]{takeda&sadakane1997}. As such, without the abundances of volatile elements, stars exhibiting $\lambda$\,Boo features and having hydrogen line types of F1 or later are usually called `late $\lambda$\,Boo candidates' and their spectral types are assigned with uncertainty, e.g. ``F2\,V\,kA5mA5~$\lambda$\,Boo?''. Unfortunately, access to lines of the desired volatile elements in the classical blue-violet spectral region is poor, and UV or red spectra are required for this. Another difficulty in early F stars is that the \specline{Mg}{ii}{4481} line becomes increasingly blended with Fe, rendering line ratios of, e.g., $\lambda4481$ to \specline{Fe}{i}{4383}, less useful.}

\blu{\subsubsection{Ages of $\lambda$\,Boo stars}}

Discussions of $\lambda$\,Boo candidates in the literature are not only limited to properties that establish their membership in the class. As an example, the age of $\lambda$\,Boo stars was uncertain for a long time, in that modellers did not know whether to use main sequence or pre-main sequence evolutionary tracks for them. \blu{We do not comment on the age of individual stars since this does not aid the arbitration of membership within the $\lambda$\,Boo group, but interested readers can refer to the work of \citet{iliev&barzova1995}, who determined ages for definite $\lambda$\,Boo stars (as classified here and elsewhere), and the review in the introduction of \citet{paunzenetal2002b} along with references therein. It has been shown that} $\lambda$\,Boo stars have ages spanning from the zero- to the terminal-age main sequence (ZAMS and TAMS, respectively), and HAeBe stars identified as $\lambda$\,Boo stars almost certainly extend that range to some of the pre-main sequence phase. For convenience, we make note of the availability (typically from \citealt{iliev&barzova1995}) of `key physical parameters', which include mass, effective temperature, radius, surface gravity, luminosity and age.

\blu{\subsubsection{Pulsation in $\lambda$\,Boo stars}}
The pulsational properties of some candidates have also been presented by various authors. \blu{In favourable circumstances, asteroseismology can distinguish between global and surface depletions of metals. The requirement of excellent asteroseismic data has led to only a few determinations being made in this way. The available pulsational information is generally insufficient for arbitrating membership in the $\lambda$\,Boo class, and not all stars have been monitored for oscillations, hence we have not generally included this information in our descriptions in \S\,\ref{sec:evaluations}. We leave the assessment of pulsations as future work.} This document hence serves as a platform for future, homogeneous studies.

\blu{\subsubsection{Stellar rotation}}
Rotation is an important property, in that rotation broadens spectral lines. At classification resolution (1.8\AA\ per 2\,px, or R$\sim$2500 at 4500\,\AA), a very fast rotator having $v\sin i > 200$\,km\,s$^{-1}$ can mimic a $\lambda$\,Boo star because the metal lines become broader and shallower. If they are sufficiently broad they can blend and appear to be part of the continuum, and hence the star appears metal weak or $\lambda$\,Boo-like. One must be very cautious in accepting a $\lambda$\,Boo star if it has been noted to have a very high $v\sin i$. In \S\,\ref{sec:table} we tabulate the rotational velocity for $\lambda$\,Boo candidates for which it has been measured.

\blu{\subsubsection{Binarity and the `composite spectrum hypothesis'}}
A discussion of binarity among $\lambda$\,Boo stars warrants a short digression. At the turn of the century, a large amount of effort was being put into finding new $\lambda$\,Boo stars while a complete understanding of the origin of their abundance anomalies was still lacking. In a series of papers, Faraggiana, Bonifacio and Gerbaldi attempted to demonstrate that the $\lambda$\,Boo stars were a ``non-homogeneous'' group and that a large number of them could be shown to be binaries (\citealt{faraggiana&bonifacio1999}; \citealt{faraggianaetal2001a}; \citealt{faraggianaetal2001b}; \citealt{faraggiana&gerbaldi2003}; \citealt{gerbaldi&faraggiana2004}). Their suggestion that the class was heterogeneous was in disagreement with \citet{paunzenetal2002a}, who showed the class to be quite homogeneous. Furthermore, the insistence that binaries be dropped from the class was in direct contradiction with the results of \citet{ilievetal2001,ilievetal2002} who showed that for two separate $\lambda$\,Boo binary systems, each member of each system (that is, all four stars) could be demonstrated to be true $\lambda$\,Boo stars. Given the importance of binary systems in providing fundamental stellar parameters such as mass and radius, the suggestion to drop them  would only hinder progress in $\lambda$\,Boo star research.

Still, the ``binary hypothesis'' -- that binarity was responsible for mimicking metal weakness -- was pursued. \citet{faraggiana&bonifacio2005} demonstrated that the characteristic solar abundance of C, N, and O in $\lambda$\,Boo stars arises naturally if a binary system is confused as a single star and labelled a $\lambda$\,Boo star. Conversely, \citet{stutz&paunzen2006} synthesised 105 hypothetical binary systems and compared the synthesised spectra with those of known $\lambda$\,Boo stars. Their results indicated that for about 90\:per\:cent of the group members, the spectroscopic binary hypothesis could not explain the observations.

The binary hypothesis requires that the radial velocity difference between the two stars in the binary causes the continuum of one star to `veil' the lines of the other, implying that the radial velocity difference should be comparable to the $v\sin i$ of the stars. \citet{griffinetal2012} sought to test the veiling mechanism and found it could produce the appearance of metal-weakness by only a factor of about 2, i.e. $[M/H] = -0.3$\,dex; it certainly could not produce metal weaknesses on the order of $-2.0$\,dex that is seen in some $\lambda$\,Boo stars.

Furthermore, \citet{faraggianaetal2004} declared many stars to have `composite spectra' based on variable radial velocities, i.e. by association in an SB1 system. Their variable radial velocities were taken from comments of $V$ or $V?$ from the Bright Star Catalogue (BSC; \citealt{warren&hoffleit1987}). \citet{griffinetal2012} demonstrated that out of a sample of 12 stars that were $\lambda$\,Boo candidates and were classified as having variable RVs in the BSC, 11 did not have any appreciable RV variability in their long-term monitoring programme. Further, they found that RV measurements of A stars ``can depend on the spectral region as well as on the technique adopted and the quality of the spectrum.'' This work abolished the composite spectrum binary hypothesis. We also note that the radial velocity variations of a pulsating ($\delta$\,Sct) star can reach several km\,s$^{-1}$ \citep{bregeretal1976}, and that the $\lambda$\,Boo stars are predicted to be less stable against pulsation than normal A stars of the same spectral type \citep{murphy2014}.

In conclusion, we do not consider binarity to be a good reason to drop a star from the $\lambda$\,Boo class, especially if the binarity is classified in the form of a `variable radial velocity' or `SB1' system. Since in A-star spectra it is not easy to distinguish between genuine line-doubling and rotational broadening if the rotational velocity is high -- around 120\,km\,s$^{-1}$ or more -- only `SB2' classifications of binarity are meaningful in this regard, and only then when they have been determined with spectra of high resolution and high signal-to-noise.

\red{None the less, it is important to be aware of close companions (closer than 2$''$) that may contaminate CCD photometry or spectroscopy. We have therefore checked each candidate for close companions in the Washington Double Star catalogue \citep[WDS,][]{masonetal2001}, and noted any matches. The Catalogue of Components of Double and Multiple stars \citep[CCDM,][]{dommagnet&nys2002} was also used for confirmation.}

\subsection{Document layout}

In \S\,\ref{sec:evaluations} we evaluate the membership of each $\lambda$\,Boo candidate. The format of \S\,\ref{sec:evaluations} is a list of stars ordered by increasing HD number, with one eponymous subsection per star\blu{\footnote{\blu{$\lambda$\,Boo candidates commonly known under other names have those names appended to the subsection title in parentheses.}}}, in which the $\lambda$\,Boo status of that star is investigated. New observations are recorded individually there. In \S\,\ref{sec:table} we summarise the properties of the candidates and the recommendations made for them. A reference list is provided at the end.

%
%
\section{Notes on members of the $\lambda$\,Boo class}
\label{sec:evaluations}

\subsection{HD\,3}
\citet{abt&morrell1995} classified this star as A0\,Vn(Lam Boo) with $v\sin i=210$\,km\,s$^{-1}$, raising the concern this might just be a rapid rotator. There is little support for $\lambda$\,Boo membership in the literature. Spectrum required.
\textbf{Recommendation:} uncertain member.

\subsection{HD\,319 (HR\,12)}
Identified as a $\lambda$\,Boo star by \citet{abt1984b}, and given the spectral type A1mA2\,Vb\,$\lambda$\,Boo by \citet{gray1988}. Gray's comments were that ``\ion{Ca}{ii} K has the strength of an A1 star, the metallic line spectrum that of an A2 star $\dots$ Hydrogen lines show peculiar profiles with cores having strengths like those of an A7 star, but with a broader, shallower profile, and extraordinarily broad but shallow wings.'' As he also noted, HD\,319 has a common-proper-motion companion 1.9$''$ away that is 5.1\,mag fainter \red{\citep[WDS,][]{masonetal2001}}. The star falls mildly among the $\lambda$\,Boo group (away from the line of normal stars) in the \citet{maitzen&pavlovski1989b} $\Delta a$ photometry study. Accepted as a $\lambda$\,Boo star in the \citet{rensonetal1990} catalogue. \citeauthor{stuerenburg1993}'s (\citeyear{stuerenburg1993}) abundance analysis shows a clear $\lambda$\,Boo pattern, with a $0.75$-dex underabundance of metals. Key physical parameters are available \citep{iliev&barzova1995}. \citet{holweger&rentzsch-holm1995} found no signature of circumstellar material around the \ion{Ca}{ii} K line of this star. \citet{paunzenetal1999a} confirmed near-solar [C/H] and [O/H] NLTE abundances. \citet{faraggianaetal2004} recorded this star as a suspected double star, based on the aforementioned common-proper-motion companion, but in reality the separation and magnitude difference are too great to cause a false positive $\lambda$\,Boo detection.
\textbf{Recommendation:} member.

\subsection{HD\,2904}
Classified as A0\,Vnn(Lam Boo) by \citet{abt&morrell1995}, with $v\sin i = 225$\,km\,s$^{-1}$. The rapid rotation calls into question the $\lambda$\,Boo classification. \citet{haucketal1998} found interstellar \ion{Ca}{ii} K line absorption, but there is no real evidence in favour of membership into the $\lambda$\,Boo class. Spectrum required.
\textbf{Recommendation:} uncertain member.

\subsection{HD\,4158}
Long-known $\lambda$\,Boo star \citep{graham&slettebak1973}. The \citet{maitzen&pavlovski1989b} $\Delta a$ photometry study gives a strong indication that this is a $\lambda$\,Boo star, which agrees well with a UV assessment \citep{faraggianaetal1990}. We have also inspected the IUE spectrum available for this star. There is not enough flux at 1600\,\AA\ to see any depression, and the 1657/1670 ratio also suffers from lack of flux, but at longer wavelengths the overall character is metal weak, but the $\lambda1937$ C line is strong. This is consistent with a $\lambda$\,Boo classification. [C] abundance is $-0.16$ \citep{paunzenetal1999a}. We classified a new spectrum of this star as F2\,V\,kA1mA1~$\lambda$\,Boo?, with the comment that \specline{Mg}{ii}{4481} is almost absent. Given the UV characteristics and near-solar C abundance, we can accept this as a $\lambda$\,Boo star.
\textbf{Recommendation:} member.

\subsection{HD\,5789}
\citet{haucketal1998} found interstellar \ion{Na}{i} D lines and \ion{Ca}{ii} K line absorption. \citet{abt&morrell1995} classified the star as B9.5\,Vnn(Lam Boo); the `nn' classification raises concerns. Nevertheless, \citet{gerbaldietal2003} found no reason to reject the star, \red{even though it is a spectroscopic binary and has $v\sin i = 300$\,km\,s$^{-1}$ \citep{slettebak1963}.} The evidence favours exclusion.
\textbf{Recommendation:} non-member.

\subsection{HD\,6173}
Among the $\lambda$\,Boo population in the \citet{maitzen&pavlovski1989} $\Delta a$ photometry study. Listed as a $\lambda$\,Boo star in the \citet{rensonetal1990} catalogue. \citet{gerbaldietal2003} found no reason to reject this star. Classification of as A0\,Vn by \citet{abt1984} and A0\,IIIn by \citet{paunzenetal2001}. Photometry suggests a late A star (roughly A7) according to both $\beta$ and $b-y$. Even if the star is a giant, $b-y$ is not strongly influenced by luminosity so the agreement suggests $\sim$A7. The $c_1$ index is high for a typical A7 star. It is at A7 that the $m_1$ index reaches its maximum, with a value of $\sim0.200$ according to Crawford's (\citeyear{crawford1975,crawford1978,crawford1979}) standard relations, yet this star has $m_1 = 0.079$ \citep{hauck&mermilliod1998} which is much weaker than even a typical A0 star. Normally $m_1$ scales with $c_1$, so that a giant luminosity class has a higher $m_1$ index, hence there is good photometric evidence for metal weakness in HD\,6173. $\lambda$\,Boo membership is therefore not ruled out but a spectrum is definitely needed.
\textbf{Recommendation:} uncertain member.

\subsection{HD\,6870}
Identified as a 15-mmag, metal-weak pulsator \citep{breger1979}, HD\,6870 falls among the $\lambda$\,Boo population in the \citet{maitzen&pavlovski1989b} $\Delta a$ photometry study. It has a clear $\lambda$\,Boo character in the UV \citep{faraggianaetal1990}, and is listed as a $\lambda$\,Boo star in the \citet{rensonetal1990} catalogue. [C/H] and [O/H] are both positive \citep{paunzenetal1999a}. Gray's classification for the \citet{paunzenetal2001} paper reads: ``HD\,6870 (kA2hA7mA2 LB PHL): this star is clearly a lambda Boo''. \citet{gerbaldietal2003} argued this star is misclassified because it has kinematics slightly deviating from those of Pop.\,I stars, and suggested it belongs the thick disc population instead. Indeed, we note a transverse velocity of 60\,km\,s$^{-1}$, easily putting it in the top decile of stars considered in this work, and note that it is called an SX\,Phe star and an `old-disc' star in the older literature. However, the solar [C/H] and [O/H] argue against a generic metal weakness expected from thick disc stars, and \citet{paunzenetal2014} warn against using velocities in $\lambda$\,Boo membership assessments. \citet{paunzenetal1999a} found discrepant $v\sin i$ values between the C and O lines; the lower (128\,km\,s$^{-1}$) agrees with the earlier value of \citet[][130\,km\,s$^{-1}$]{rodgers1968}.
\textbf{Recommendation:} member.

\subsection{HD\,7908}
Classified as hF0mA3\,LB by \citet{paunzenetal2001}, for which Gray commented that the spectrum was `clearly $\lambda$\,Boo'. It was accepted as a bona-fide $\lambda$\,Boo star in the results paper of that series \citep{paunzen2001}. \citet{gerbaldietal2003} found no reason to reject this star from the $\lambda$\,Boo class.
\textbf{Recommendation:} member.

\subsection{HD\,9100}
This star is listed by \citet{abt&morrell1995} as an A2.5\,V star (with $v\sin i = 110$\,km\,s$^{-1}$). It had been earlier classified as A3\,Vb $\lambda$\,Boo PHL? by \citet{gray&garrison1989b}. To be specific, those authors said ``The hydrogen lines of this star show very broad, but rather shallow, wings and are reminiscent of the hydrogen lines of the PHL $\lambda$\,Boo stars of \citet{gray1988}. The $\beta$ index of this star is surprisingly low for hydrogen lines with such broad wings, a characteristic which also recalls the PHL $\lambda$\,Boo stars. The $\lambda$4481 line is slightly, but not obviously, weak. In the $b-y$, $m_1$ plane, this star lies within the $\lambda$\,Boo `distribution'.'' We obtained two spectra of this star with two different instruments, both of which we (independently) classified as A3\,IVs~(4481-wk). In both spectra the \specline{Mg}{ii}{4481} line is quite weak, but the other metal lines show weaknesses of no more than half a spectral subclass (i.e.\ A2.5). This classification differs considerably from the classification of \citet{gray&garrison1989b}. Could this star be a spectrum variable? It is noteworthy that it is a $\delta$\,Sct star with 20-mmag light variation, and was at one time the bluest $\delta$\,Sct star known \citep{breger1969a}. There are many papers on its $\delta$\,Sct nature.
\textbf{Recommendation:} non-member.

\subsection{HD\,11413}
First identified by \citet{abt1984b} and later classified by \citet{gray1988} as `A1\,Va $\lambda$\,Boo', whereupon it was described as having \ion{Ca}{ii} K line and metallic-line types ``intermediate to the A0 and A1 standards, with weak $\lambda4481$. Hydrogen lines show broad, but rather shallow wings, but peculiarly weak cores, similar to the cores of an F0 star.'' This star lies among the $\lambda$\,Boo population in the \citet{maitzen&pavlovski1989b} $\Delta a$ photometry study, and appears in the \citet{rensonetal1990} $\lambda$\,Boo catalogue. The abundance analysis carried out by \citet{stuerenburg1993} shows solar [C/H] but most metals are between $-1.0$ and $-1.5$\,dex. Key physical parameters are available \citep{iliev&barzova1995}. \citet{holweger&rentzsch-holm1995} detected \textit{infalling} circumstellar gas around this star's \ion{Ca}{ii} K line. [C/H] and [O/H] are $-0.25$ and $-0.11$\,dex, respectively \citep{paunzenetal1999a}, which alongside the abundances from \citet{stuerenburg1993} for the metals confirms a $\lambda$\,Boo pattern. \citet{koenetal2003} conducted a dedicated analysis of the pulsational content and stellar parameters from spectroscopy, which confirmed the star's $\lambda$\,Boo nature.
\textbf{Recommendation:} member.

\subsection{HD\,11502}
HD\,11502 ($\gamma^1$\,Ari) and HD\,11503 ($\gamma^2$\,Ari) are in a double system separated by 8.51$''$. The former has been reclassified for this work as A0\,IV-V(n)\,kB8; the latter is an $\alpha^2$\,CVn star (and thus not a $\lambda$\,Boo star), with spectral type `knA0hA3\,(IV)\,SiSr' \citep{gray&garrison1989b}. The spectrum of HD\,11502 is metal weak except for the $\lambda4481$ line, and the H and He lines confirm the temperature subclass A0. It is not a $\lambda$\,Boo star.

However, confusion has arisen in the literature, as indicated by the fact that an essential note on SIMBAD states that HD\,11503 is \textit{not} HR\,545; HR\,545 is the designation of HD\,11502. Such confusion is exemplified in \citet{abt&cardona1984}, and corroborated by the fact that literature values of $v\sin i$ for HD\,11503 are wildly discordant, and much too high for an $\alpha^2$\,CVn star.
\textbf{Recommendation:} non-member.

\subsection{HD\,11503}
See entry for HD\,11502.
\textbf{Recommendation:} non-member.

\subsection{HD\,11905}
Misclassified, according to \citet{rensonetal1990}. Classified as B8\,III by \citet{cowleyetal1969}, and as B9\,HgMn in the \citet{renson&manfroid2009} catalogue. Rejected by \citet{paunzenetal1997a} for the Ap/Bp classification.
\textbf{Recommendation:} non-member.

\subsection{HD\,13755}
\citet{paunzenetal2001} classified this star as hF2mA5\,V~LB. Gray's notes to Paunzen for this star read: `HD\,13755 (kA5hF2mA5\,V\,LB): H-lines and metallic-line morphology match the F2\,V std quite well.  However, metals are clearly weak, at about A5.  Hence, clearly metal-weak. hF2mA5\,V.' We have observed and classified HD\,13755 for this work and give the spectral type F1\,V\,kA9mA6~($\lambda$\,Boo), as thus accept it as a mild member. An abundance analysis should be performed to confirm membership.
\textbf{Recommendation:} member.

\subsection{HD\,15164}
See entry for HD\,15165. \textbf{Recommendation:} member.

\subsection{HD\,15165 (VW\,Ari)}
We note this is a visual triple system. We give component A the spectral type F2\,V\,kA2mA2~$\lambda$\,Boo? (i.e. this is a `late $\lambda$\,Boo candidate', see \S\,\ref{sssec:spectral-range}). Component B is in SIMBAD as HD\,15164, to which we give the spectral type F1\,V\,kA7mA6~($\lambda$\,Boo)?, and component C has the designation HD\,15165C that we classify as a normal K2\,V star. The stars have proper motions that are the same to within one or two mas/yr, so they are a physical triple. \citet{paunzenetal2002b} did an abundance analysis, but did not specify which component was analysed. We therefore assume it was component A, which is the brightest by two magnitudes. They found that C, N, O and S are approximately solar while Mg, Si and Ca are 1-dex underabundant, which is a typical $\lambda$\,Boo abundance pattern. \citet{andrievskyetal1995} published a paper on the composition of components A and B, which include a 2.2- and a 1.8-M$_{\odot}$ star. The secondary is described as being of solar composition but the primary has [Fe/H] of $-0.46$. Their leading hypothesis for the differing composition was stellar capture. The primary, they say, is the $\delta$\,Sct (or SX\,Phe) star and the secondary is non-variable. There are many $\delta$\,Sct-themed papers on this star. We disagree that the secondary can have solar abundances \blu{given the appearance of the classification spectrum}; a self-consistent abundance analysis of all three components, with NLTE abundances for the volatile elements, is highly desired.
\textbf{Recommendation:} member.

\subsection{HD\,16811}
\citet{haucketal1998} noted \ion{Na}{i} absorption for this star, which is of uncertain membership in the \citet{rensonetal1990} catalogue. The \citet{maitzen&pavlovski1989b} $\Delta a$ photometry study shows it to be rather mild, not lying far from the line of normal stars. \citet{gray&garrison1987} classified it as A0\,IVn, with the comment `classified as a marginal $\lambda$\,Boo star by Abt (1984). It does not look peculiar in our one (slightly underexposed) spectrum. Occultations indicate this star is a triple system.'
\textbf{Recommendation:} non-member.

\subsection{HD\,16955}
Very near the line of normal stars in the \citet{maitzen&pavlovski1989b} $\Delta a$ photometry study. \citet{andrillatetal1995} stated `according to all indications, this is a normal star'. A star with circumstellar \ion{Ca}{ii} K line absorption \citep{haucketal1998}. $uvby\beta$ observations \citep{hauck&mermilliod1998} indicate a normal A3 star.
\textbf{Recommendation:} non-member.

\subsection{HD\,16964}
HD\,16964 was recorded as the brighter star in a visual double with HD\,16965 and given a spectral type of A1 by \citet{bidelman1943}. HD\,16964 was classified as an A0\,V~$\lambda$\,Boo star on a 1.2-mm wide spectrum by \citet{abt1988} and is clearly designated there as the fainter of the pair, but by only 0.3\,mag in V. Their separation is 15.8$''$. No details were given on the reason for classification as $\lambda$\,Boo, so we have reobserved both members. We give the spectral type A0\,IV-V to component A, and A0.5\,IVn to component B. The \ion{Ca}{ii} K line in component B is slightly peculiar, but neither component is a $\lambda$\,Boo star.
\textbf{Recommendation:} non-member.

\subsection{HD\,17138 (RZ\,Cas)}
\citet{narusawaetal2006a} determined this eclipsing algol-type binary to have $\lambda$\,Boo-like abundances, and \citet{narusawaetal2006b} presented evidence suggestive of circumstellar matter in the system. \citet{richards&albright1999} previously observed a gas stream for RZ\,Cas. The A3\,V spectral type of the primary of this system is agreed upon in the literature, the secondary being K0\,IV. The abundance analysis of \citet{narusawaetal2006a} showed ``definite under-abundances (by $-0.45$\,dex or more) of Mg, Si, Ti, Cr, and Fe $\dots$ On the other hand, light elements C and O show[ed] solar abundances and Ca [was] slightly under-abundant ($\sim -0.3$\,dex).'' The mild $\lambda$\,Boo abundance pattern favours acceptance, but its existence in a short-period binary (P = 1.1953\,d) raises suspicions. The RZ\,Cas system is well parametrized in, for example, \citet{narusawaetal2006a}.
\textbf{Recommendation:} probable member.

\subsection{HD\,21335}
Classified by \citet{gray&garrison1989b} as an A3\,IVn star with the comment: `Metals may be slightly weak, but $\lambda4481$ is quite normal. Abt (1984) classifies this star as a marginal $\lambda$\,Boo star. High $v\sin i$.'  Falls between the true $\lambda$\,Boo and the normal stars in the \citet{maitzen&pavlovski1989b} $\Delta a$ photometry study. Tagged `LB' in the \citet{andrillatetal1995} IR study of $\lambda$\,Boo stars, with the comment `Ca is only slightly deficient', indicating no sign of a shell. \citet{gerbaldietal2003} found no cause for rejection. Likely a composite spectrum according to \citet{faraggianaetal2004} because Speckle interferometry indicated an object separated by less than 0.1$''$ but with no magnitude difference recorded. \red{The 0.1$''$ separation is confirmed in the WDS catalogue \citep{masonetal2001}}.
\textbf{Recommendation:} non-member.

\subsection{HD\,22470}
Rejected by \citet{gray1988}. \blu{Has a mutually exclusive spectral type: B9\,Si \citep{renson&manfroid2009}. First appeared in papers on magnetic chemically peculiar stars four decades ago \citep{landstreetetal1975}, and its average surface field strength is measured to be 2350\,G \citep{glagolevskij2011}.} \red{Has a companion at 0.2$''$ that is 0.4\,mag fainter \citep[WDS,][]{masonetal2001}.}
\textbf{Recommendation:} non-member.

\subsection{HD\,23258}
First classified as $\lambda$\,Boo by \citet{abt&morrell1995} (A0\,Vp(Lam Boo)), but was described as a `borderline case' by \citet{paunzen&gray1997}, ultimately being classified as `A0\,Vb\,(sl wk metals)'. \citet{paunzenetal2001} put it on a list of stars that were `not classical $\lambda$\,Boo stars', but of which some `may be metal weak'. It did not make the \citet{paunzen2001} `new and confirmed' list of $\lambda$\,Boo stars. \citet{andrievskyetal2002} found `the typical abundance pattern (C and O solar whereas Mg, Si, Cr and Fe are moderately underabundant)'.
Listed as a mild $\lambda$\,Boo star on Gray's website, with spectral type A3\,V\,kB9.5mB9.5\,(Boo).
\textbf{Recommendation:} member.

\subsection{HD\,23392}
Presented as a new $\lambda$\,Boo star with a spectral type A0\,Va$^{-}$ ($\lambda$\,Boo) by \citet{paunzen&gray1997}, with the comment `mild $\lambda$ Bootis star, hydrogen lines show broad wings.' We reclassified this star for this work, giving the spectral type A0\,Va$^{-}$\,kB8.5~($\lambda$\,Boo), and note that the \ion{He}{i} and H-lines clearly fix the spectral type at A0, but the $\lambda$4481 and K-line are weak. Photometrically constant at the 2-mmag limit \citep{paunzenetal1998c}. \citet{gerbaldietal2003} called this star's UV flux `inconsistent' because it was too high for them to fit, and included the note `bin?', implying they suspected the star of binarity, but \citet{marchettietal2001} found no evidence of binarity in their speckle interferometry survey.
\textbf{Recommendation:} member.

\subsection{HD\,24712}
Misclassified according to \citet{rensonetal1990}. Recorded as an $\alpha^2$\,CVn star on SIMBAD, with spectral type `A9\,Vp SrEuCr' \citep{abt&morrell1995}, necessitating rejection.
\textbf{Recommendation:} non-member.

\subsection{HD\,24472}
Classified as `hF2mA5\,V~LB' \citep{paunzenetal2001, paunzen2001}. \citet{gerbaldietal2003} noted `inconsistent UV flux (bin?)'. An uncertain late $\lambda$\,Boo star on Gray's website, \citet{gray1989} has previously classified it as F3\,V~m-2, and in his notes to Paunzen on the aforementioned classification, Gray wrote ``Is it a Lambda Boo, or a Pop.\,II?'' NLTE abundances of volatile elements would be required to confirm.
\textbf{Recommendation:} probable member.

\subsection{HD\,26801}
An uncertain member of the class according to the \citet{rensonetal1990} catalogue, owing to a classification by \citet{abt1985} of `A0\,III lambda Boo?'. The `doubtful' infrared IRAS excess described by \citet{king1994} \red{is most likely associated with a neighbour at 38$''$.}
No strong support for $\lambda$\,Boo membership in the literature.
\textbf{Recommendation:} non-member.

\subsection{HD\,27404}
Appears in Hauck's (\citeyear{hauck1986}) `Search for $\lambda$\,Boo candidates' paper, but is listed with a classification of Ap~Si. \citet{rensonetal1990} rejected this star from their catalogue. \citet{renson&manfroid2009} gave a spectral type A0\,Si. \blu{It is a known magnetic chemically peculiar star, with longitudinal field strength on the order of a kG \citep{kudryavtsevetal2006}.} Definitely not a $\lambda$\,Boo star.
\textbf{Recommendation:} non-member.

\subsection{HD\,30422}
First identified as a $\lambda$\,Boo star by \citet{gray&garrison1989b}, with the spectral type A3\,Vb\,$\lambda$\,Boo. Key physical parameters are available \citep{iliev&barzova1995}. NLTE abundances for [C/H] and [O/H] are $-0.27$ and $-0.25$, respectively \citep{paunzenetal1999a}. \citet{gerbaldietal2003} found no cause for rejection, but \citet{faraggianaetal2004} called it a `probable double' implying a composite spectrum, based on the radial velocities of \citet{grenieretal1999}; we see no evidence for binarity there, and note that only `SB2' would be convincing evidence of binary contamination. We reobserved this target and classify it as A7\,V\,kA3mA3~($\lambda$\,Boo).
\textbf{Recommendation:} member.

\subsection{HD\,30739}
Spectral type of A0\,Vp(Boo)n in \citet{abt&morrell1995}. Comment of `inconsistent UV flux (solar ab.)' in \citet{gerbaldietal2003}. Classified as a normal star (A0.5\,IVn) by \citet{gray&garrison1987}. No real evidence in favour of $\lambda$\,Boo membership.
\textbf{Recommendation:} non-member.

\subsection{HD\,31293}
\citet{gray&corbally1998} classified HD\,31293 (AB\,Aur) as `A0\,Vaer\,Bd$<$\,Nem5' but made no mention of $\lambda$\,Boo properties and were explicitly looking for young $\lambda$\,Boo stars. Identified by \citet{acke&waelkens2004} to have deficient Fe and enhanced N and O. \citet{folsometal2012} called the \citet{acke&waelkens2004} abundance analysis into question, by pointing out the remarkably high $\log g$ value of $5.0$ that was used. \citet{folsometal2012} favoured a more solar-like Fe abundance. They also provided key physical parameters. The star is a dusty HAeBe star, with an unconfirmed planetary system \citep{oppenheimeretal2008, hashimotoetal2011}. We have eight spectra, and the \specline{Mg}{ii}{4481} line is prominent in all of them. It is a spectrum variable, but at no phase does it appear like a $\lambda$\,Boo star. In one of our spectra it looks like a shell star, whereas in the others the hydrogen lines are in emission.
\textbf{Recommendation:} non-member.

\subsection{HD\,31295}
One of the first $\lambda$\,Boo stars identified \citep{slettebak1954}. Shown to have small space velocities \citep{hauck&slettebak1983}, and particularly strong $\lambda$\,Boo characteristics in the UV \citep{bascheketal1984} -- equivalent to or stronger than $\lambda$\,Boo itself. The UV properties were later confirmed by \citet{faraggianaetal1990}. \citet{sadakane&nishida1986} reported a 60\,$\upmu$m excess, and included it in a list of Vega-like stars, implying the presence of cool circumstellar material. \citet{gray1988} included it in his list of $\lambda$\,Boo stars, with the spectral type A0\,Va\,$\lambda$\,Boo, and the comment: ``Very weak K line and $\lambda$4481. Hydrogen lines are strong, like those of an early A type dwarf, and show broad, well-developed wings. Weak metallic lines are visible at [\ion{Fe}{i}] $\lambda$4046 and [\ion{Fe}{i}] $\lambda$4064.'' In this work, we re-observed the star and give the classification kA0hA3mA0\,Va$^{-}$~$\lambda$\,Boo. HD\,31295 falls mildly off the line of normal stars (towards $\lambda$\,Boo stars) in the \citet{maitzen&pavlovski1989} $\Delta a$ photometry study. Accepted as a definite member of the $\lambda$\,Boo class in the \citet{rensonetal1990} catalogue. \citet{iliev&barzova1993b} inspected the hydrogen lines of this and other $\lambda$\,Boo stars, and proclaimed HD\,31295 is an `excellent example of a $\lambda$ Bootis type star with normal hydrogen lines.' \citet{stuerenburg1993} confirmed a $\lambda$\,Boo abundance pattern, with $-1$\,dex metals, which was supported by later NLTE abundance analyses that showed almost solar C, N, O and S \citep{paunzenetal1999a,kampetal2001} and by an LTE analysis of metals \citep{paunzenetal1999b}. Key physical parameters are available \citep{iliev&barzova1995}. \citet{gerbaldietal2003} found no reason to reject this star, but \citet{faraggianaetal2004} called it a probable double based on a variable RV that could not be traced in the cited text. This is one of the best-studied, bona-fide $\lambda$\,Boo stars.
\textbf{Recommendation:} member.

\subsection{HD\,34787}
\blu{\citet{paunzenetal1997a} rejected HD\,34787 because it did not meet the UV criteria of $\lambda$\,Boo stars in the IUE analyses by \citet{bascheketal1984} and \citet{faraggianaetal1990}. However, according to \citet{gerbaldietal2003}, HD34787 was never mentioned in either of those two papers and has never been observed by the IUE satellite. Thus \citet{gerbaldietal2003} rejected this star due to former misclassification, whereas they should have reobserved and/or reclassified the star based on the available literature.}

\citet{abt&morrell1995} gave the spectral type B9.5\,Vp(Lam Boo)n, but we have reobserved the star and give the spectral type A0\,IIIn. We think the rapid rotation gave the appearance of weakness in the 4481 line that  \citet{abt&morrell1995} detected. Evidence for an expanding gas shell was presented by \citet{haucketal1998}, which suggests the weak 4481 line may also have a shell origin, though we did not see this ourselves.
\textbf{Recommendation:} non-member.

\subsection{HD\,34797}
Proposed by \citet{slettebak1963} as an uncertain member of the $\lambda$\,Boo group. Membership can be excluded on the grounds of its $\alpha^2$\,CVn nature, and having a spectral type `B7\,Vp\,He\,wk' \citep{abt&cardona1983}. Meets none of the UV criteria for membership in \citet{faraggianaetal1990}. Listed as misclassified in the \citet{rensonetal1990} catalogue.
\textbf{Recommendation:} non-member.

\subsection{HD\,35242}
Although already classified as A1\,Vp(4481 weak) by \citet{abt&morrell1995}, the $\lambda$\,Boo nature of the object was first noted in the \citet{paunzenetal2001} survey, for which Gray's notes on the spectrum read: ``This is clearly a Lambda Boo star$\dots$ H-lines are an excellent match to Bet Leo [A3\,Vas], but metals are clearly weaker, including $\lambda$4481. Metals are closest to A0.5; K-line is about A1.'' The spectral type thus assigned was A3mA1\,Va~Lam\,Boo.
\textbf{Recommendation:} member.

\subsection{HD\,36496}
A mild $\lambda$\,Boo star according to \citet{abt1984}, but \citet{abt&morrell1995} classified it as `A5\,Vn'. Its distance from the line of normal stars in the $\Delta a$ photometry study of \citet{maitzen&pavlovski1989} is also consistent with mild $\lambda$\,Boo peculiarity. No evidence for a shell in the IR \citep{andrillatetal1995}. \citet{faraggianaetal2004} listed the star as having a composite spectrum; \red{in the WDS this object is one of a pair of stars separated by 0.2$''$ with a magnitude difference of 1.0}. $uvby$ photometry is consistent with a star around A8 with metals around A1. A new spectral classification is needed.
\textbf{Recommendation:} uncertain member.

\subsection{HD\,36726}
A new $\lambda$\,Boo star in the \citet{paunzen&gray1997} survey paper, where its spectrum is described as being very similar to HD\,31295 -- one of the best studied $\lambda$\,Boo stars. They gave a spectral type of `kA0hA5mA0\,V $\lambda$\,Boo'. The abundances show solar C and O, but underabundances of refractory elements of between 0.5 and 1.0\,dex \citep{andrievskyetal2002}. We reobserved this target and classified it as hA4\,Vb\,kA0.5mA0.5~$\lambda$\,Boo, noting the particularly broad H wings.
\textbf{Recommendation:} member.

\subsection{HD\,37411}
A Herbig Ae star, and dubbed a ``fully-fledged $\lambda$\,Boo star'' by \citet{gray&corbally1998}, where it is given the spectral type `kA0hA3mA0\,Va(e)~$\lambda$\,Boo' and is discussed at length. We also obtained a new spectrum for this work and give the classification hA2\,Vae\,kB8mB8~$\lambda$\,Boo, noting clear emission in H\,$\beta$.
\textbf{Recommendation:} member.

\subsection{HD\,37886}
This star was described in a paper on very young (1.7\,Myr) HgMn stars \citep{woolf&lambert1999}, and it is this classification that leads us to reject the star.
\textbf{Recommendation:} non-member.

\subsection{HD\,38043}
Lies among the $\lambda$\,Boo population in the $\Delta a$ photometry study of \citet{maitzen&pavlovski1989b}, and is a $\lambda$\,Boo star according to the \citet{rensonetal1990} catalogue. It is recorded in the first volume of the Michigan Catalog \citep{houk&cowley1975} as `hF0kA4/7 metals very weak or invisible'. The discrepancy in the $b-y$ and $m_1$ indices \citep{eggen1984} amounts to metal weakness of an entire spectral class. \citet{paunzenetal2005} did not label the star as $\lambda$\,Boo because it was classified earlier as a blue giant. We observed this star and give the spectral type F1\,V\,kA5mA3~$\lambda$\,Boo?. The 4481 line is very weak, but in early F stars this is blended with lines of Fe (see \S\,\ref{sssec:spectral-range}), and so an NLTE abundance analysis showing solar abundances of C, N, O and/or S is required to distinguish it from Pop.\,II stars. We do note that radial velocity and proper motions are very small, and thus tentatively accept it as a $\lambda$\,Boo star for now.
\textbf{Recommendation:} member.

\subsection{HD\,38545 (HR\,1989; 131\,Tau)}
Identified as a $\lambda$\,Boo star by \citet{gray&garrison1987}, with a classification `A2\,Va+\,$\lambda$\,Boo', and thus included as a $\lambda$\,Boo star by \citet{gray1988}. It does not lie far from the line of normal stars in the \citet{maitzen&pavlovski1989b} $\Delta a$ photometry study. An abundance analysis \citep{stuerenburg1993} shows very mild metal weakness; at around $-0.5$\,dex, the abundances are only 1--2\,$\sigma$ from solar for most refractories. Key physical parameters are available \citep{iliev&barzova1995}. Its $\lambda$\,Boo nature has been discussed by \citet{bohlenderetal1999}, who reviewed the literature and suggested it may instead be an Ae or shell star, though we note that those classes do not \textit{ipso facto} exclude membership in the $\lambda$\,Boo group. Indeed, \citet{bohlender&walker1994} considered it a $\lambda$\,Boo star with a shell. It was included in the \citet{paunzenetal1997a} consolidated catalogue of $\lambda$\,Boo stars but was later dropped \citep{paunzen2000} because of recently detected binarity. \red{The WDS catalogue \citep{masonetal2001} lists a companion at 0.1$''$ that is only 0.5\,mag fainter.} The star's rapid rotation (191\,km\,s$^{-1}$; \citealt{royeretal2007}) may also have had some role in early identifications as a $\lambda$\,Boo star, especially when combined with mild metal weakness. Our cursory inspection of an IUE UV spectrum shows that this star is normal, according to the UV criteria outlined by \citet{faraggianaetal1990}. The evidence argues against a $\lambda$\,Boo classification.
\textbf{Recommendation:} non-member.

\subsection{HD\,39283}
One of the first $\lambda$\,Boo stars recorded \citep{slettebak1952}, but it is actually a normal star -- classified as `A1\,Va' \citep{gray&garrison1987}, it lies on the line of normal stars in the $\Delta a$ photometry study of \citet{maitzen&pavlovski1989} and according to \citet{faraggianaetal1990} the UV character of the star does not warrant inclusion in the $\lambda$\,Boo group. They added that it might have `some slight Al deficiency' and that it is a `possible weak-line star'. It was thus given a misclassified (`/') designation in the \citet{rensonetal1990} catalogue. Strong evidence for a shell was presented by \citet{andrillatetal1995}, who also called the star $\lambda$\,Boo but possibly only in reflection of the literature. \citet{gerbaldietal2003} wrote `inconsistent UV flux (solar ab.)', thus suggesting rejection.
\textbf{Recommendation:} non-member.

\subsection{HD\,39421}
Falls among the $\lambda$\,Boo population in the $\Delta a$ photometry study of \citet{maitzen&pavlovski1989b}. Described as an A2\,Vp star with strong \ion{O}{i} and \ion{S}{i} star in the IR study of \citet{andrillatetal1995}. Classified as A1\,Va (wk 4481) by \citet{paunzenetal2001}, but not as a $\lambda$\,Boo star, even though the star was included in the consolidated list of \citet{paunzenetal1997a}. \citet{abt&morrell1995} also noted the $\lambda4481$ weakness. $uvby\beta$ photometry \citep{hauck&mermilliod1998} indicates a hydrogen type about A4 from both $\beta$ and $b-y$, for which the $m_1$ index of 0.161 is only slightly small. UV spectrum analysis shows a $\lambda$\,Boo nature (this work). However, our re-investigation of the star's optical spectrum yielded the spectral type A1\,Vn, i.e.\ not $\lambda$\,Boo. Earlier detections of a weak 4481 lines might be due to the rapid rotation. In light of the mixed evidence, we leave the membership as uncertain. \textbf{Recommendation:} uncertain member.

\subsection{HD\,40588}
\citet{abt&morrell1995} noted a weak $\lambda4481$ line in this star, and \citet{paunzenetal2001} classified it as `A1\,Va (sl weak 4481)' but neither paper refers to the star as a $\lambda$\,Boo star. \citet{andrievskyetal2002} confirmed membership in the $\lambda$\,Boo group with an LTE abundance analysis, showing approximately solar C and O, but deficient Mg and $-1$\,dex abundances of Fe and Cr. We reobserved this target and give the spectral type A3\,V\,kA0.5mA0~$\lambda$\,Boo. \textbf{Recommendation:} member.

\subsection{HD\,41580}
Classified `A0\,Vp lambda Boo' by \citet{abt1985}, and appears in the \citet{rensonetal1990} catalogue. Not suggested for rejection by \citet{gerbaldietal2003}. No real evidence in favour of inclusion into the class, and our new observations indicate the spectral type A1\,IIIp~Si.
\textbf{Recommendation:} non-member.

\subsection{HD\,42503}
An A2\,V $\lambda$\,Boo star \citep{paunzenetal2002a}. $v\sin i$ was not determined, but $\log g = 3.1$ is very low for a $\lambda$\,Boo star. $[Z]$ was given as $-0.83\pm0.20$. Handler's (\citeyear{handler1999}) $uvby\beta$ photometry gives consistent hydrogen line strengths from $b-y$ and $\beta$ that suggest an $\sim$A8 dwarf, with an $m_1$ index about as small as an A0 star. We observed this star and give the spectral type hA9\,Vn\,kA2mA2~$\lambda$\,Boo.
\textbf{Recommendation:} member.

\subsection{HD\,47152}
The essential notes on SIMBAD for this object, 53\,Aur, indicate it is a $\lambda$\,Boo star. This is incorrect. It was classified as a $\lambda$\,Boo star by \citet{abt&morrell1995} but later dismissed by \citet{haucketal1998} because of its "well-known" identification as an Hg star (they referenced \citealt{osawa1965}). \citet{zverkoetal2008} revisited this system to give a definitive classification. They determined B9\,Mn for the primary, and F0m for the secondary, i.e., neither object is a $\lambda$\,Boo star.\footnote{A recommendation has been made that the SIMBAD administrators remove the note that states this star is a $\lambda$\,Boo star.}
\textbf{Recommendation:} non-member.

\subsection{HD\,54272}
Classified by \citet{paunzenetal2001} as kA3hF2mA3\,V~LB and included in the results paper of that series with the comment ``very metal-weak star'' \citep{paunzen2001}. No $\delta$\,Sct pulsational variability at the 1.4-mmag level \citep{paunzenetal2002a}. \citet{gerbaldietal2003} noted `inconsistent UV flux (bin.?)'. On Gray's list of late $\lambda$\,Boo candidates, though in his notes to Paunzen, Gray writes ``FBS?? (This is one of Olsen's FBS).'' [FBS = field blue straggler.] Proper motions (0.8 and $-49.3$\,mas/yr) are large in the declination direction (only).

Previously classified as an RRab pulsator, with 0.1-mag light variation and 0.779-d pulsation period \citep{szczygiel&fabrycky2007, skarka2013}, but recently overturned in favour of $\gamma$\,Dor pulsation by \citet{paunzenetal2014a} who gave a new spectral type of kA3hF2mA3\,V LB, and found $v\sin i$ = $250\pm25$\,km\,s$^{-1}$ and [M/H]$=-0.8$ to $-1.1$\,dex. The quoted $v\sin i = 250$\,km\,s$^{-1}$ is very rapid for an F star. From three spectra taken with two different instruments, we find that HD\,54272 is {\it not} a rapid rotator. \blu{We could not find any literature $v\sin i$ values against which to compare.} Yet the spectrum that \citet{paunzenetal2014a} presented does look broad-lined. We \blu{tentatively} suggest that the spectrum they gave is misidentified.

Our new spectra do show clear metal weakness, but there is no evidence that this is a $\lambda$\,Boo star ($\lambda$4481 is borderline normal, and is blended with Fe in F2 stars; see \S\,\ref{sssec:spectral-range}). We give the spectral type F1\,V\,kA6mA3. Since this is not a rapid rotator, a high-resolution spectrum including lines of volatile elements could provide a definitive membership assessment.
\textbf{Recommendation:} uncertain member.

\subsection{HD\,56405}
This star was given an uncertain designation in the \citet{rensonetal1990} catalogue, presumably because it had only been identified as a $\lambda$\,Boo candidate from Abt's (\citeyear{abt1984b}) work. \citet{gerbaldietal2003} rejected the star because of `inconsistent UV flux (bin.?)' and noted RV variability. Rejected earlier by \citet{paunzenetal1997a} because it was identified by \citet{gray&garrison1987} as a high $v\sin i$ A2\,Va standard.
\textbf{Recommendation:} non-member.

\subsection{HD\,64491}
First classified as $\lambda$\,Boo by \citet{abt&morrell1995}, its $\lambda$\,Boo nature was confirmed by \citet{paunzen&gray1997} who gave the spectral type `kA3hF0mA3\,V $\lambda$\,Boo.' \citet{kampetal2001} derived [N] and [S] abundances of $-0.30$ and $-0.09$, respectively, when considering the star as single. However, in a dedicated paper where this star's literature coverage is documented in detail, \citet{faraggiana&gerbaldi2003} presented evidence that HD\,64491 is not a single object of [M/H] $\sim-1.5$\,dex, but is better explained as two slightly metal poor objects ($\sim-0.5$\,dex) of similar temperature and different $v\sin i$. Their poor temporal coverage leaves their SB2 detection as a borderline one in significance, and depending on the exact metallicity of the two objects, they may have found that this object is composed of two mild $\lambda$\,Boo stars rather than one single one. We obtained two spectra of this star and each has the spectral type F1\,Vs\,kA3mA3~$\lambda$\,Boo.
\textbf{Recommendation:} member.

\subsection{HD\,66684}
Close visual binary, primary is B9\,Va and secondary A1\,IVn \citep{gray&garrison1987}. \citet{abt&morrell1995} classified the B component as A0\,Vp (4481 wk)n and the A component as B9.5 Vp(Lam Boo). \citet{paunzenetal2001} gave the classification as B9.5\,Va (sl wk met) but did not state whether this was the A or B component. LTE abundance analysis revealed underabundance of 1\,dex in Fe, but $\lambda$\,Boo membership could not be ascertained \citep{andrievskyetal2002}. Ironically, \citet{gerbaldietal2003} rejected HD\,66684 because of `discordant $v\sin i$ values', not because of binarity. Our own classification is based on one new spectrum when the seeing was not good enough to resolve the two components (the separation is $\sim2.5''$). We give the spectral type B9.5\,Va~(wk 4481), but we do not consider it a $\lambda$\,Boo star.
\textbf{Recommendation:} non-member.

\subsection{HD\,66920}
This star is allegedly misclassified as a $\lambda$\,Boo star \citep{paunzenetal2002a}, since \citet{paunzenetal2001} gave the non-peculiar classification A3\,V. $\Delta a$ photometry \citep{vogtetal1998} does not support a $\lambda$\,Boo classification.
\textbf{Recommendation:} non-member.

\subsection{HD\,68695}
Spectral type `A0\,V' according to \citet{houk1978}. \citet{folsometal2012} described this Ae star as having `clear $\lambda$\,Boo peculiarities', with C, N, O and S being 1$\sigma$ above solar, and Fe-peak elements being 2--3$\sigma$ below solar. We obtained two spectra at different times for this work, from which we conclude the star is probably a spectrum variable; our best spectral type is A3\,Ve\,kA1mA0, in which we note strong emission in H\,$\beta$, but it does not appear to be a $\lambda$\,Boo star generally. We leave the membership as uncertain, and suggest an investigation of the UV spectrum to shed further light. \textbf{Recommendation:} uncertain member.

\subsection{HD\,68758}
Very rapid rotator ($\geq 270$\,km\,s$^{-1}$); only [Fe/H] could be determined ($-0.58$) by \citet{solanoetal2001}, who also gave a mass of 2.4\,M$_{\odot}$ and an age of 427\,Myr. It was given the spectral type A1\,IVp by \citet{paunzenetal2001}, but the exact peculiarity is not specified. If it were a $\lambda$\,Boo star, that would have been indicated. Probably just a fast rotator.
\textbf{Recommendation:} non-member.

\subsection{HD\,73210}
\citet{pavlovskietal1993} used the $\Delta\,a$ photometry method to find three new $\lambda$\,Boo stars in the Praesepe cluster: KW\,50, KW\,114 and KW\,375, which are HD\,73210, 73345 and 73872. Each of the three stars was refuted later by \citet{gray&corbally2002}. HD\,73210 was rejected because it was a A5\,IIIs standard \citep{gray&garrison1989b}. The other two stars have spectral types A7\,V\,kA8 and kA6hA8mA6\,V(n), respectively \citep{gray&corbally2002}, i.e., none is a $\lambda$\,Boo star.
\textbf{Recommendation:} non-member.

\subsection{HD\,73345}
See HD\,73210.
\textbf{Recommendation:} non-member.

\subsection{HD\,73872}
See HD\,73210.
\textbf{Recommendation:} non-member.

\subsection{HD\,74873}
Identified as a 4481-weak star by \citet{abt&morrell1995}, and considered as a $\lambda$\,Boo star by \citet{kuschnigetal1996a}. \citet{paunzen&gray1997} included this as one of their new $\lambda$\,Boo stars and gave it a more precise spectral type of `kA0.5hA5mA0.5\,V $\lambda$\,Boo.' There is disagreement in the literature over this star's $v\sin i$ value (see \citealt{heiter2002} for a discussion). \citet{heiter2002} determined $v\sin i = 130$\,km\,s$^{-1}$ from high-resolution spectroscopy, and performed an abundance analysis showing moderate (0.5\,dex) underabundances of metals and overabundances of volatiles, i.e., HD\,74873 has a $\lambda$\,Boo abundance pattern. \citet{paunzenetal2003} noticed an infrared excess that they attributed to dust. \citet{gerbaldietal2003} noted variable RVs had been reported for this star, though speckle interferometry had yielded no companion. They reported an inconsistent UV flux, in that solar abundances fit better than $\lambda$\,Boo-like abundances in the UV, which \citet{faraggianaetal2004} interpreted as evidence of a companion (despite one not being found).
\textbf{Recommendation:} member.

\subsection{HD\,74911}
Candidate $\lambda$\,Boo star in \citet{paunzen&gray1997}, where its spectral type was given as A2\,IV (wk 4481). Classified as A3\,IV by \citet{paunzenetal2001}, who were specifically looking for $\lambda$\,Boo stars, but \citet{andrievskyetal2002} `confirmed' its $\lambda$\,Boo membership with an abundance analysis showing metal deficiency ($-0.8$\,dex) alongside a [C] abundance of $-0.25$\,dex. Str\"omgren photometry is consistent with a mid-A star with mild metal weakness. Our new spectrum yields the spectral type A8\,Vn\,kA1mA1, from which we also conclude this star is just metal weak -- it is not a $\lambda$\,Boo star. \textbf{Recommendation:} non-member.

\subsection{HD\,75654}
Serendipitously discovered to be variable by \citet{balona1977} with 11.5\,d$^{-1}$ variability, and later confirmed as a multi-periodic $\delta$\,Sct star by \citet{paunzenetal2002a}. \citet{hauck1986} adopted HD\,75654 as a $\lambda$\,Boo candidate upon noticing it meets the photometric conditions for a $\lambda$\,Boo star -- a fact corroborated by \citet{maitzen&pavlovski1989b} who showed it to lie among the $\lambda$\,Boo stars in a $\Delta a$ photometry. Included as a $\lambda$\,Boo star in the \citet{rensonetal1990} catalogue. [C], [N], [O], and [S] abundances are $-0.44$, $+0.30$, $+0.14$, $-0.10$, respectively \citep{paunzenetal1999a, kampetal2001}, which alongside abundances of metals of $-1$\,dex in table\:4 of \citet{solanoetal2001} support a $\lambda$\,Boo classification. Included in \citeauthor{paunzen2001}'s (\citeyear{paunzen2001}) results paper as a confirmed $\lambda$\,Boo star with spectral type `hF0mA5\,V $\lambda$\,Boo.' Mentioned as `a probable double' by \citet{faraggianaetal2004}. A $\lambda$\,Boo star on Gray's website with spectral type `A7\,kA3mA3\,V $\lambda$\,Boo.'
\textbf{Recommendation:} member.

\subsection{HD\,78316}
Misclassified as a $\lambda$\,Boo star, according to \citet{rensonetal1990}. It is a well known Ap star, with spectral type `B8IIIp\,HgMnEu~MgII\,4481A\,weak' \citep{abt&morrell1995}. \red{The WDS catalogue \citep{masonetal2001} lists a companion at 0.2$''$ that is 3.5\,mag fainter.}
\textbf{Recommendation:} non-member.

\subsection{HD\,78661}
An uncertain member of the $\lambda$\,Boo group \citep{rensonetal1990}, this star was rejected from the \citet{paunzenetal1997a} consolidated catalogue because it is an F2 star. IRAS infrared excess detected \citep{king1994}. \red{According to the WDS catalogue \citep{masonetal2001}, this is a suspected occultation binary.} The spectral type given by \citet{abt&morrell1995} is `F2\,V\,mA8.' $\beta$ and $b-y$ are in rough agreement with each other, and support the F2 type, but $m_1$ indicates a metal type about 4 spectral subclasses earlier. This is in agreement with the metallicity of [Fe/H]$\:=-0.55$ in \citet{nordstrometal2004}, and with our classification of a new spectrum, which is F2\,V\,kA7mA7. We noted that the spectrum is metal-weak, but $\lambda4481$ is not unusually weak. \textbf{Recommendation:} non-member.

\subsection{HD\,79025}
Classified as A9\,Vn by \citet{paunzenetal2001}, who were specifically looking for $\lambda$\,Boo stars. Among the normal (or even enhanced metals) stars in a ($b-y, \Delta a$) diagram, based on values catalogued by \citet{vogtetal1998}.
\textbf{Recommendation:} non-member.

\subsection{HD\,79108}
Not far from the line of normal stars in the \citet{maitzen&pavlovski1989b} $\Delta a$ photometry study. Included in the \citet{rensonetal1990} catalogue, and classified as A0\,V~$\lambda$\,Boo by \citet{abt&morrell1995}. An RV variable according to \citet{gerbaldietal2003}, and listed as a star with a composite spectrum by \citet{faraggianaetal2004}, where in their section\:3.2 their conclusion is that this star is `in reality a complex system composed of at least two stars of similar luminosity and a third less luminous component.' In any case, the star is only slightly metal weak ([Fe/H]$\:=-0.07$; \citealt{saffeetal2008}) and some doubt has arisen over the $v\sin i$ value, in that those authors tabulate only 10.2\,km\,s$^{-1}$ and their reference does not follow up.
\textbf{Recommendation:} non-member.

\subsection{HD\,79469}
One of the first $\lambda$\,Boo stars recorded \citep{morganetal1943}, but classified as `B9.5\,IV~(CII)' by \citet{gray&garrison1987}. Falls exactly on the line of normal stars in the \citet{maitzen&pavlovski1989b} $\Delta a$ photometry study. \citet{faraggianaetal1990} concluded this is not a $\lambda$\,Boo star because it meets none of the UV requirements for membership. It is `misclassified' according to the \citet{rensonetal1990} catalogue.
\textbf{Recommendation:} non-member.

\subsection{HD\,80081}
A lack of $\lambda$\,Boo character in the UV led \citet{paunzenetal1997a} to declassify this star from a $\lambda$\,Boo member, in accordance with the papers on the UV spectra of $\lambda$\,Boo stars by \citet{bascheketal1984} and \citet{faraggianaetal1990}. Lies directly on the line of normal stars in the \citet{maitzen&pavlovski1989} $\Delta a$ photometry study. Classified as `A1\,V' by \citet{abt&morrell1995} and as A2\,IV$^{-}$ by \citet{gray&garrison1987}. \red{There are indications of multiplicity in the WDS catalogue \citep{masonetal2001}.}
\textbf{Recommendation:} non-member.

\subsection{HD\,81104}
\citet{abt1985} gave the spectral type A2\,Vn~$\lambda$\,Boo, but \citet{bidelmanetal1988} did not note peculiarity in their spectral type of A3\,Vn and \citet{paunzenetal2001} recorded a spectral type A3\,Van, i.e.\ not $\lambda$\,Boo although they were specifically looking for $\lambda$\,Boo stars. Rejected by the \citet{paunzenetal1997a} catalogue, for inferences made in that survey.
\textbf{Recommendation:} non-member.

\subsection{HD\,81290}
Among the $\lambda$\,Boo stars in in the \citet{maitzen&pavlovski1989b} $\Delta a$ photometry study. Classified as `kA5hF3mA5\,V LB' by \citet{paunzenetal2001}. Metal deficiencies are at about $-1$\,dex \citep{solanoetal2001}. From our new spectrum we give the spectral type hF2\,V\,kA3mA3~($\lambda$\,Boo?), meaning that it is a candidate as a mild $\lambda$\,Boo star, but the late H line type is concerning (see \S\,\ref{sssec:spectral-range}). An NLTE analysis of the volatile elements is necessary to eliminate the possibility that this is a field horizontal branch star.
\textbf{Recommendation:} probable member.

\subsection{HD\,82573}
Claimed to be misclassified by \citet{paunzenetal2002a} because it had been classified by \citet{paunzenetal2001} as a normal star (A3\,V). It is also normal according to \citet[][A7\,V]{abt&morrell1995}. $uvby\beta$ photometry \citep{hauck&mermilliod1998} indicates a slightly high $c_1$ index, while the $m_1$ index is what one expects given the $\beta$ and $b-y$ indices.
\textbf{Recommendation:} non-member.

\subsection{HD\,83041}
Very far from the line of normal stars in the \citet{maitzen&pavlovski1989b} $\Delta a$ photometry study, and a late but distinct $\lambda$\,Boo member according to their figure\:1. Spectral type of kA2hF2mA2\,V LB \citep{paunzenetal2001}. Metal weakness of over 1\,dex \citep{solanoetal2001}. Rejected by \citet{gerbaldietal2003} because it was considered to be a field horizontal branch star with $\lambda$\,Boo properties by \citet{corbally&gray1996}, who actually classified it as `F2\,V\,kA3mA2 lam Boo PHL'. The measured radial velocity there was 46\,km\,s$^{-1}$. \citet{grayetal1996} claimed this star lies significantly below the ZAHB and suggested it is a field blue straggler. We reobserved this star and give the spectral type F1\,V\,kA3mA2~$\lambda$\,Boo, but note that due to the late hydrogen line type, an NLTE abundance analysis should be carried out to ensure a $\lambda$\,Boo abundance pattern is shown, and that this is not just a metal weak or horizontal branch star.
\textbf{Recommendation:} probable member.

\subsection{HD\,83277}
\citet{paunzenetal2001} gave the spectral type kA3hF2mA3V LB. \citet{gerbaldietal2003} noted an inconsistent UV flux and possible binarity. We observed this star for this work and give the spectral type F1.5\,V\,kA3mA3~$\lambda$\,Boo?, indicating it is either a late $\lambda$\,Boo star or Pop.\,II. A full abundance analysis including volatile elements is required.
\textbf{Recommendation:} probable member.

\subsection{HD\,84123}
A abundance analysis \citep{heiteretal1998} confirms this star is a member of the $\lambda$\,Boo group with $[Z]=-1.0$. $T_{\rm eff}=6800$\,K and $\log g=3.5$. It was given the spectral type `kA6hF1mA6\,V (LB)' by \citet{paunzenetal2001}. \citet{gerbaldietal2003} argued misclassification because of high space velocities, but the abundance pattern \citep{heiteretal1998} is truly $\lambda$\,Boo, with [C], [N], [O] = $-0.1$, $0.0$ and $-0.3$\,dex, respectively. The radial velocity is only 16.3\,km\,s$^{-1}$ and the proper motions, -17.34 and -86.27\,mas/yr, are large in the declination direction (only). Our new spectrum yields the spectral type hF2\,V\,kA6mA6~$\lambda$\,Boo.
\textbf{Recommendation:} member.

\subsection{HD\,84948}
Among the $\lambda$\,Boo population in the \citet{maitzen&pavlovski1989} $\Delta a$ photometry study. \citet{andrillatetal1995} noted `LB, shell' in their infrared study, where in their figure\:9 they show the asymmetric Paschen lines and redshifted Ca and O lines, but this is probably due to the binary nature. \citet{heiter2002} performed an abundance analysis, finding $\sim1$\,dex metal underabundances for the A component, and slightly smaller underabundances for the B component. The atmospheric parameters used were those of \citet{paunzenetal1998a} which were challenged by \citet{faraggianaetal2001a}, which might explain the discrepant abundances for the two members, but which might also point to a triple system. \citet{paunzenetal2001} gave the spectral type `kA7hF1mA6\,V (LB)', and from our new spectrum we give F1.5\,Vs\,kA5mA5~$\lambda$\,Boo?. NLTE abundances of the volatile elements would help arbitrate between a true $\lambda$\,Boo and a Pop.\,II star.
\textbf{Recommendation:} probable member.

\subsection{HD\,87271}
\citet{handler1999} took $uvby\beta$ photometry for HD\,87271. $\beta$ ($=2.775$) and $b-y$ ($=0.151$) are in agreement and correspond to a star with a hydrogen line type around A9, but the $m_1$ index is remarkably low, even lower than that expected of a B8 dwarf. \citet{handleretal2000} followed up on the object with classification-resolution spectroscopy and described it as one of the most extreme classical $\lambda$\,Boo stars known. They derived the spectral type A9\,kA0mA0\,V~$\lambda$\,Boo; we arrived at the same spectral type from our new spectrum.
\textbf{Recommendation:} member.

\subsection{HD\,87696}
\citet{abt&morrell1995} gave the spectral type A9\,Vp (Boo; met: A5). According to \citet{griffinetal2012} this star `appears to be a single, rotationally-broadened late-A star and has been widely classified as A7\,V, with near solar metallicity'.
\textbf{Recommendation:} non-member.

\subsection{HD\,89239}
Given the spectral type of A2\,Vp(Boo; met:B9.5) in the \citet{abt&morrell1995} catalogue. No other evidence in favour. The $uvby\beta$ data on SIMBAD show discordant Str\"omgren indices, but our new spectrum suggests this is a normal B9.5\,V star.\textbf{Recommendation:} non-member.

\subsection{HD\,89353}
Post-AGB star \citep{kohoutek2001}, discussed here because it was in the \citet{rensonetal1990} catalogue as an uncertain member. $uvby\beta$ photometry \citep{hauck&mermilliod1998} clearly confirms this is not an A-type dwarf.
\textbf{Recommendation:} non-member.

\subsection{HD\,90821}
Presented as a new $\lambda$\,Boo star with spectral type `kA2hA7mA2\,Vn~$\lambda$\,Boo' and a comment `normal hydrogen lines' by \citet{paunzen&gray1997}, and included in the results paper of that series \citep{paunzen2001}. \citet{andrievskyetal2002} performed an LTE abundance analysis and wrote that HD\,90821 ``can be definitely ruled out as being [a member] of the $\lambda$ Bootis group'', adding ``the abundances derived for seven elements do not show any significant deviation from solar values.'' The solar metallicity is corroborated by a [Fe/H] value of $-0.03$ in the PASTEL catalogue \citep{soubiranetal2010}. We have obtained a new spectrum and classify HD\,90821 as a normal A3\,IV-V star.
\textbf{Recommendation:} non-member.

\subsection{HD\,91130}
HD\,91130A = HR\,4124. \citet{abt&morrell1995} classified this star as `A0\,Vp(Lam Boo)', which was confirmed by \citet{paunzen&gray1997} with the spectral type `A0\,Va$^{-}$ $\lambda$\,Boo (PHL)' and a comment that its hydrogen lines are shallow and broad. Gray has reclassified that spectrum for this work, giving the spectral type `kA0hA1mA0\,Va~($\lambda$\,Boo),' and he noted a very weak $\lambda$4481 line compared to the A0 standard. NLTE abundances for [N], [O] and [S] are typical for $\lambda$\,Boo stars at $-0.30$, $0.14$ and $0.18$, respectively \citep{paunzenetal1999a, kampetal2001}. LTE abundances of refractory elements from \citet{andrievskyetal2002} are around $-1$\,dex, with $-1.7$\,dex for [Fe].
\textbf{Recommendation:} member.

\subsection{HD\,97411}
\citet{abt&morrell1995} described HD\,97411 (HR\,4347) as A0 IVp (4481-wk) with $v\sin i = 25$\,km\,s$^{-1}$. \citet{paunzenetal1997a} rejected HD\,97411 based on \citeauthor{gray1988}'s (\citeyear{gray1988}) comments. \citeauthor{gray1988} wrote: ``The 4481 line is present and appears to be of quite normal strength. Hence there appears to be no compelling reason to include this star in the $\lambda$\,Boo class.'' He added that any peculiarities could be produced by the close binary nature (sep.\,0.2'', $\Delta V=0.4$\,mag) of this star.
\textbf{Recommendation:} non-member.

\subsection{HD\,97773}
Intermediate between the $\lambda$\,Boo stars and normal stars in the \citet{maitzen&pavlovski1989} $\Delta a$ photometry study. Membership is uncertain according to the \citet{rensonetal1990} catalogue. In their infrared study, \citet{andrillatetal1995} listed a spectral type of A8\,Vwl, with a metallic line type of A3, and `LB?', and no indication of a shell. \citet{gerbaldietal2003} wrote `composite, Hipparcos' next to this star, and it was therefore recorded by \citet{faraggianaetal2004} as a $\lambda$\,Boo star with a composite spectrum. \red{The WDS catalogue \citep{masonetal2001} lists a companion at 0.3$''$ of similar magnitude.} We obtained a new spectrum \red{that did not resolve the binarity and that we found difficult} to classify. We give the type F1\,Vn\,kA7mA5, with the comment: `a very rapid rotator, which causes some uncertainty in the metal line type, even when comparing to A5\,Vn and A7\,Vn standards. The \specline{Mg}{ii}{4481} line looks weak even compared to the rapidly rotating standards, but an artificially broadened spectrum of the F0\,Vs standard, HD\,23585, up to $v\sin i = 250$\,km\,s$^{-1}$ has approximately the same \specline{Mg}{ii}{4481} strength.' Its $\lambda$\,Boo nature remains uncertain.
\textbf{Recommendation:} uncertain member.

\subsection{HD\,97937}
Lies among the $\lambda$\,Boo stars in the \citet{maitzen&pavlovski1989b} $\Delta a$ photometry study, but was given an uncertain membership likelihood in the \citet{rensonetal1990} catalogue. Classified as A9\,Vp 4481-wk \citep{abt&morrell1995}. Rejected by \citet{paunzenetal1997a} because it was allegedly classified by \citet{gray1989} as an evolved star, but inspection of Gray's work shows a metallicity class $-1.5$, but a spectral type of F0\,V, i.e.\ not evolved. No abundances, but $uvby\beta$ photometry \citep{hauck&mermilliod1998} shows the $\beta$ and $b-y$ indices are in agreement with an early F star and the $c_1$ index is consistent with a luminosity class V object there. Then the $m_1$ index suggests metal weakness by about 10 spectral subclasses. We obtained a new spectrum and give it the spectral type F1\,V\,kA9mA6, but we note that the discrepancy between the K-line and metallic-line types suggest this is a composite spectrum. We assign the star to the `uncertain' membership category, pending a full abundance analysis.
\textbf{Recommendation:} uncertain member.

\subsection{HD\,98353 (55\,UMa)}
\citet{abt1984b} noted the peculiarity of this object, that was later accepted as a $\lambda$\,Boo star in the \citet{rensonetal1990} catalogue. Earlier investigations into its UV properties revealed no peculiarities; a spectral type of A2 was derived \citep{cucchiaroetal1980}. \citet{andrillatetal1995} presented substantial evidence for a shell. Long known to be a spectroscopic binary system, characterised by \citet{battenetal1978} as having an orbital period of 2.5\,d. Its binarity has been the subject of dedicated papers; \citet{lloyd1981} confirmed the orbital period and noted a high eccentricity ($e=0.43$), which is very odd given the short orbital period -- \citeauthor{lloyd1981} calculated that the orbit should have circularised within $\sim5$\,Myr. \citeauthor{lloyd1981} concluded that at least one component of the system appears to have a UV excess, consistent with a $\lambda$\,Boo star, but given the estimated mass ratio of the objects of 0.9, it was not possible to say which. \citet{gray&garrison1987} classified this star as A1\,Va with the comment ``Cores of the hydrogen lines look slightly washed out. A composite spectrum?'' In a later works, \citet[and references therein]{hornetal1996} developed the picture of a triple system for 55\,UMa, where the third component has a 1870-d orbital period. With synthetic spectra they obtained a best fit of $T_{\rm eff}=9500$\,K and $\log g = 4.5$ for both the primary and tertiary components, with $v\sin i = 33$ and $55$\,km\,s$^{-1}$, respectively, but the secondary could only be determined to be of spectral type A. In a tomographic separation of the constituent spectra, \citet{liuetal1997} refined the $T_{\rm eff}$ values of each member to $9230\pm230$, $8810\pm250$ and $9290\pm190$ for components Aa, Ab and B respectively. Their synthetic spectrum analysis showed that all the components, but particularly Aa and Ab of the close binary, had enhanced metal contents, and were consistent with the class of marginal Am stars. \blu{This directly} conflicts with a $\lambda$\,Boo classification.
\textbf{Recommendation:} non-member.

\subsection{HD\,98772}
Included in the \citet{rensonetal1990} catalogue, but determined not to be a $\lambda$\,Boo star by \citet{andrievskyetal2002} because it had only been described as a rapidly rotating normal or 4481-weak star in the past, and they found the abundances were not typical for a $\lambda$\,Boo star -- the abundances of Mg and Si are solar. In fact, it was in the consolidated catalogue of $\lambda$\,Boo stars \citep{paunzenetal1997a} but not the following paper in the series \citep{paunzenetal2001}, where its spectral type was A1\,Va.
\textbf{Recommendation:} non-member.

\subsection{HD\,100546}
``A clear $\lambda$\,Boo star'' \citep{acke&waelkens2004}. Also one of the closest and brightest HAeBe objects, and a $\beta$\,Pic-like object. Classified as `B9\,Vne' by \citet{houk&cowley1975}, and our new spectrum has the type A0\,Vae\,kB8, with clear emission in H\,$\beta$ and almost absent metals. It does have a $\lambda$\,Boo character, in that the $\lambda$4481 line is weak, but due to the near-absence of metal lines in the classification spectrum with which to compare the $\lambda$4481 weakness, we cannot rule out that the star is just metal-weak overall, rather than a $\lambda$\,Boo star. We have also inspected the available IUE spectrum, which looks normal. Specifically, there is no \mbox{1600-\AA} flux depression. \citet{muldersetal2013} ruled on the sub-stellar companion associated with this star, concluding it is a $60^{+20}_{-40}$-M$_{\rm Jup}$ brown dwarf.
\textbf{Recommendation:} uncertain member.

\subsection{HD\,100740}
Classified in \citet{paunzenetal2001} as kA4hA8mA5\,V, but not as $\lambda$\,Boo even though mild metal weakness is apparent. It is not classified as a $\lambda$\,Boo anywhere else, and our new spectrum shows that the star has a normal $\lambda4481$ line. It is slightly metal weak though, even after accounting for the very rapid rotation. We give the spectral type A3\,IVnn\,kA2.5mA2.5.
\textbf{Recommendation:} non-member.

\subsection{HD\,101108}
One of \citeauthor{hauck&slettebak1983}'s (1983) nine official $\lambda$\,Boo stars. Falls among the $\lambda$\,Boo population in the $\Delta a$ photometry study of \citet{maitzen&pavlovski1989}. Matches all of the UV criteria for membership in \citet{faraggianaetal1990}; our re-investigation of an IUE spectrum agrees. \citet{iliev&barzova1993b} confirmed the PHL profile with a 8000-K core and an 8400-K wing. They also said of the spectrum that the \ion{Ca}{ii} K line was strong but metallic lines were otherwise weak. No shell recorded by \citet{andrillatetal1995}. \citet{paunzenetal2001} did not classify it as a $\lambda$\,Boo star, though, rather as A3\,IV (wk 4481). Strangely, it was then adopted in the results paper \citep{paunzen2001} as `A3\,IV-V ($\lambda$\,Boo)', but without a special note other than to say it was discussed in their outline paper \citep{paunzen&gray1997}, where the comment given was that it could not be classified as a $\lambda$\,Boo star at first sight. \citet{heiter2002} later found a mild $\lambda$\,Boo abundance pattern. The hydrogen lines in our spectrum match the A3\,IV standards better than the A7\,V ones, and then the metal lines are not generally peculiarly weak. The $\lambda4481$ line is still weak in that case, though. We give the spectral type A3\,IV~(4481-wk), and conclude that this may be an extremely mild $\lambda$\,Boo star.
\textbf{Recommendation:} probable member.

\subsection{HD\,101412}
\citet{cowleyetal2010} studied HD\,101412 and noted it to be somewhere between a $\lambda$\,Boo star and a Vega-like object. The $\lambda$\,Boo characteristics were confirmed by \citet{folsometal2012}, who also summarised literature magnetic field measurements. This star is unusual in that it is a magnetic $\lambda$\,Boo star, with a longitudinal field of $500\pm100$\,G. It shows depletions in the intermediate volatile zinc, which has bearing on the gas-grain separation discussed in $\lambda$\,Boo formation theory. We have classified a new spectrum of HD\,101412 as A3\,Va(e)\,kA0mA0~$\lambda$\,Boo, noting an emission notch in H\,$\beta$ and a weak $\lambda4481$ line. \textbf{Recommendation:} member.

\subsection{HD\,102541}
Str\"omgren colours suggest HD\,102541 is a metal-weak dwarf \citep{gray&olsen1991}, contrary to the SIMBAD spectral type (A3\,II/III, \citealt{houk1982}). \citet{kuschnigetal1997} took a spectrum and derived a spectral type kA3hA5mA3\,V (LB), noting the \specline{Mg}{ii}{4481} line is normal for A3 and not remarkably weak. Confirmed as a $\lambda$\,Boo star by \citet{paunzen2001} with a spectral type. `kA5hF0mA5\,V\,LB', and in this work we give the spectral type A9\,V\,kA4mA4~($\lambda$\,Boo), with the parentheses denoting its mild $\lambda$\,Boo character. A note in \citet{gerbaldietal2003} reads `inconsistent UV flux (bin.?)' but no firm evidence to suggest binarity exists. \blu{This star does have an unusually large transverse velocity, which cannot be explained by large uncertainties in the parallax or proper motion measurements. A full abundance analysis should be carried out to be sure the abundance pattern matches that of the $\lambda$\,Boo stars.}
\textbf{Recommendation:} \blu{probable} member.

\subsection{HD\,103483}
The spectral type, kA2hA5mA3\,V, given by \citet{paunzenetal2001} is consistent with a mild $\lambda$\,Boo star, but they did not identify it as such. \citet{andrievskyetal2002} performed an abundance analysis for this star, showing metal underabundances of $\sim0.6$\,dex, but since oxygen was equally underabundant, they rejected it from the $\lambda$\,Boo group. Also a detached eclipsing binary of the Algol type. In fact, it is part of a sextuple system \citep{zascheetal2012}.
\textbf{Recommendation:} non-member.

\subsection{HD\,105058}
Its $\lambda$\,Boo status was established by \citet{slettebaketal1968}, who noted that the ratio of the \ion{Ca}{ii} K line with the hydrogen lines corresponded to a spectral type near A2, but \specline{Mg}{ii}4481 and other metallic lines were weak for this type. \citet{bascheketal1984} investigated the UV spectrum for this star, which has a moderate $\lambda$\,Boo character, as confirmed by \citet{faraggianaetal1990}. Falls firmly in the $\lambda$\,Boo group in the \citet{maitzen&pavlovski1989} $\Delta a$ photometry study. Adopted as a definite member of the $\lambda$\,Boo group in the \citet{rensonetal1990} catalogue. Its hydrogen lines are of the PHL type \citep{iliev&barzova1993b}, and show `emission-like' features \citep{iliev&barzova1998}. \citet{andrillatetal1995} had nothing peculiar to report in the infrared. \citet{paunzen&gray1997} gave a spectral type `kA1hA7mA1\,V $\lambda$\,Boo'. \citet{andrievskyetal2002} found $-1$\,dex metal underabundances in an LTE analysis, but concluded its $\lambda$\,Boo nature is ambiguous because no abundances for C, N, O or S are available. From our spectrum we obtain the spectral type hA8\,V\,kA0.5mA0.5~$\lambda$\,Boo. \textbf{Recommendation:} member.

\subsection{HD\,105199}
\citet{paunzen&gray1997} discussed this star. With a spectral type there of `kA0.5hF0mA3\,V $\lambda$\,Boo:' they wrote ``[t]his would be the first $\lambda$\,Boo star where the K-line type is in gross disagreement with the metallic-line spectrum.'' In his `Miscellaneous spectroscopic notes', \citet{bidelman1988} entered the note `early Am' for this star, in direct contravention of the expected metal weakness of $\lambda$\,Boo stars. In this work, a \mbox{3.6-\AA} spectrum of this star was obtained that shows a clear, mild Am character with the spectral type kA2hA5mA5 (IV), with mildly enhanced \specline{Sr}{ii}{4077} (typical of Am stars), and a mild anomalous luminosity effect. The only deviation from an Am star is a slightly weak \specline{Mg}{ii}{4481} line. Definitely not a $\lambda$\,Boo star.
\textbf{Recommendation:} non-member.

\subsection{HD\,105260}
An uncertain member of the class \citep{rensonetal1990}, and classified by \citet{abt1984} as F0\,V\,wl(met:A5). A reinvestigation of the spectrum \blu{is required.} \textbf{Recommendation:} \blu{uncertain} member.

\subsection{HD\,105759}
Despite being 6th magnitude, HD\,105759 was not discovered to be variable until 1991, with a 0.045-d period \citep{colombaetal1991}. Further frequencies were found and Str\"omgren photometry obtained by \citet{koenetal1995}; Str\"omgren photometry indicates a metal-poor star near A9. It was described as a multi-periodic $\delta$\,Sct star with $[Z] = -1$\,dex in a dedicated paper \citep{martinezetal1998}, though no abundances for C, N, O or S were determined. We refer the reader there for more information, including their justification for acceptance of this star into the $\lambda$\,Boo group. We accept this star as a member based on their recommendation, but a spectrum is desired.
\textbf{Recommendation:} member.

\subsection{HD\,106223}
\citet{slettebak1968} described this object as having Balmer lines near F0, but `all of the metallic lines, including the \ion{Ca}{ii} K line line, are exceedingly weak for this spectral type, which is reflected in the very low $m_1$ index'. One of the stars furthest from the line of normal stars in the $\Delta a$ photometry study of \citet{maitzen&pavlovski1989}, and the reddest of the stars they studied. It was described as a cool $\lambda$\,Boo star by \citet{faraggianaetal1990}, but was too cool for their UV membership criteria to be applied. It was accepted into the \citet{rensonetal1990} catalogue, and \citet{iliev&barzova1993b} confirmed a peculiar hydrogen line profile, whose cores matched a 6900-K spectrum but whose wings matched a 7300-K spectrum. They argued the hydrogen lines did not agree with suggestions that this is a field horizontal branch (FHB) star. No evidence for a shell in IR observations was found by \citet{andrillatetal1995}. Observed to be constant at the 3-mmag level \citep{paunzenetal1998c}. Given the spectral type `kA3hF3mA3\,V LB' by \citet{paunzenetal2001}, where an FHB nature was again ruled out. An abundance analysis confirmed $\sim2$\,dex underabundances of metals \citep{andrievskyetal2002}. A spectrum obtained for this work shows a $\lambda$\,Boo character, with spectral type F4\,V\,kA1.5mA1~$\lambda$\,Boo. We noted a weak G band and peculiar H lines.
\textbf{Recommendation:} member.

\subsection{HD\,107223}
Examined by \citet{paunzenetal2002a} for pulsation, but was found to be constant at the 1.9-mmag level. That is the only literature reference for this star on SIMBAD, so we are forced to consider it may have been a typo of the confirmed member HD\,107233 (see next subsection) or HD\,106223 (above). Str\"omgren photometry \citep{hauck&mermilliod1998} shows $\beta=2.9$, which is as high as values of $\beta$ reach. However, the $m_1$ index ($=0.11$) is discordant with that and would imply a metal weakness. This is another potential explanation for its appearance in the\citet{paunzenetal2002a} paper. There is no evidence to suggest membership, and our spectrum shows this star to be an excellent match to the A1\,IVs standard, $\rho$\,Peg.
\textbf{Recommendation:} non-member.

\subsection{HD\,107233}
\citet{gray&olsen1991} took Str\"omgren photometry for this star in their work on supergiants -- HD\,107233 is classified as such on SIMBAD. The measurements were $b-y = 0.192$, $m_1 = 0.110$, $c_1 = 0.710$ and $\beta = 2.743$. Collectively, $b-y$, $c_1$ and $\beta$ are consistent with a dwarf near F1, but for such a star the $m_1$ index is exceedingly weak (though no note to that effect is provided \textit{ibid.}). The next time HD\,107233 appears in the literature is in a table of confirmed $\lambda$\,Boo stars (table\:2 of \citealt{gray&corbally1993}), with a spectral type kA1hF0mA1\,Va~$\lambda$\,Boo, which was later repeated by \citet{paunzenetal2001}. Key physical parameters are available from \citet{iliev&barzova1995}, who found this star to be both the lowest mass (1.6\,M$_{\odot}$) and the oldest (1090\,Myr) $\lambda$\,Boo star then known. \citet{solanoetal2001} found a temperature of 6950\,K and $v\sin i = 110$\,km\,s$^{-1}$. \citet{heiter2002} preferred 6800\,K and $v\sin i = 80$\,km\,s$^{-1}$, and confirmed $\lambda$\,Boo abundances with only slightly deficient [C] ($-0.3$\,dex) alongside average metal deficiencies of $-1.4$\,dex. If the typo of HD\,107223 (see previous subsection) was meant to be 107233 in the \citet{paunzenetal2002a} paper, then we conclude non-variability at the 1.9-mmag level.
\textbf{Recommendation:} member.

\subsection{HD\,108283}
A $\lambda$\,Boo star according to \citet{abt&morrell1995}, but \citet{haucketal1998} rejected it because its $\Delta m2$ index is positive, with the same value as that of an Am or Ap star, and it has been classified as an HgMn star. Indeed, \citet{gray&garrison1989} classified it as A9\,IVnp\,Sr\,II.
\textbf{Recommendation:} non-member.

\subsection{HD\,108714}
Listed as an uncertain member of the $\lambda$\,Boo group in the \citet{rensonetal1990} catalogue. Falls among the population of normal stars in the extended $\Delta a$ photometry study of \citet{maitzen&pavlovski1989b}. $uvby\beta$ photometry shows no discordance between the $m_1$ index and the $\beta$ and $b-y$ indices.
\textbf{Recommendation:} non-member.

\subsection{HD\,108765}
\citet{paunzenetal2001} gave a 3-part spectral type in their observational paper of `kA3hA3mA0\,V', while \citet{andrievskyetal2002} stated this star can definitely be ruled out as being a member of the $\lambda$\,Boo group because the abundances were, within the errors, solar.
\textbf{Recommendation:} non-member.

\subsection{HD\,109738}
A comment on this star in Vol.\,I of the Michigan Catalog \citep{houk&cowley1975} describes the \ion{Ca}{ii} K line and metal lines as very weak. Consequently HD\,109738 appeared in Hauck's (\citeyear{hauck1986}) list of $\lambda$\,Boo candidates. It appears well amongst the $\lambda$\,Boo stars in the extended $\Delta a$ photometry study of \citet{maitzen&pavlovski1989b}. A member of the $\lambda$\,Boo group in the \citet{rensonetal1990} catalogue. Given a spectral type of `kA1hA9mA1\,V LB' in \citeauthor{paunzen2001}'s (\citeyear{paunzen2001}) `new and confirmed' list of $\lambda$\,Boo stars. (\blu{As an aside, contrary} to \citeauthor{paunzen2001} in that 2001 article, this should have been a `confirmed' rather than a `new' $\lambda$\,Boo star). A new spectrum obtained for this work confirms the $\lambda$\,Boo membership with the spectral type A9\,Vn\,kB9.5mA0~$\lambda$\,Boo. \citet{solanoetal2001} derived metal underabundances of $-1$\,dex, but no abundances for volatile elements are available.
\textbf{Recommendation:} member.

\subsection{HD\,109980}
\citet{abt&morrell1995} classified this star as A6\,Vp(Lam Boo) with an uncertain $v\sin i=255$\,km\,s$^{-1}$. The rapid rotation may be responsible for the perceived metal weakness. Spectral type in \citet{grayetal2001a} is `A8\,Vnn\,kA6', i.e.\ it was not identified them as a $\lambda$\,Boo star.
\textbf{Recommendation:} non-member.

\subsection{HD\,110377}
Classified as A6\,Vp(Lam Boo) by \citet{abt&morrell1995}, but with $v\sin i = 160$\,km\,s$^{-1}$. Subject of a dedicated paper into its $\delta$\,Sct variability \citep{bartolinietal1980}, where the multiperiodicity was interpreted as radial pulsation. Well covered in the $\delta$\,Sct literature. $uvby\beta$ photometry \citep{hauck&mermilliod1998} suggests a normal star around A5. No abundances to suggest membership.
\textbf{Recommendation:} non-member.

\subsection{HD\,110411 ($\rho$\,Vir)}
\citet{gray1988} classified HD\,110411 as A0\,Va ($\lambda$\,Boo) with the comment ``$\lambda$4481 very weak compared to the metallic-line spectrum. Hydrogen-line profiles are nearly normal, except wings appear shallow. This star is a variable with an amplitude of 0.02\,V. It is also a spectroscopic binary.'' He has reclassified the star for this work, with the spectral type `kA0hA3mA0\,Va~$\lambda$\,Boo,' and noted the $\lambda$4481 line is weak compared to the A0 standard. HD\,110411 falls far from the line of normal stars in the \citet{maitzen&pavlovski1989b} $\Delta a$ photometry study. Meets all the UV criteria for $\lambda$\,Boo membership \citep{faraggianaetal1990}. A $\lambda$\,Boo member in the \citet{rensonetal1990} catalogue. LTE metal abundances for measured elements and ions sit at $-1$\,dex \citep{stuerenburg1993}. Key physical parameters are available from \citet{iliev&barzova1995}, but their suggested age of 200\,Myr is inconsistent with the $500\pm200$\,Myr age recorded by \citet{vican2012}. Described as a dusty system by \citet{chengetal1992}, but \citet{holweger&rentzsch-holm1995} found no evidence of circumstellar dust in the \ion{Ca}{ii} K line, and concluded that any circumstellar component present at the time of observation must have been weaker than 3\,m\AA. However, substantial evidence for a debris disc does exist \citep{boothetal2013}. Although HD\,110411 appears in the online data of the second paper \citep{paunzenetal2001} of the series of papers by Paunzen, which searched for new $\lambda$\,Boo stars, it is curiously not included in the `new and confirmed' list of $\lambda$\,Boo stars in the third \citep{paunzen2001}. \blu{The reason is that known $\lambda$\,Boo stars were observed as comparison stars for the project; these stars were not included in print, but did appear in the online data of the second paper in the series (Paunzen, private communication).} \citet{heiter2002} confirmed a $\lambda$\,Boo abundance pattern, with both [C] and [O] being super-solar, though it was noted that strangely Mg and Si were not as underabundant as other metals. Although some evidence is conflicting, there is enough to support $\lambda$\,Boo membership.
\textbf{Recommendation:} member.

\subsection{HD\,111005}
\citet{solanoetal2001} gave the spectral type hF0mA3\,V, with abundances showing only mild ($\sim$0.3\,dex) metal deficiency, but only Fe and Ti were measured. \citet{kampetal2001} added Ca, finding [Ca/H]$\:=-0.40$ with an LTE analysis. \citet{paunzenetal2001} classified this star as kA3hF0mA3\,V LB. Composite spectrum reported by \citet{gerbaldietal2003}. Recorded as an (uncertain) Am star in the \citet{renson&manfroid2009} catalogue with kA4mF2. Our new spectrum fixes the spectral type at F2\,V\,kA5mA5~$\lambda$\,Boo?, where the late hydrogen line type necessitates abundances of C or O to distinguish this object from the intermediate Pop.\,II stars.
\textbf{Recommendation:} probable member.

\subsection{HD\,111164}
The first spectral types assigned to this star were as an early A star: A3\,V \citep{appenzeller1967} and A3\,Vn \citep{jackisch1972}, while the first classifiers to assign a $\lambda$\,Boo classification were \citet{abt&morrell1995} with `A3\,Vp(Lam Boo)'. We obtained a new spectrum and saw a normal A3\,IV-V star. Unfortunately, HD\,111164 does not have wide coverage in the $\lambda$\,Boo literature and an abundance analysis is lacking, though a high $v\sin i$ value (175\,km\,s$^{-1}$, \citealt{abt&morrell1995}; 191\,km\,s$^{-1}$, \citealt{royeretal2002b}) might preclude one. The high $v\sin i$ might also have swayed \citet{abt&morrell1995} towards a mild $\lambda$\,Boo classification. \citet{hauck&mermilliod1998} took $uvby\beta$ measurements for this star, and Johnson $B$ and $V$ values are available. $B-V$, $b-y$, $\beta$ and $c_1$ collectively agree on a dwarf near A4. However an enhanced $m_1$ index indicates \textit{strong} metals, near A7.
\textbf{Recommendation:} non-member.

\subsection{HD\,111604}
Various consistent sources of $uvby\beta$ photometry suggest this is a star near A7, where the $m_1$ index suggests the star is metal weak. Hence the IUE proposal \citep{faraggiana1990} with which data were obtained under the title ``The Lambda Boo Stars'' before this star was confirmed as a class member. \citet{rensonetal1990} adopted this star as an uncertain member of the class, but \citet{weissetal1994} subsequently criticised this classification because the latter authors \textit{mistakenly} associated HD\,111604 with the Nova DQ\,Her. Based on the UV criteria for membership that were established by \citet{faraggianaetal1990}, \citet{gerbaldi&faraggiana1993} admitted this star to the $\lambda$\,Boo group. \citet{abt&morrell1995} independently classified this star as `A5\,Vp lambda Boo'. \citet{andrievskyetal2002} carried out an LTE abundance analysis, finding [C/H]$ = -0.25$ and metal deficiencies of $1$\,dex, confirming its $\lambda$\,Boo status. A spectral type of `A8\,Vn\,kA1.5\,mA1~$\lambda$\,Boo' was given by \citet{griffinetal2012}, who, like \citet{stutz&paunzen2006}, refuted the claims that this star's spectrum is a composite, made by \citet{gerbaldietal2003} and \citet{faraggianaetal2004}.
\textbf{Recommendation:} member.

\subsection{HD\,111786 (HR\,4881)}
\citet{andersen&nordstrom1977} provided the first $\lambda$\,Boo classification, but with a slightly contradictory statement: ``probably $\lambda$\,Boo star: neutral lines strong, \ion{Ca}{ii} K and \ion{Mg}{ii}$\lambda$4481 weak.'' 

There is a lot of evidence in favour of membership in the $\lambda$\,Boo group. \citet{bascheketal1984} noted moderate $\lambda$\,Boo characteristics in the UV from low-resolution IUE spectra; these characteristics were confirmed and emphasised by \citet{faraggianaetal1990}. \citet{maitzen&pavlovski1989b} included this star in their $\Delta a$ photometry study, wherein it lies clearly among the $\lambda$\,Boo population. The many investigations before 1990 led \citet{rensonetal1990} to include HD\,111786 as a definite member of the class. \citeauthor{stuerenburg1993}'s (\citeyear{stuerenburg1993}) LTE abundance analysis showed metal depletions of $-1.5$\,dex and a nearly-solar C abundance; he calculated NLTE corrections for each element which revises the metal weakness to around $-1.3$\,dex. It was independently classified as $\lambda$\,Boo by \citet{abt&morrell1995}. \blu{We note that HD\,111786 did not appear in the paper version of \citeauthor{paunzen2001}'s (\citeyear{paunzen2001}) `new and confirmed' list of $\lambda$\,Boo stars because it was observed as a known $\lambda$\,Boo (i.e. as a comparison star), but it is there in the online data in the second paper of that series.}

\citet{gray1988} classified HD\,111786 as `A1.5\,Va$^{-}$~$\lambda$\,Boo' , adding that $\lambda$4481 was very weak compared to the metallic-line spectrum, and that a weak, narrow, absorption core could be seen in the \ion{Ca}{ii} K line, which he attributed to circumstellar material.  With higher resolution spectra, \citet{holweger&stuerenburg1991} described that `weak, narrow, absorption core' as ``extremely conspicuous'', as illustrated in their figure\:3. Narrow absorption components were also noted for Na\,D lines. \citet{holweger&rentzsch-holm1995} noted the radial velocity of the circumstellar feature at \ion{Ca}{ii} K line to vary with time. In addition to the \ion{Ca}{ii} K line feature, \citet{gray1988} also noted peculiar hydrogen lines, which \citet{iliev&barzova1998} later investigated to show substantial residuals in the core of the Balmer lines.

Both the circumstellar features and the peculiar hydrogen lines have been explained as arising from an SB2 system \citep{faraggianaetal1997}, where the primary is a broad-lined and the secondary a narrow-lined A star. \citet{faraggianaetal2001a} discovered that HD\,111786 is, in fact, a multiple system having (probably) 5 members, as illustrated using the \ion{O}{i}\,7772--7775 feature in their figure\:15. It appears to be composed of one broad-lined star, and four narrow lined stars each having similar physical parameters (luminosity and temperature). We acknowledge their statement that ``[i]f we accept the definition of $\lambda$\,Boo stars as single objects with peculiar atmospheric abundances then the classification of this complex object as a $\lambda$\,Boo star must be definitely rejected.''

However, we are not ready to reject this classic $\lambda$\,Boo star just yet. Let us consider the \ion{Ca}{ii} K line, for which the `extremely conspicuous' absorption feature remains conspicuous in the binary hypothesis. The \ion{Ca}{ii} K line, for late-A stars and especially for early-F stars, is intrinsically broad -- no amount of slow rotation produces a narrow absorption component. We also look to the recent paper by \citet{griffinetal2012}, which rejects the composite-spectrum hypothesis as producing the $\lambda$\,Boo phenomenon. They state that $\lambda$\,Boo stars cannot be dismissed by ``that route''.

It is, incidentally, a 6-mmag pulsator with a frequency of 30\,d$^{-1}$ \citep{kuschnigetal1994a} that was later found to be multi-periodic \citep{paunzenetal1998b} and pulsating non-radially \citep{bohlenderetal1999}. If the pulsations can be shown to originate in one star, asteroseismic modelling may be able to shed light on the metal weakness. Key physical parameters are available \citep{iliev&barzova1995} to facilitate this.

We obtained a new spectrum for this work, and classify it as F0\,V\,kA1mA1~$\lambda$\,Boo. It looks extremely metal weak and like a classic $\lambda$\,Boo star. \specline{Mg}{ii}{4481} is barely a notable feature of the spectrum. H cores are sl. weak.

\textbf{Recommendation:} member.

\subsection{HD\,111893}
HD\,111893 is designated an uncertain member in the \citet{rensonetal1990} catalogue. Rapid rotation was noted in the spectral type (A5\,IV-Vnn) given by \citet{gray&garrison1989b}. Not identified as $\lambda$\,Boo by \citet{abt&morrell1995}, either. IRAS infrared excess detected \citep{king1994}. It is not clear why \citet{rensonetal1990} ever considered this a $\lambda$\,Boo candidate.
\textbf{Recommendation:} non-member.

\subsection{HD\,112097}
Classified as `F0\,Vp(Boo, met:A7) by \citet{abt&morrell1995}, but the \citet{skiff2013} catalogue states it has been classified as Am: in the past. It is an Am star in the \citet{renson&manfroid2009} catalogue (\ion{Ca}{ii} K line: A6, metal lines F1), but inspection of $uvby\beta$ photometry suggests hydrogen lines of about F0 with substantial metal weakness. \citet{grayetal2001a} gave a spectral type of kA7\,hF0\,mF0\,(V), supporting an Am: type.
\textbf{Recommendation:} non-member.

\subsection{HD\,113848}
\citet{cowley&bidelman1979} gave the classification F3\,pec with the note `H $\sim$ F8; \ion{Ca}{ii} $\sim$ F0; no G-band; metals $\sim$ F3.' \red{The fact that this is a known double star with sub-arc-second separation \citep[WDS,][]{masonetal2001} could explain the peculiar spectrum.} Designated as uncertain in the \citet{rensonetal1990} catalogue. IRAS infrared excess detected \citep{king1994}. SIMBAD [Fe/H] values show only mild deficiency. Probably too late to be a $\lambda$\,Boo star. A new and preferably high-resolution spectrum might provide a definitive assessment.
\textbf{Recommendation:} non-member.

\subsection{HD\,114879}
Designated as uncertain in the \citet{rensonetal1990} catalogue. Classified by \citet{paunzenetal2001} as A3\,V, with no comment suggesting a $\lambda$\,Boo character, when they were looking for $\lambda$\,Boo stars. Thus we might assert it is not a $\lambda$\,Boo star. \citeauthor{hauck&mermilliod1998}'s (1998) $uvby\beta$ data show agreement in the $\beta$ and $b-y$ indices that this star is about A5/6, and the $c_0$ index matches A6 for a luminosity class V object. Then the $m_1$ index suggests only mild metal weakness of 1--2 spectral subclasses. Reinvestigation of spectrum advised.
\textbf{Recommendation:} non-member.

\subsection{HD\,114930}
Designated as uncertain in the \citet{rensonetal1990} catalogue. Classified by \citet{paunzenetal2001} as F0\,IV, with no comment suggesting a $\lambda$\,Boo character, when they were looking for $\lambda$\,Boo stars. Thus, as for HD\,114879, we might assert it is not a $\lambda$\,Boo star. However, inspection of the \citet{hauck&mermilliod1998} $uvby\beta$ photometry suggests hydrogen lines of about F0/1 but metals of strength about A1. We have obtained and classified a new spectrum of this star, which warrants its rejection: F1\,Vs, definitely not metal weak. The Ti/Fe\,{\sc ii}\,$4172$--$9$ blend clearly points to luminosity class V.
\textbf{Recommendation:} non-member.

\subsection{HD\,118623}
\citet{abt&morrell1995} classified HD\,118623 as `F0\,Vp(Lam Boo)n' with $v\sin i =190$\,km\,s$^{-1}$. Classified as F0\,Vnn\,kA8 by \citet{grayetal2001a}, and Gray recommended it be dropped from the $\lambda$\,Boo class for this reason, that is, for very rapid rotators $\lambda$\,Boo classifications should be treated with skepticism. Previously classified by \citet{jaschek&jaschek1980} as `UV: abn. CrEuSr $\dots$ MK: A7p (Sr, Cr) $\dots$ notes: $v\sin i$ = 204\,km\,s$^{-1}$. Others: A7\,III.' \red{HD\,118623 has one close companion at 1.8$''$ east, which is 2\,mag fainter (WDS, \citealt{masonetal2001}).} \citet{gerbaldietal2003} noted `composite, Hipparcos' and this candidate was therefore described as a $\lambda$\,Boo star with a composite spectrum by \citet{faraggianaetal2004} because it is less than 2$''$ from another star that is less than 2.2\,mag fainter. In this work, we have inspected the UV spectrum against the $\lambda$\,Boo criteria in the UV given by \citet{faraggianaetal1990}; it does not have a $\lambda$\,Boo character.
\textbf{Recommendation:} non-member.

\subsection{HD\,119288}
Uncertain member in the \citet{rensonetal1990} catalogue. F3\,Vp star with [Fe/H]$\:=-0.46$, $T_{\rm eff}=6600$\,K and $\log g=4.03$ \citep{cenarroetal2007}. Classified as `F5\,V ((metal-weak))' by \citet{grayetal2001a}. IRAS infrared excess detected \citep{king1994}. Reported long-period roAp star \citep{matthews&wehlau1985}, where its spectrum was described as being atypical for a magnetic Ap star since Ca is enhanced and metals slightly underabundant, but follow-up observations could not confirm variability \citep{heller&kramer1988}. No particular evidence in favour of $\lambda$\,Boo membership except slight metal weakness; a high-resolution spectrum should easily confirm our recommendation for rejection.
\textbf{Recommendation:} non-member.

\subsection{HD\,120500}
Presented as a new $\lambda$\,Boo star by \citet{paunzen&gray1997}, who gave a spectral type of `kA1.5hA5mA1.5\,V~($\lambda$\,Boo)' and a comment ``mild $\lambda$ Bootis star.'' NLTE abundances for [N], [O], and [S] are $-0.30$, $+0.19$ and $-0.13$, respectively \citep{paunzenetal1999a, kampetal2001}. LTE abundances for metals \citep{andrievskyetal2002} suggest only a mild $\lambda$\,Boo character, which is why \citeauthor{andrievskyetal2002} described their results as `ambiguous' in evaluating this star's $\lambda$\,Boo membership. We accept HD\,120500 as a mild $\lambda$\,Boo star.
\textbf{Recommendation:} member.

\subsection{HD\,120896}
Classified as kA6F0mA6\,V~LB in the \citet{paunzenetal2001} spectroscopic survey and given in the results paper \citep{paunzen2001} as a new $\lambda$\,Boo star. Str\"omgren indices from \citet{hauck&mermilliod1998} ($b-y = 0.166$, $\beta=2.674$) agree with the hydrogen line type, for which $m_1$ ($= 0.150$) reveals a marked metal weakness. This is a mild $\lambda$\,Boo star.
\textbf{Recommendation:} member.

\subsection{HD\,123299}
This star, $\alpha$\,Dra, was adopted by \citet{gray&garrison1987} as one of their A0\,III low $v\sin i$ standards, and is upheld as such by \citet{gray&corbally2009}. This is sufficient reason to reject the star from the $\lambda$\,Boo group. Uncertain designation in the \citet{rensonetal1990} catalogue. Rejected by \citet{paunzenetal1997a} because of its identification by \citet{gray&garrison1987} as a non-$\lambda$\,Boo star.
\textbf{Recommendation:} non-member.

\subsection{HD\,125162}
This is $\lambda$\,Boo itself, the group prototype. Obviously a member. The spectral type on Gray's website is `A3 Va kB9mB9 Lam Boo'. \textbf{Recommendation:} member.

\subsection{HD\,125489}
\citet{abt&morrell1995} gave the spectral type `F0\,Vp(Boo, met: A5)'. \citet{gerbaldietal2003} gave the remark: `inconsistent UV flux (bin.?)'. RVs at different wavelengths in \citet{griffinetal2012} show discrepancies, but they described it as `a single, late-A dwarf or sub-giant with slightly enhanced metal abundances of 0.20\,dex.' \red{The WDS does list a companion at 1.8$''$, but it is 6.5\,mag fainter.} So although binary contamination is not an issue, metal enhancement means it is not $\lambda$\,Boo star.
\textbf{Recommendation:} non-member.

\subsection{HD\,125889}
One of the confirmed $\lambda$\,Boo stars of \citet{paunzen2001} with a spectral type `kA4hF2mA4\,V ($\lambda$\,Boo)'. We confirm the weak 4481 line in our spectrum, with the type F1\,Vs\,kA4mA4~($\lambda$\,Boo), but given the late hydrogen line type, an abundance analysis is required to distinguish from an intermediate Pop.\,II star.
\textbf{Recommendation:} probable member.

\subsection{HD\,128167 ($\sigma$\,Boo)}
An uncertain member of the $\lambda$\,Boo group according to the \citet{rensonetal1990} catalogue. A fairly late $\lambda$\,Boo candidate, with $B-V = 0.36$, classified by \citet{grayetal2001a} as `F4\,V~kF2mF1'. \citet{paunzenetal1997a} rejected it for this reason. Abundances from \citet{adelmanetal1997} show metal deficiencies of about 0.4\,dex, but the same is true for [C], hence the $\lambda$\,Boo pattern is not exhibited. IRAS infrared excess detected \citep{king1994}. NLTE abundances would help to more firmly reject this candidate.
\textbf{Recommendation:} non-member.

\subsection{HD\,130158}
Misclassified according to the \citet{rensonetal1990} catalogue. Called `A0\,IIIp(Lam Boo)' later by \citet{abt&morrell1995}. Identified as an Ap star: `A0\,II-IIIp (Si)' in the observational paper of \citet{paunzenetal2001} and B9\,Si in the \citet{renson&manfroid2009} catalogue. Physical parameters of this magnetic Ap star can be found in \citet{wraightetal2012}.
\textbf{Recommendation:} non-member.

\subsection{HD\,130767}
\citet{paunzenetal2001} gave the spectral type `A0\,Va $\lambda$\,Boo' in their observational paper, with the additional comment in the results paper that this star is very similar to $\lambda$\,Boo itself \citep{paunzen2001}. Key physical parameters are available \citep{paunzenetal2002b}. No published high-resolution spectroscopy / abundance analysis is available.
\textbf{Recommendation:} member.

\subsection{HD\,138527}
\citet{abt&morrell1995} classified HD\,138527 as `B9.5\,Vp(Boo: Ca, 4481 wk)' but were unable to determine $v\sin i$. Adaptive optics observations revealed this star to have a companion \citep{gerbaldietal2003}, but RVs are constant within the errors (table\:3 of \citealt{griffinetal2012}). The latter authors also wrote `metals are sub-solar by 0.8\,dex and its high rotational velocity [135\,km\,s$^{-1}$] accentuates the weakness of its lines', and that an interstellar \ion{Ca}{ii} K line feature was seen (see their figure\:7). They acknowledge the double star nature claimed by \citet{gerbaldietal2003}, but assert the spectra show no RV shift or other signature of an SB2 system, and that the magnitude difference ($\Delta m = 3.2$\,mag in the IR, and greater in the blue) is too large for the `contaminating' star to veil the lines of the brighter component. An NLTE abundance analysis of volatile elements, if possible given the $v\sin i$, would arbitrate membership into the $\lambda$\,Boo group.
\textbf{Recommendation:} probable member.

\subsection{HD\,139614}
Although \citet{acke&waelkens2004} found an overall metal deficiency, they did not see selective $\lambda$\,Boo depletion. \citet{folsometal2012} did, however, and describe this star as having a `clear $\lambda$\,Boo pattern'. Inspection of their abundances show this is very mild. Volatile and refractory elements are, on average, about 0.1 and 0.5\,dex below solar, respectively. Mg is 0.23\,dex ($=3\sigma$) below solar. Our spectral type is F0\,Vse\,kA4mA6~($\lambda$\,Boo). The $\lambda$4481 weakness is mild, and so we accept this as a very mild $\lambda$\,Boo star. \red{The WDS catalogue \citep{masonetal2001} lists a companion at 1.4$''$, but $\Delta m = 3.5$\,mag, so any contamination is small.} No IUE spectrum is available to investigate the UV properties.
\textbf{Recommendation:} member.

\subsection{HD\,141569}
\citet{gray&corbally1998} grouped this star into a set that were metal weak, without displaying all $\lambda$\,Boo features, and commented that these stars are probably metal weak due to accretion of metal-poor material. \citet{folsometal2012} `found $\lambda$\,Boo peculiarities' in this star through an abundance analysis. Although C is approximately solar, N and O are underabundant by 0.3\,dex each (1 and 3\,$\sigma$, respectively), Mg is underabundant by $0.4$\,dex (4\,$\sigma$), and Fe and Ti are both $0.7$\,dex (2\,$\sigma$) underabundant. This is consistent with a mild $\lambda$\,Boo star. Our spectral type is A2\,Ve\,kB9mB9~$\lambda$\,Boo. We noted infilling of H\,$\beta$, H\,$\gamma$ and H\,$\delta$. $\lambda$4481 is weak, and the spectrum is generally metal weak. \red{The WDS catalogue \citep{masonetal2001} paints a picture of multiplicity: companions of estimated masses 0.45, 0.22 and 1\,M$_{\odot}$ for components B, C and D were determined by \citet{feigelsonetal2003}. Their separations from HD\,141569\,A are 7.6, 8.1 and 1.5$''$, respectively.}
\textbf{Recommendation:} member.

\subsection{HD\,141851}
Identified as a $\lambda$\,Boo star by \citet{abt1984b} and included in the \citet{rensonetal1990} catalogue. \citet{paunzenetal1999a} found NLTE [C] and [O] abundances of $-0.81$ and $-0.21$, respectively, and $v\sin i$ in excess of 200\,km\,s$^{-1}$. \citet{kampetal2001} did an LTE abundance analysis and found [Ca/H]$\:=-1.30$\,dex but reliable NLTE abundances for N and S could not be determined. \citet{paunzenetal2001} gave a spectral type `A2\,Van', but no mention of $\lambda$\,Boo. \citet{paunzenetal2002a} argued this star is misclassified \blu{(that is, not a $\lambda$\,Boo star)}, citing the results paper of \citet{paunzen2001}. \blu{No mention of this star is actually made in the latter paper, but \citet{paunzenetal2001} did clearly state that this is not a $\lambda$\,Boo star.} \citet{andrievskyetal2002} found [Fe/H]$\:=-0.70$ and [Si/H]$\:=-0.65$ with [Na/H]$\:=+0.60$\,dex, but were not able to decide if this star was a $\lambda$\,Boo star. Indeed, the overall picture from abundance analyses is that of either a metal poor or mild $\lambda$\,Boo star. \citet{gerbaldietal2003} recalled the existence of a companion at 0.1$''$ and claimed to measure a magnitude difference but did not provide it. A comment of `detection of a star 3 times fainter 0.15 arcsec (east)' is given in their adaptive optics summary table (their table\:3). The contamination is discussed in more detail by \citet{faraggianaetal2004}. \red{The WDS catalogue \citep{masonetal2001} lists a separation of 0.4$''$ and $\Delta V= 2.6$\,mag.}

In our own spectrum of this star, we note that the H lines do not match particularly well at any type, but the spectrum is consistent with a normal A2\,IVn star. The H-line cores are shallower than in $\beta$\,Ser (the A2\,IVn standard), even though $v\sin i$ is about the same, but the match is otherwise good. The $\lambda4481$ line is still weak, but in rapidly rotating early A stars, $\lambda$4481 does look weak \citep{gray1986}.

In light of the new spectral type, the mild metal underabundances, the non-solar volatile element abundances, \blu{the rejection from earlier catalogues}, and the complexities introduced by the double/multiple star nature, we reject this candidate.
\textbf{Recommendation:} non-member.

\subsection{HD\,142666}
Has a spectral type of A8\,Ve \citep{moraetal2001}. \citet{folsometal2012} found C, N and O to have solar abundance while the Fe-peak elements are 0.5\,dex below solar (Fe itself is at $-0.25$\,dex), concluding this is a weak $\lambda$\,Boo star. We re-observed this target, whose spectrum suggests it is a shell star: `F0\,Vs~shell?', and perhaps a spectrum variable. In our spectrum the H lines have very strong cores and the metal lines have inconsistent morphologies with one another.
Despite the mildly anomalous abundances reported by \citet{folsometal2012}, the morphology is consistent with a shell star but not a $\lambda$\,Boo star, and so in the absence of a UV spectrum we are forced to reject this star.
\textbf{Recommendation:} non-member.

\subsection{HD\,142703}
Suggested candidate by \citet{hauck1986} and listed in the \citet{rensonetal1990} catalogue. PHL with 7200-K cores but 7700-K wings \citep{iliev&barzova1993b}. \citet{gray&corbally1993} classified the star as `kA1hF0mA1\,Va $\lambda$\,Boo'. Key physical parameters are available \citep{iliev&barzova1995}.. \citet{heiter1998} listed a cool $T_{\rm eff}$ of 7000\,K and found $\log g = 3.7$,  [Z]$\:=-1.5$ and $v\sin i = 100\pm10$\,km\,s$^{-1}$. \citet{iliev&barzova1998} saw emission in H$\gamma$ (see their figure\:1). NLTE abundances of [C], [N], [O] and [S] are $-0.52$, $-0.60$, $-0.19$ and $-0.52$, respectively \citep{paunzenetal1999a, kampetal2001}, and the latter authors also measured an LTE Ca abundance of $-1.40$\,dex. For refractory metals (Mg through to Zn) abundances are on the order $-1$\,dex \citep{solanoetal2001, heiteretal2002a}. No remarks to justify rejection in \citet{gerbaldietal2003}, but \citet{faraggianaetal2004} claimed it is a probable multiple system with no further comment.
\textbf{Recommendation:} member.

\subsection{HD\,142944}
Described in \citet{paunzenetal2002a} as a pulsating $\lambda$\,Boo star with $v\sin i=180$\,km\,s$^{-1}$, [Z]$\:=-0.91$ and a period of 8.1\,d$^{-1}$, but the low Q value on the pulsation implies this is not the fundamental radial mode. Key physical parameters available from \citet{iliev&barzova1995} and \citet{paunzenetal2002b} are in agreement. The MK type of A0\,V \citep{houk1982} is much earlier than the $B-V$ measurement of 0.17\,mag \citep{slawsonetal1992} would suggest. No argument in favour of $\lambda$\,Boo membership, based on spectra, can be found in the literature. No $uvby\beta$ photometry exists. Our conclusion is that its entry was a mistake (typo) on the part of \citet{paunzenetal2002a, paunzenetal2002b}, and really they meant to write HD\,142994. The latter is confirmed as a $\lambda$\,Boo star in the following subsection. Given the faintness of HD\,142944, and lack of published Str\"omgren photometry, we conclude it did not fall within Paunzen's survey boundaries. \blu{E.~Paunzen has confirmed in a private communication that HD\,142944 was a typo in the data tables (it should have read HD\,142994), but the main text remains accurate.}
\textbf{Recommendation:} non-member.

\subsection{HD\,142994}
\citet{gray1988} claimed this is a genuine $\lambda$\,Boo star with a spectral type `A3\,Va $\lambda$\,Boo $\dots$ PHL' and a note: `the K line and metallic-line spectrum are similar in strength to those of the A3\,IV standard $\beta$\,Eri, but $\lambda$4481 is very weak, but present. The hydrogen lines have weak cores, with strengths similar to those of F0 stars, but with broad and very shallow wings. Moderate $v\sin i$. Photometry from \citet{olsen1983} who reported possible variability in the $c_1$ index. Photometry of this star in May 1987 shows similar residuals in $c_1$'.


Key physical parameters are available \citep{iliev&barzova1995}. \citet{paunzenetal1999b} could only gather LTE abundances for three elements: [Mg/H]$\:=-0.2$, [Ti/H]$\:=-1.5$ and [Fe/H]$\:=-1.5$\,dex. High $v\sin i$ (180\,km\,s$^{-1}$; \citealt{paunzenetal2003}) inhibits abundance analyses, but none the less those authors recorded the presence of circumstellar gas, citing \citet{bohlenderetal1999}. No evidence of a companion (e.g. \citealt{solanoetal2001}; \citet{gerbaldietal2003}). From existing and new spectra, we determine this is a bona fide $\lambda$\,Boo star with spectral type F0\,V\,kA3mA3~$\lambda$\,Boo.
\textbf{Recommendation:} member.

\subsection{HD\,143148}
Given as `$\lambda$\,Boo(?)' by \citet{paunzenetal1996a}. \citet{paunzenetal2002a} argued this star to have been misclassified, since \citet{paunzenetal2001} found no $\lambda$\,Boo characteristics, giving a classification of A7\,IVn. \blu{We suspect} this star was originally a $\lambda$\,Boo candidate sometime after \citet{olsen1979} published Str\"omgren indices. He only identified two $\lambda$\,Boo candidates, and this was not one of them. Rather, it was identified as a field horizontal branch star. But if one studies carefully the $uvby\beta$ photometry available in the literature \citep{hauck&mermilliod1998}, one can infer hydrogen strengths near F0 but metal strengths near A0. Given the lack of evidence in favour of a $\lambda$\,Boo classification, we reject the star for now, but it would be worthwhile to obtain a spectrum. \textbf{Recommendation:} non-member.

\subsection{HD\,144708}
\citet{abt&morrell1995} classified HD\,144708 as B9\,Vp(Lam Boo)nn with a $v\sin i$ measurement of `255:' i.e.\ given with uncertainty. Not discredited by \citet{gerbaldietal2003}, despite the `essential note' on SIMBAD that this object includes another component. \citet{faraggiana&bonifacio1999} pointed out the duplicity of this star, with $P_{\rm orb} = 4.02$\,d. We find the high $v\sin i$ for what must be a very strongly tidally braked system to be suspicious (see discussion in \citealt{murphy2014}, Ch\,2). Highly discordant $uvby\beta$ photometry. No good evidence in favour of $\lambda$\,Boo membership.
\textbf{Recommendation:} non-member.

\subsection{HD\,145782}
Of the five spectral types available in the \citet{skiff2013} catalogue, not one suggests $\lambda$\,Boo membership. Indeed, \citet{paunzenetal2001} classified the star as A3\,V, and they were specifically looking for $\lambda$\,Boo stars. This is probably why \citet{paunzenetal2002a} rejected this star, which was labelled as `$\lambda$\,Boo(?)' by \citet{paunzenetal1996a}.
\textbf{Recommendation:} non-member.

\subsection{HD\,148638}
\citet{solanoetal2001} \red{had this target in their sample, but noting it was a binary they chose not to} derive the abundance pattern. \red{In fact, the separation is 21.9$''$ \citep[WDS,][]{masonetal2001} so there should have been no problem.} \citet{kampetal2001} did, however, proceed to analyse the spectrum and determined super-solar N and S abundances using NLTE, along with [Ca/H]$\:=-1.20$ using LTE. Their figure\:2 does appear to show large residuals after fitting the synthetic spectrum, possibly due to the binarity. \citet{paunzenetal2001} classified the star as `hA7mA2\,V LB', where the wide visual binary nature is acknowledged as a note in the result paper. Metal abundances have not been determined. $uvby\beta$ photometry \citep{oblak1978} is available; the $b-y$ and $\beta$ indices suggest a late A star, but then the $m_1$ index suggests considerable metal weakness, and the $c_1$ index is high. From our new spectrum we classify the star as A2\,IV$^-$n~(4481-wk), but note that the spectrum is otherwise normal. Unfortunately the substantial rotation will hinder high-resolution spectroscopic analyses, but a new analysis that considers the possibility of a hotter and less peculiar star would perhaps shed more light on this system.
\textbf{Recommendation:} uncertain member.

\subsection{HD\,149130}
Classified as kA7hF0mA7\,V LB by \citet{paunzenetal2001}, with the note in \citet{paunzen2001} that the spectrum is only mildly metal weak and that the Michigan Catalog gives the spectral type as A8wl. This star is mentioned in many papers on weak-lined stars. Abundance analysis desired to confirm membership.\textbf{Recommendation:} probable member.

\subsection{HD\,149303}
Appears in the \citet{rensonetal1990} catalogue. NLTE [O/H]$\:=-0.14$ and LTE [Ca/H]$\:=-0.50$ \citep{kampetal2001}. Abundance analysis in \citet{paunzen2000} exists only for Mg ($-0.2$), Ti ($-1.3$), Fe ($-0.9$) and Cr ($-0.5$\,dex), indicating a mild $\lambda$\,Boo star. Classified as \mbox{A3\,IV-V~(wk~4481)} by \citet{paunzenetal2001}, but not as $\lambda$\,Boo, hence it was rejected by \citet{paunzenetal2002a}. Binarity not recorded by \citet{gerbaldietal2003} even though it was detected as an SB system by \citet{paunzenetal1999a}. \red{The separation is large, at 16.3$''$ \citep[WDS,][]{masonetal2001}.} \citet{grayetal2001a} gave the secondary a spectral type of F9\,V, but did not observe the primary. We observed both stars for this work. We give the spectral type A2\,IV-Vn (normal) for the primary, and F9\,V for the secondary. The secondary shows emission reversals in the \ion{Ca}{ii} K and H lines, and is therefore an active star. Age-activity relations will allow an age for the system to be determined.
\textbf{Recommendation:} uncertain member.

\subsection{HD\,153747}
The results paper of \citet{paunzen2001} lists the `estimated' spectral type as hA7mA0\,V LB, where its $\delta$\,Sct variability is also noted; \citet{desikachary&mcinally1979} reported multiperiodicity in the light curve. \citet{paunzenetal2002a} gave [Z/H]$ = -0.86\pm0.20$. The pulsational nature of this star has led to wide coverage in the literature, but little appears to be said of its $\lambda$\,Boo character. We obtained a new spectrum to rectify this, but the classification was difficult. The hydrogen lines fit best at A7, taking into account the enormous difference in line blanketing, but a match near A1\,V$^-$ cannot be excluded, although the core is a bit too deep and narrow for this 
match. In either case, the star is metal-weak, and would be classified as Lambda Boo under both interpretations. The photometry really does not help to decide between the two, because of the possibility of reddening, and the fact that a very metal-weak A7 star would be quite blue. The dominance of \specline{Ca}{i}{4226} in its region and complete lack of Ti/Fe\,{\sc ii}\,$4172$--$9$ \specline{Fe}{ii}{4233} argues for the cooler type. Our final spectral type is A7\,V\,kA0mA0~$\lambda$\,Boo. \textbf{Recommendation:} member.

\subsection{HD\,153808}
\citet{abt&morrell1995} gave the classification A0\,IVp ($\lambda$\,Boo) with $v\sin i = 50$\,km\,s$^{-1}$, whereas \citet{gray&garrison1987} gave the classification A0\,IV$^{+}$. The star's duplicity (multiplicity) was discussed by \citet{faraggianaetal2001a}, where the contamination is demonstrated to be large. There is no evidence, other than a contaminated and uncorroborated classification by \citet{abt&morrell1995}, to suggest we confer $\lambda$\,Boo membership.
\textbf{Recommendation:} non-member.

\subsection{HD\,154153}
\citet{rensonetal1990} listed this star as misclassified, but it did not appear to be decisively dropped by others; \citet{paunzenetal1997a} rejected it because it is an evolved star, while the online data for the observational paper of \citet{paunzenetal2001} has the spectral type `A3\,V LB', and the results paper \citep{paunzen2001} gave the spectral type `hF0mA5 ($\lambda$\,Boo)'. In their investigation into the period-luminosity-colour relation in pulsating $\lambda$\,Boo stars, \citet{paunzenetal2002a} treated this star as a $\lambda$\,Boo star and found it to be non-variable. We observed this target and give the spectral type F1\,V\,kA3mA3~$\lambda$\,Boo?, where the late hydrogen line type has prevented distinction between a $\lambda$\,Boo star and a Pop.\,II star until NLTE abundances of volatile elements are available. The $\lambda4481$ line is weak.
\textbf{Recommendation:} probable member.

\subsection{HD\,156954}
The remark for this star in \citet{paunzenetal1997a} reads `\ion{Mg}{ii} too weak, otherwise normal metal spectrum', but \citet{paunzen2001} later gave the spectral type `hF1mA5\,V ($\lambda$\,Boo)' in his results paper. \citet{solanoetal2001} provided physical parameters. Metal deficiencies there are \mbox{$\sim0.5$} to $1.0$ dex, with $<$[Fe/H]$>\:=-0.66$\,dex. \citet{gerbaldietal2003} noted an inconsistent UV flux and suspected binarity. We have classified a new spectrum of this star as F1\,Vs\,kA5mA4, with the comment `H cores are shallow. Not Boo, just metal weak -- $\lambda4481/\lambda4383$ ratio is the same as at F1, or perhaps very marginally weak, and at F1 this is a blend with Fe.' With the low $v\sin i$ of 51\,km\,s$^{-1}$ \citep{heiteretal2002a} an NLTE abundance analysis of the volatile elements would provide a decisive assessment.
\textbf{Recommendation:} uncertain member.

\subsection{HD\,159082}
Classified as A0\,IVp $\lambda$\,Boo by \citet{abt&morrell1995}. \citet{haucketal1998} found a circumstellar \ion{Na}{i}\,D signature, but noted this star was classified as a HgMn star (cf.\ entry in the \citealt{renson&manfroid2009} catalogue as A0\,HgMn) and should thus be rejected. \citet{gerbaldietal2003} said the UV flux was inconsistent and this star is a suspected binary; indeed, \citet{bidelman1988} described it as a 6.80-d binary and noted strong \ion{Hg}{ii}. \textbf{Recommendation:} non-member.

\subsection{HD\,160928}
Listed in the \citet{rensonetal1990} catalogue. A comment in the \citet{paunzenetal1997a} consolidated catalogue of $\lambda$\,Boo stars notes that \ion{Mg}{ii} is too weak, but the spectrum is otherwise normal. \citet{paunzenetal2001} later classified the star as A2\,IV (wk met), i.e.\ not as $\lambda$\,Boo. This, in itself, is a good reason to exclude the star here. Fulfils the criteria of \citet{faraggianaetal2004} for having a composite spectrum, because it is in the Washington Double Star Catalog as having a separation of less than 2$''$ from a star with a magnitude difference not larger than 2.2\,mag. Indeed, this is evident in our classification spectrum, in that the H cores appear weak. We give the spectral type A2\,IV$^-$n, and note a normal $\lambda4481$ line, given the rapid rotation.
\textbf{Recommendation:} non-member.

\subsection{HD\,161223}
Studied among other pulsating $\lambda$\,Boo stars by \citet{gopkaetal2007}, where its chemical composition was provided for the first time and the authors described it as ``a new $\lambda$\,Boo star''. However, it was already classified as a ``mild but bona fide $\lambda$\,Boo star'' by \citet{gray&corbally2002}, with a spectral type of A9 V kA5 mA5 ($\lambda$\,Boo). It has, in the past, been classified as an SX\,Phe star (a population II $\delta$\,Sct star), but the pulsation frequencies quoted by \citet{gopkaetal2007} are quite low for such an object. Furthermore, its proper motions (-2.7, 5.2\,mas/yr; no parallax available) are much smaller than one expects for a population II star. \citet{gopkaetal2007} found $\log g = 3.3$, indicating this is quite an evolved $\lambda$\,Boo star, and they state ``the classiÞcation of V2314\,Oph as a $\lambda$\,Boo type star seems reliable.'' Clarification with a UV spectrum should be sought, and an abundance analysis is desirable.
\textbf{Recommendation:} member.

\subsection{HD\,161868}
Classified as `A0\,Vp (4481 wk)n' by \citet{abt&morrell1995}, with $v\sin i = 185$\,km\,s$^{-1}$. \citet{gray&garrison1987} had earlier classified this as A0\,Van, and \citet{grayetal2003} classified this star as A1\,Vn\,kA0mA0 with $T_{\rm eff} = 8951$\,K, $\log g=4.03$ and [M/H]$= -0.81$. \citet{haucketal1998} reported an expanding circumstellar gas from the \ion{Ca}{ii} K line of this star (see also their figure\;1.h). Perhaps it was the lack of `$\lambda$\,Boo' in the \citet{gray&garrison1987} classification led \citet{paunzenetal1997a} to reject this star from the $\lambda$\,Boo group when making their consolidated catalogue.
\textbf{Recommendation:} non-member.

\subsection{HD\,168740}
Proposed as a `good $\lambda$\,Boo candidate' by \citet{hauck1986}. \citet{paunzenetal1999a} found NLTE [C] and [O] abundances of $-0.42$ and $-0.03$, respectively. \citet{solanoetal2001} provided key physical parameters and gave a spectral type of hA7mA2\,V and [Fe/H]$=-0.73$. The abundance analysis in \citet{heiter2002} agrees with the $\sim1$\,dex underabundance of metals, and the abundance pattern was described there as being typical of $\lambda$\,Boo stars. From our new spectrum we give the spectral type A9\,V\,kA2mA2~$\lambda$\,Boo.
\textbf{Recommendation:} member.

\subsection{HD\,168947}
The results paper of \citet{paunzen2001} lists this as a new $\lambda$\,Boo star with a spectral type of kA3hF0mA3\,V $\lambda$\,Boo. We confirm this spectral type with our new spectrum. In their analysis of the \citet{rodriguezetal2000} catalogue of $\delta$\,Sct stars, \citet{rodriguez&breger2001} tabulated this star as a 20-mmag $\delta$\,Sct star with an oscillation frequency of 17\,d$^{-1}$ (its variability was first detected by \citealt{paunzenetal1994}). Str\"omgren photometry from \citet{gray&olsen1991} supports the spectral type.
\textbf{Recommendation:} member.

\subsection{HD\,169009}
Spectral type of A1\,V $\lambda$\,Boo according to \citet{abt&morrell1995}, but a normal A0\,IV star in the \citet{paunzenetal2001} observational data. Since they were looking for $\lambda$\,Boo stars, and did not classify this as $\lambda$\,Boo, we can assert for now that it is not a member. This assertion is apparently justified, as in their extensive notes on this star, \citet{griffinetal2012} discussed the peculiarity of this star as `B9\,V~He-wk', and presented no evidence supporting a $\lambda$\,Boo classification.
\textbf{Recommendation:} non-member.

\subsection{HD\,169022}
Originally classified as `$\lambda$\,Boo or composite' by \citet{slettebak1975}, but \citet{gray1988} argues misclassification. \citet{faraggianaetal1990} confirmed this star meets none of the UV criteria for $\lambda$\,Boo membership, except from a negative $\Delta m_1$ index. In the \citet{rensonetal1990} catalogue it is listed as misclassified, and \citet{gerbaldietal2003} wrote the spectral type as `A0\,II$^{-}$(n)shell $\dots$ misclassified', where that spectral type is the one assigned by \citet{gray&garrison1987}. High proper motion but small tangential velocity.
\textbf{Recommendation:} non-member.

\subsection{HD\,169142}
\citet{paunzenetal2001} classified HD\,169142 as `F0\,Ve'. \citet{folsometal2012} found `a clear $\lambda$\,Boo pattern of abundances' in this star. `C and O are solar, and S is only slightly underabundant, while iron peak elements are underabundant by between 0.5 and 1\,dex.' We have obtained two spectra of this target, which we classify as F1\,V\,kA3mA3 and F1\,V\,kA4mA5. Both spectra show shallow H cores, but neither has a typical $\lambda$\,Boo morphology. The metal lines have inconsistent strength, being stronger in the violet than the blue. The shallow H cores could be from emission, though we see no clear emission lines, or due to binarity. \blu{Indeed, weak emission in the H\,$\alpha$ \citep{dunkinetal1997} and the \specline{O}{i}{6300} line \citep{ackeetal2005} are recorded. The former gave the classification A5\,Ve, and noted `an absence of any significant depletion of Ca, Ti or Fe.'} We infer HD\,169142 is a spectrum variable.
\textbf{Recommendation:} uncertain member.

\subsection{HD\,170000}
A0\,IIIp($\lambda$\,Boo) in the \citet{abt&morrell1995} catalogue, with $v\sin i = 65$\,km\,s$^{-1}$. \citet{garrison&gray1994} gave the spectral type `kB9hB9HeA0\,V (Si)', in agreement with the \citet{renson&manfroid2009} entry `A0\,Si'. \blu{HD\,170000 was even used as an Ap standard in testing of the spectral classification software, {\sc mkclass} \citep{gray&corbally2014}}. \red{The WDS catalogue \citep{masonetal2001} lists a companion at 0.5$''$ with $\Delta V = 1.4$\,mag. A probable orbital period of 27\,d is given there.}
\textbf{Recommendation:} non-member.

\subsection{HD\,170680}
Has a spectral type of A0\,Van ($\lambda$\,Boo) NHL in \citet{paunzen&gray1997}, with the additional note `classified as A0\,Vp (Ca, Mg wk) by \citet{abt1984b}, confirmed as $\lambda$\,Boo star in the UV by \citet{baschek&slettebak1988}.' Gray has reclassified that spectrum for this work, as `A0\,Van\,kB9~($\lambda$\,Boo).' NLTE abundances for [C] and [O] show solar values, at $-0.06$ and $-0.07$\,dex, respectively \citep{paunzenetal1999a}. Abundance analysis of four metals shows mild weakness \citep [$\sim0.4$\,dex,][]{heiter2002}, but $v\sin i$ is recorded there as 200\,km\,s$^{-1}$, making abundance analysis difficult. \red{Listed as a double star with 0.1$''$ separation in the WDS catalogue \citep{masonetal2001}, but with no matching entry in the CCDM \citep{dommagnet&nys2002}.}
\textbf{Recommendation:} member.

\subsection{HD\,171948\,A}
Classified as A0\,Vp $\lambda$\,Boo by \citet{abt1985}. Has a spectral type from \citet{paunzen&gray1997} of A0\,Vb $\lambda$\,Boo NHL. Gray has reclassified that spectrum for this work as `A3\,Va$^{-}$\,kB8.5~$\lambda$\,Boo', and commented this is an extreme $\lambda$\,Boo star with a very weak $\lambda$4481 line. \citet{ilievetal2002} conducted a full abundance analysis on this binary system and determined that both stars are true $\lambda$\,Boo stars, and that the system is quite young (10--100\,Myr). They found $T_{\rm eff}=9000\pm200$\,K, $\log g =4.0 \pm 0.15$ and $v_{\rm mic} = 2\pm0.5$\,km\,s$^{-1}$ for both components, with $v\sin i = 15$ and 10\,km\,s$^{-1}$ for the A and B components, respectively. Each star has a mass listed there as $2.0\pm0.1$\,M$_{\odot}$.
\textbf{Recommendation:} member.

\subsection{HD\,171948\,B}
See entry for HD\,171948\,A.
\textbf{Recommendation:} member.

\subsection{HD\,172167}
This star, Vega, was noted to have $\lambda$\,Boo like abundances by \citet{baschek&slettebak1988} and is listed as an uncertain member of the $\lambda$\,Boo group in the \citet{rensonetal1990} catalogue. Section\:6.5 of \citet{holweger&rentzsch-holm1995} is dedicated to the discussion of Vega and $\beta$\,Pic as $\lambda$\,Boo stars, but for Vega this is largely focussed on its rotation velocity and inclination angle. It is listed there with the `dusty, normal stars'. Its abundances are like those of a mild $\lambda$\,Boo star, with C and O being solar while metals are about 0.5\,dex below solar (see \citealt{hekkeretal2009} for a literature comparison). The question of whether Vega should be adopted was addressed by \citet{ilijicetal1998}. They decided it should be based on abundances. The metal weakness is also observed in the UV, but there is no published application of $\lambda$\,Boo UV criteria to Vega. A cursory examination of one of many of the IUE spectra of Vega for this work showed no 1600-\AA\ flux depression, and a normal C/Al ratio. It thus remains unclear whether Vega should be accepted as a (mild) $\lambda$\,Boo star.
\textbf{Recommendation:} probable member.

\subsection{HD\,174005}
Described as an A0+A2 spectroscopic binary in the online catalogue of \citet{paunzenetal2001}, it was given the spectral type A7\,V\,kA2\,mA2 $\lambda$\,Boo by \citet{grayetal2001a}, with the comment that it is a `classic $\lambda$\,Boo star'. \red{\citet{faraggianaetal2004} included HD\,174005 in their list of $\lambda$\,Boo stars with composite spectra, but the maximum separation is 37.9$''$ \citep{solanoetal2001,masonetal2001},} which is rather far for contamination to be an issue. We note from our own observations that this is the brightest star around, and all other noticeable but fainter stars within 1$'$ have their own HD designations.
\textbf{Recommendation:} member.

\subsection{HD\,175445}
Spectral type of hA5mA2\,V $\lambda$\,Boo in the results paper of \citet{paunzen2001}, with the note that this star is very similar to HD\,120500. Not much information is available on this star; Str\"omgren photometry indicates a spectral type around A2/3 with slightly weak metals, but reddening is unknown. There is no abundance analysis or UV spectroscopy to evaluate membership, but our new spectrum shows this is a normal star, to which we give the spectral type A1.5\,Van.
\textbf{Recommendation:} non-member.

\subsection{HD\,177120}
Originally termed $\lambda$\,Boo by \citet{abt1985}, it was classified as `A0.5\,IV (shell)' in the online data of the observations paper of \citet{paunzenetal2001}, but importantly, not classified as $\lambda$\,Boo, indicating the star should not be listed as a member of the $\lambda$\,Boo group. This means it was `downgraded' by them from $\lambda$\,Boo, on account of its appearance in the earlier paper of \citet{paunzenetal1997a} as a $\lambda$\,Boo star. This is explained by the (unpublished) classification notes that Gray wrote for that paper: ``The hydrogen lines are certainly peculiar; their profiles are best matched at A0/A1, but they are shallow in the cores, and the wings are roughly III-IV, although this is a function of the hydrogen line. For instance, at H\,$\gamma$, the wings are best matched at III-IV, but at H\,$\delta$ they are closer to IV, and at the higher [blue-ward] H-lines they agree well with the A1\,IV std. The discrepancy at the core is most pronounced for H\,$\gamma$, but is present for all hydrogen lines. The K-line has a curious profile, having broad wings and a shallow, pointed core. All of these details suggest a shell phenomenon. Thus, A0.5 IV (shell).'' We obtained a new spectrum, which shows exactly the same properties. Although HD\,117120 is potentially an astrophysically interesting object worthy of spectroscopic follow-up, there is no evidence in favour of it being a $\lambda$\,Boo star. Although we reject this candidate, an abundance analysis and would be valuable, as would a UV spectrum, but no IUE data exist for this star.
\textbf{Recommendation:} non-member.

\subsection{HD\,177756}
An \citet{abt&morrell1995} $\lambda$\,Boo star that was suspected of having a composite spectrum on account of its variable RV \citep{faraggianaetal2004}, yet RV measurements in the \citet{griffinetal2012} debunking paper show constant RV within the errors. Those authors claimed any RV variation could be explained by rapid rotation and an oblate spheroidal stellar structure. The variations are not consistent with a SB, they said. They commented the spectrum is around B8 or B9, with a metal deficiency of about 0.5\,dex. Spectral type given there was `B9\,IV (He-wk)', and it was noted that the Si lines looked normal.

We investigated this star's IUE spectrum for this work, but the results are inconclusive: the C/Al ratio ($\lambda$1657/$\lambda$1671) would suggest a $\lambda$\,Boo-like character in a cooler star, but at such an early spectral type the behaviour of these lines is unknown. Contrary to typical $\lambda$\,Boo spectra in the UV, there is no 1600-A depression. Our spectral type based on a new blue-violet spectrum is of a normal B8.5\,Vn star, though the $\lambda4481$ line is weak. In reality, HD\,177756 is just too early for a $\lambda$\,Boo star.
\textbf{Recommendation:} non-member.

\subsection{HD\,179218}
Spectral type of A0\,IVe \citep{moraetal2001}. Quoting \citet{folsometal2012}: ``We found a solar He abundance, abundances of C, O and Na that are marginally enhanced relative to solar, and N and S abundances that are uncertain, but apparently enhanced relative to solar. Iron peak elements are depleted relative to solar by $\sim$0.5 dex. We conclude that HD\,179218 displays $\lambda$\,Boo peculiarities.'' Additional membership criteria, such as investigation of the UV spectrum, should be evaluated in light of the apparently mild peculiarity.
\textbf{Recommendation:} probable member.

\subsection{HD\,179791}
Uncertain classification in the \citet{rensonetal1990} catalogue. Hydrogen-line profile was determined to be normal \citep{iliev&barzova1993a}, where they described the \specline{Mg}{ii}{4481} line to be strong. Described as misclassified by \citet{paunzenetal2002a}; indeed, \citet{paunzenetal2001} gave it the spectral type A2\,IV in their search for $\lambda$\,Boo stars. \citet{gerbaldietal2003} said the UV spectrum showed solar abundances.
\textbf{Recommendation:} non-member.

\subsection{HD\,181470}
A `significantly evolved star' according to table\:1 of \citet{king1994}, and thus not pursued further by him. \citet{narusawaetal2006a} called this star a `metal-poor A star' and provided abundances from the literature showing metal deficiencies of between 0.4 and 0.8\,dex alongside a sub-solar C abundance of -0.36\,dex. \citet{cowley1991} made reference to this star's low $v\sin i$, putting it in a category alongside Vega as a suspected $\lambda$\,Boo star seen pole-on. \citet{abt&morrell1995} did actually classify this star as `A0\,V', which is the same type as Vega. There is little evidence suggesting HD\,181470 is a $\lambda$\,Boo star. We therefore observed this star at classification resolution and found it to be distinctly non-$\lambda$\,Boo in nature. We give the spectral type kA0hA1mA1\,III-IVs, which is typical of a mild (hot) Am star. The $\lambda4481$ line is normal. We also looked at an available IUE spectrum, which showed this star is clearly not a $\lambda$\,Boo star (and is perhaps a mild Am star); in $\lambda$\,Boo stars the \specline{C}{i}{1657} / \specline{Al}{ii}{1670} ratio is usually enhanced, but in this star it is considerably weak. \red{In our observations we did not detect a binary nature, but the WDS catalogue \citep{masonetal2001} records a separation of 0.1$''$ and notes that this is the spectroscopic binary U\,Sge. \citet{ibanogluetal2006} determined the absolute parameters of this binary and gave spectral types B7.5\,V + G4\,III-IV.}
\textbf{Recommendation:} non-member.

\subsection{HD\,183007}
Erroneously dubbed a probable Am star by \citet{abt&morrell1995} with the classification `Am (A1/A4:/A3)', but the early K and metal lines are actually consistent with the $\lambda$\,Boo type. Entry in \citet{paunzenetal2001} online data is just `A1\,IV'. Gray's notes on Paunzen's spectrum \blu{(private communication)} read: ``Yes, this is a mild Lambda Boo star$\dots$ hA7mA3 V Lam Boo'', yet strangely, \citet{paunzen2001} did not include it in his `new and confirmed' list. Known binary in a 165-d orbit \citep{budaj1997}. \citet{cucchiaroetal1979} noted that if their UV classification scheme is applied to this star, it has an Am character in the UV. High proper motion with a transverse velocity of 43.5\,km\,s$^{-1}$.

We assign the spectral type A8\,Vs\,kA2(p)mA3~$\lambda$\,Boo with the following comments: The \ion{Ca}{ii} K line is peculiar, with a narrow core but deep wings. This is common in magnetic Ap stars, but this spectrum shares no other properties with Ap stars. Instead, the peculiar K line is probably indicative of a composite spectrum. Indeed, we also note that the metal lines are stronger in the violet, where they have a strength of around A5, than in the blue where they are at A2 or A3. The metals are still quite weak compared to the H line type, especially the \specline{Mg}{ii}{4481} line, which is much weaker than for an A8 star. This indicates a $\lambda$\,Boo nature.

In conclusion, the spectrum shows a variety of evidence consistent with the composite nature reported in the literature. HD\,183007 could be a mild LB star with an Am companion, but that hypothesis is very difficult to evaluate.

\textbf{Recommendation:} uncertain member.

\subsection{HD\,183324}
\citet{gray1988} classified the spectrum as A0\,Vb $\lambda$\,Boo NHL, with the note `very weak K line, metallic-line spectrum featureless. $\lambda4481$ is only marginally present. Hydrogen lines show very broad wings. Hydrogen line profiles are quite normal, except that the cores may be slightly weak for A0.' He has reclassified the star for this work to give a K-line and metallic line type, as `kB9hA3mB9\,Va~$\lambda$\,Boo,' and commented that this is an extreme $\lambda$\,Boo star. \citet{rensonetal1990} described this as one of only six `definite' members of the $\lambda$\,Boo class.  One of the more extreme $\lambda$\,Boo stars, with metal underabundances of 1 to 2\,dex, with normal C and Na \citep{stuerenburg1993}. \citet{iliev&barzova1993b} observed the hydrogen lines, confirming the NHL profile with $T_{\rm eff} = 9100$\,K. Possibly variable, according to \citet{kuschnigetal1994b}. The IR study of \citet{andrillatetal1995} resulted in the note `LB shell? The \ion{O}{i} lines are sharp, although not strong and the \ion{C}{i} feature at 10684[\AA] is resolved into two components as in other shells.' Key physical parameters are available \citep{iliev&barzova1995}. Volatiles are near-solar ([C]$ = -0.14$, [N]$ = -0.30$ and [S]$ = -0.13$; \citealt{kampetal2001}). RV variable \citep{gerbaldietal2003}. Ambiguous evidence for circumstellar gas, and no indication of dust \citep{paunzenetal2003}.
\textbf{Recommendation:} member.

\subsection{HD\,184190}
No metal deficiency seen \citep{solanoetal2001}, contradicting the classification of A7/9s wl ``no metals'' by \citet{houk1978}. Thus there is no good evidence for $\lambda$\,Boo status, but a lack of metal deficiency at high resolution is sufficient to rule it out.
\textbf{Recommendation:} non-member.

\subsection{HD\,184779}
The spectral type in \citeauthor{paunzen2001}'s (2001) list of $\lambda$\,Boo stars is kA4hF1mA4\,V ($\lambda$\,Boo), where it is described as a `very cool $\lambda$\,Boo star similar to HD\,107223.' \citet{houk1978} gave the spectral type A3/5\,IIp, with the qualifier `probably early weak lined rather than luminous (narrow H lines)'. Appears in the paper on field horizontal branch stars by \citet{philip&hayes1983} but is labelled there as Pop.\,I. We have reobserved this star and classify it as F0.5\,V\,kA5mA5~($\lambda$\,Boo). We accept it as a mild $\lambda$\,Boo star, but abundances are desirable for confirmation.
\textbf{Recommendation:} member.

\subsection{HD\,187949}
Identified by \citet{hauck1986} as a metal weak binary in a search for new $\lambda$\,Boo stars, with spectral type A0\,V + F8\,IV, but that spectral type actually comes from the Bright Star Catalogue \citep{hoffleit&jaschek1982}. Eclipsing binary of the Algol type (\blu{detached}). \citet{faraggianaetal1990} rejected the star because the only evidence of $\lambda$\,Boo character they found was a slight discrepancy between the optical and UV spectral types, and on account of the binarity. Our new spectrum has shallow H cores, indicative of the binarity, and we give the spectral type A2\,IV(n).
\textbf{Recommendation:} non-member.

\subsection{HD\,188164}
\citet{paunzenetal1996a} concluded this `$\lambda$\,Boo(?)' star was non-pulsating at the 4-mmag level. Spectral type of A3\,V in \citet{paunzenetal2001}, i.e.\ they did not label this star $\lambda$\,Boo, and they were specifically looking for $\lambda$\,Boo stars. We have reobserved this target and give the spectral type \mbox{A5\,IV-V\,kA2mA3~($\lambda$\,Boo)}, but we note that there is degeneracy in the hydrogen line profile with A7\,V being an equal (and good) match. The $\lambda$\,Boo peculiarity is only mild, and we recommend a full abundance analysis.
\textbf{Recommendation:} member.

\subsection{HD\,188728}
Spectral type of A1- in the \citet{renson&manfroid2009} catalogue, where the `-' character is used to denote an Am star. Temperature subclass is in agreement with other literature classifications. \citet{lemke1989} determined [Fe/H] was $\sim$0.5\,dex above solar. One of the few stars appearing \textit{above} the line of normal stars in the $a$ vs.\ $b-y$ plot of \citet{maitzen&pavlovski1989}, whereas $\lambda$\,Boo stars fall below that line. Misclassified according to the \citet{rensonetal1990} catalogue.
\textbf{Recommendation:} non-member.

\subsection{HD\,191850}
Spectral type of kA4hF0mA4\,V $\lambda$\,Boo in the results paper of \citet{paunzen2001}, where its entry in the Michigan Catalogue as A2\,II/III was described as an indication of probable membership in the $\lambda$\,Boo group. Atmospheric parameters listed by \citet{paunzenetal2002a}, including [Z]$=-0.96\pm0.30$. We re-observed the star, and give the spectral type F0\,V\,kA3mA3~$\lambda$\,Boo.
\textbf{Recommendation:} member.

\subsection{HD\,192424}
Described by \citet{abt1985} as `A2\,Vp $\lambda$\,Boo' and the southern of a 6$''$ common proper motion pair. This star is listed as a $\lambda$\,Boo star in the \citet{rensonetal1990} catalogue, but \citet{paunzenetal2002a} state this star is misclassified. \citet{gerbaldietal2003} found the best fit to the star's spectrum was with solar metallicity.
\textbf{Recommendation:} non-member.

\subsection{HD\,192640 (29\,Cyg)}
Discovered by \citet{slettebak1952}, this is one of the best studied $\lambda$\,Boo stars. It has a spectral type of A0.5\,Va$^{-}$ $\lambda$\,Boo PHL \citep{gray1988}. Additional comments there were ``$\lambda$4481 extremely weak. The K line and the metallic-line spectrum are intermediate in strength to the A0 and A1 standards. The hydrogen lines have very broad, shallow wings and very shallow, weak cores $\dots$ many lines of \ion{Fe}{ii} are either missing or very weak. The \ion{Si}{ii} doublet $\lambda\lambda4128-30$ is also extremely weak. $\delta$\,Sct star with an amplitude of 0.03\,V.'' We reclassified the star for this work, as `kA1.5hA7mA0.5~$\lambda$\,Boo.' Doubly periodic \citep{paunzen&handler1996}. Well below the line of normal stars (and therefore among the $\lambda$\,Boo stars) in the $a$ vs.\ $b-y$ plot of \citet{maitzen&pavlovski1989}. Meets all the UV criteria for membership \citep{faraggianaetal1990}. One of only six stars with the certain member designation in the \citet{rensonetal1990} catalogue. PHL confirmed in \citet{iliev&barzova1993a}, where a strong \ion{Ca}{ii} K line was noted. \citet{stuerenburg1993} conducted an abundance analysis that showed metals are 1.0--1.5\,dex below solar and [C/H]\:$\sim-0.1$. A later abundance analysis shows C, N, O and S are approximately solar (within 1 or 2 sigma) and almost all metals studied are 1--2\,dex below solar \citep{heiteretal1998}. An NLTE abundance analysis of [N], [O] and [S] yielded -0.55, -0.15, -0.39 \citep{kampetal2001}. The low $v\sin i$ (35\,km\,s$^{-1}$, \citealt{abt&morrell1995}) favours abundance analyses, though interestingly \citet{paunzenetal2003} found a higher $v\sin i$ of 80\,km\,s$^{-1}$ when looking in the IR spectrum of this star, and found evidence for interstellar gas and ambiguous evidence for circumstellar dust. \citet{iliev&barzova1998} observed H$\gamma$ emission which they describe as `a shell seen nearly pole-on', but \citet{paunzenetal2003} ruled out an active accretion disk. Key physical parameters are available \citep{iliev&barzova1995}.
\textbf{Recommendation:} member.

\subsection{HD\,193063}
Noted as a $\lambda$\,Boo star in a wide binary with spectral type A0\,III by \citet{abt1985}. Also in the \citet{rensonetal1990} catalogue. A double/multiple star with a magnitude difference $\Delta V = 0.62\pm0.01$ and separation 5.38$''$ \citep{nakosetal1995}. There was not much evidence for or against $\lambda$\,Boo membership in the literature, so we reobserved this target. We give the spectral types for HD\,193063A and B as B9\,III and B9\,IIIa, respectively. The spectra of the two stars are very similar, with the H\,$\gamma$ line slightly narrower in B. \textbf{Recommendation:} non-member.

\subsection{HD\,193256}
Spectral type of A2\,Va $\lambda$\,Boo PHL in \citet{gray1988}. The comments on this object recorded there are extensive, since this star is in a widely separated (27.5$''$) double with HD\,193281, with the latter being 1.1\,mag brighter. Perhaps the binarity is what led \citet{rensonetal1990} to record this star's membership in the $\lambda$\,Boo group as uncertain. \citet{stuerenburg1993} carried out an abundance analysis of both objects, where for both stars [C] is normal, [Na] is $+1$\,dex and most metals are at $-0.5$ to $-1$\,dex. The pattern is not the same for each metal for each star. Unusually, Mg is not weak for either star, in contradiction with Gray's (\citeyear{gray1988}) assessment. Key physical parameters are available for both stars \citep{iliev&barzova1995}. \citet{paunzenetal2003} looked for gas and dust around these stars, finding only interstellar gas and no evidence of dust. They recorded $v\sin i$ for each star, obtaining 250 and 95\,km\,s$^{-1}$, respectively. \citet{soubiranetal2010} listed $T_{\rm eff}=7860$, $\log g=3.74$ and [Fe/H] $=-0.95$ for HD\,193256. We give the spectral type A9\,Vn\,kA2mA2~$\lambda$\,Boo to HD\,193256 from our new spectrum.
\textbf{Recommendation:} member.

\subsection{HD\,193281}
See also entry on HD\,193256. Classified as `A3mA2Vb lambda Boo' \citep{gray&garrison1987}. This is one of those $\lambda$\,Boo candidates where the chosen luminosity class has a large bearing on the inferred peculiarity. For instance, the spectral types `A7\,Vn\,kA2mA2' and `A2\,IVn' both describe the spectrum well, but since the latter does not require the star to be peculiar, it is preferred. The spectrum is a good match to the A2\,IVn standard ($\beta$\,Ser), with possibly a slightly weak $\lambda$4481 line, but the rotational broadening causes uncertainty. Given literature abundance analyses that suggest this is a $\lambda$\,Boo star, the evidence is in conflict.
\textbf{Recommendation:} uncertain member.

\subsection{HD\,193322\,D}
\citet{rensonetal1990} designated the membership of this star in the $\lambda$\,Boo group as uncertain. Its spectral type there is recorded as B8\,V~Si, and there appears to be no independent suggestion that it should be a member of the $\lambda$\,Boo group.\\
\textbf{Recommendation:} non-member.

\subsection{HD\,196821}
Classification as a HgMn star led \citet{haucketal1998} to reject this star from the $\lambda$\,Boo group, though interestingly the \citet{renson&manfroid2009} catalogue does not list it as such, but rather as B9-. Its spectral type from \citet{abt&morrell1995} is A0\,IIIp ($\lambda$\,Boo)s, with $v\sin i = 10$\,km\,s$^{-1}$. \citet{haucketal1998} also flagged this star as `SB?', perhaps because of the low $v\sin i$; the radial velocity from the bright star catalogue is -37\,km\,s$^{-1}$. It occurs in \citeauthor{faraggianaetal2004}'s (2004) list of $\lambda$\,Boo stars with composite spectra. The work of \citet{griffinetal2012}, refuting the composite spectrum hypothesis, includes this star with the spectral type A0\,III$^{+}$p\,kB8mA1 (CP). They showed this star is \textit{not} a spectroscopic binary. They note changes in line profiles and depths that are `unmistakeable', but state the overall envelope of a line does not move, which they interpret as a variable surface abundance rather than an SB2 spectrum. They confirm the spectral peculiarities but not of the $\lambda$\,Boo kind.
\textbf{Recommendation:} non-member.

\subsection{HD\,198160}
\citet{gray1988} gives the spectral type as A2\,Vann wk4481, with the note `Forms a close visual binary with HD\,198161 [separation 2.4$''$]. The very broad, shallow hydrogen-line wings seen in this star may be an effect of rapid rotation (cf.\ GG), but $\lambda$4481 is extremely weak. \citet{corbally&garrison1980} classify this star as A2\,Vn with A3\,Vn for HD\,198161. \citet{corbally1984} lists a spectral type for the pair as A2\,III and A3\,III, but notes that his (underexposed) 67\AA/mm spectra give a type of V for each, and that isochrones put the stars near the main sequence. Note the extremely red $b-y$ color [=0.108] for an A2 star.'

The abundance analysis by \citet{stuerenburg1993} shows solar C, Na, and Mg, and moderate underabunances ($-0.5$ to $-1$\,dex) of most metals for both stars, under the assumption that they are twins (with the same $T_{\rm eff}$ and $\log g$), though \citet{gerbaldietal2003} later argued this assumption to be invalid based on magnitude differences of 0.35 and 0.39 in V and B, respectively. \citet{iliev&barzova1995} provided key physical parameters (for HD\,198160 only). Circumstellar absorption in the \ion{Ca}{ii} K line was reported as `probable' by \citet{holweger&rentzsch-holm1995}. NLTE abundances of [C] and [O] are $-0.16$ and $-0.18$, respectively \citep{paunzenetal1999a}, but these are quoted for HD\,198160/1, i.e. not as separate entities. Evidence for dust causing an infrared excess was presented by \citet{paunzenetal2003}, where uncertain circumstellar gas absorption was also noted.

The observation of substantial metal deficiency for each star is in favour of membership, but solar NLTE abundances of volatile elements are not confirmed for each individual star. The abundances are suggestive of the typical $\lambda$\,Boo abundance pattern, so we tentatively accept both stars as members.
\textbf{Recommendation:} member.

\subsection{HD\,198161}
See entry on HD\,198160.
\textbf{Recommendation:} member.

\subsection{HD\,200841}
This star was first `estimated' to be a $\lambda$\,Boo star by \citet{olsen1980}, and was subsequently shown to lie well below the line of normal stars in the \citet{maitzen&pavlovski1989} $\Delta a$ photometry study. However, it was classified by \citet{abt1984} as `A0\,Vn', which led \citet{rensonetal1990} to call it an uncertain $\lambda$\,Boo star in their catalogue. We have obtained a new spectrum of this star which we classify as A0\,III-IV(n). The spectrum is a good match to the A0\,IVn standard, HR\,8451, with the exception of slightly narrower H lines in HD\,200841.
\textbf{Recommendation:} non-member.

\subsection{HD\,201019}
This star's late spectral type [F3w (for metal weak), \citealt{houk&cowley1975}], and red colour ($B-V = 0.34$), make it an uncertain member of the $\lambda$\,Boo group, as reflected in the \citet{rensonetal1990} catalogue.

\citet{gray1988} commented on this star, that its membership as a horizontal branch star seems to be ruled out by low space velocities and by a $\delta c_1$ (=0.21) index in common with main-sequence stars. He suggested the late nature of this star might indicate it is evolved, but that it `lies on the extension of the ``$\lambda$\,Bootis distribution'' in the $b-y$, $m_1$ plane'' (see figure\:2, there). He recommended a high-resolution spectrum for further classification. \citet{king1994} recorded a `doubtful' detection of an \textit{IRAS} infrared excess. We have reobserved this star and give the spectral type F2\,V\,kA7mA6~$\lambda$\,Boo?, where an abundance analysis of volatile elements is required to distinguish between a true $\lambda$\,Boo or a Pop.\,II star.
\textbf{Recommendation:} probable member.

\subsection{HD\,201184}
The identification of this star as a $\lambda$\,Boo member is ambiguous according to \citet{andrievskyetal2002}. Their spectroscopic analysis provided $T_{\rm eff} = 9970$\,K and $\log g = 4.2$, which agrees with the spectral type of A0\,V \citep{houk&smith-moore1988} and $B-V=0.01$. \citet{andrievskyetal2002} recorded $v\sin i = 200$\,km\,s$^{-1}$ and found only mild metal underabundances (e.g.\ [Fe/H] = -0.41) but gave no C or N abundances.
When investigating the observational properties of $\lambda$\,Boo stars, \citet{paunzenetal2002b} even used HD\,201184 as a normal comparison star, presumably because they were unaware of its slight metal weakness. $uvby\beta$ observations \citep{hauck&mermilliod1998} suggest the strongest of hydrogen lines (i.e.\ around A2\,V), but the $m_1$ and $c_1$ indices are consistent with an A0\,V star, and thus point to a mild metal weakness. Cross-referencing the \citet{eggleton&tokovinin2008} catalogue of multiplicity among bright stars shows this is a single star.

In the absence of firm evidence of this being a $\lambda$\,Boo star, we reobserved and classified its spectrum. It has the spectral type A0.5\,Vn, and is therefore not a $\lambda$\,Boo star.
\textbf{Recommendation:} non-member.

\subsection{HD\,204041}
Classified as `A1 Vb $\lambda$\,Boo PHL' \citep{gray1988}, with the following comment: `The metallic-line spectrum and the K line are similar to the A1 standards. $\lambda$4481 is weak for that spectral type. The hydrogen-line profiles are peculiar with broad wings but weak cores.' Str\"omgren colours and an LTE abundance analysis were given in \citet{stuerenburg1993}, where normal C and Na are shown alongside $\sim1$\,dex underabundances of most metals. A later NLTE abundance analysis \citep{paunzenetal1999a} for C and O revealed [C/H]$ = -0.81$ and [O/H]$ = -0.38$, and for N and S \citep{kampetal2001} revealed [N/H]$=-0.35$ and [S/H]$ = -0.17$. Key physical parameters are available \citep{iliev&barzova1995}. \citet{gerbaldietal2003} remarked an `inconsistent UV flux' (see their figure\:9). HD\,204041 was also included in a study of $\lambda$\,Boo stars compared with dusty, normal A stars by \citet{holweger&rentzsch-holm1995}, where its $v\sin i$ is listed as 68\,km\,s$^{-1}$. They reported no evidence of circumstellar material in the \ion{Ca}{ii} K line core. \citet{solanoetal2001} remarked this star shows no sign of a companion in the spectrum, and provided an abundance analysis, yet \citet{faraggianaetal2004} suspected HD\,204041 is a double star.

The abundances do not distinguish unambiguously in favour of $\lambda$\,Boo classification, but do hint at it. \textbf{Recommendation:} member.

\subsection{HD\,204754}
A known variable star, with a spectral type of B8\,III \citep{hube1970}, and an uncertain member of the $\lambda$\,Boo class according to \citet{rensonetal1990}. $T_{\rm eff}$ and $\log g$ have been recorded in \citet{cenarroetal2007} as 12923\,K and 3.50, respectively, making this a very hot $\lambda$\,Boo candidate. \citet{cayreldestrobeletal1997} listed [Fe/H] as $-0.28$, which is not non-solar to any high significance. We obtained a new spectrum for this star, which we classify as B5\,III-IVs. We note that the \ion{Ca}{ii} K line is quite strong, probably indicating a substantial interstellar component, and that \ion{Si}{ii} and some \ion{Ti}{ii} lines are enhanced. It is not a $\lambda$\,Boo.
\textbf{Recommendation:} non-member.

\subsection{HD\,204965}
In the \citet{rensonetal1990} catalogue as a misclassified member, probably because \citet{hauck1986} rejected it as a $\lambda$\,Boo member for having a high luminosity class (III). \citet{abt&morrell1995} then classified the star as `A2\,Vp \specline{Mg}{ii}{4481}-wk'. It was listed as a spectroscopic binary by \citet{gerbaldietal2003}, but this is not recorded on SIMBAD. $uvby\beta$ photometry \citep{hauck&mermilliod1998} puts the star near A3, for which the $m_1$ value indicates metal weakness but the $c_1$ value is high, implying an evolved star or perhaps a binary. There is little evidence in favour of membership, and a new spectrum obtained for this work has spectral type A2\,IV~4481-wk; it is not a $\lambda$\,Boo star.
\textbf{Recommendation:} non-member.

\subsection{HD\,207978}
Listed in the \citet{rensonetal1990} catalogue as a misclassified member. We found no source that recommended membership. This star has a late spectral type (F6IV-V\,w\,var, \citealt{barry1970}) for a $\lambda$\,Boo star, though some spectral types are earlier with no indication of metal weakness (F2\,V, \citealt{abt&morrell1995}). It appears that only the metal-weak classification would have led to a suggestion of this star as a member in the first place.
\textbf{Recommendation:} non-member.

\subsection{HD\,210111}
\citet{gray1988} gave the spectral type ``A2hA7mA2\,Vas~$\lambda$\,Boo $\dots$ PHL'', and the comment: ``hydrogen lines show very weak cores like those of an A7 or possibly F0 star, but broad and very shallow wings. The K line and the metallic-line spectrum are similar to the A2 standards, with $\lambda$4481 weak with respect to that type. The spectrum of this star is very similar to that of HD\,161817, a well-known blue horizontal-branch star, except for the very broad wings.'' The abundance analysis by \citet{stuerenburg1993} showed $-1$\,dex metals, while NLTE abundances \citep{kampetal2001} for [C] and [O] were $-0.45$ and $-0.20$, respectively. Key physical parameters are available \citep{iliev&barzova1995}. \citet{solanoetal2001} confirmed the low metal abundances. Known SB, and the subject of a dedicated paper \citep{paunzenetal2012}.
\textbf{Recommendation:} member.

\subsection{HD\,210418}
\citet{andrillatetal1995} studied this star's infrared spectrum and remarked that it appears to be a normal star. \citet{gray&garrison1987} gave the spectral type A2mA1\,IV-V  with the note `SB2, and therefore the spectrum may be composite and not actually metal-weak'. High proper motion, with a transverse velocity of 38.1\,km\,s$^{-1}$. We reject the star due to lack of evidence in favour of membership, but UV observations and an abundance analysis could rule out membership definitively.
\textbf{Recommendation:} non-member.

\subsection{HD\,212061}
A binary star and one of the first stars identified as a $\lambda$\,Boo star \citep{parenago1958}, but there is no trace of the original report of an abnormal spectrum. One of the few stars in the \citet{maitzen&pavlovski1989} $\Delta a$ photometry study that lies with the normal stars rather than the $\lambda$\,Boo stars. It shows none of the UV characteristics of $\lambda$\,Boo stars \citep{faraggianaetal1990}, and was described as `misclassified' in the \citet{rensonetal1990} catalogue. Indeed, \citet{gray&garrison1987} gave the normal spectral type `B9.5\,III-IV'.
\textbf{Recommendation:} non-member.

\subsection{HD\,212150}
Classified as `A0\,Vp $\lambda$\,Boo' by \citet{abt&morrell1995}, but this is not supported elsewhere. \citet{gerbaldietal2003} reported an inconsistent UV flux and the star is in the list compiled by \citet{faraggianaetal2004} of stars that likely display composite spectra, but with no specific comment. Not much is known about this star, and there is no compelling evidence to admit membership at all. Our spectrum of HD\,212150, with spectral type B9\,IIInp\,kA0, shows that this is a peculiar star, but it is not a $\lambda$\,Boo star. $\lambda4481$ is normal.
\textbf{Recommendation:} non-member.

\subsection{HD\,213669}
HD\,213669 was given the spectral type kA1hF0mA1\,V $\lambda$\,Boo by \citet{paunzenetal2002a}, who noted that bona-fide status was pending a full abundance analysis. Atmospheric parameters are also given there, but it is the only paper reporting on this star's $\lambda$\,Boo status. We therefore obtained a new spectrum, for which our spectral type is F0.5\,V\,kA2.5mA2.5~$\lambda$\,Boo. Abundances would still be nice. No IUE spectrum is available.
\textbf{Recommendation:} probable member.

\subsection{HD\,214454}
\citet{abt&morrell1995} gave the spectral type `F0\,Vp ($\lambda$\,Boo; metals A6)'. SB according to \citet{gerbaldietal2003}, but most probable configuration is single star according to the \citet{eggleton&tokovinin2008} catalogue of multiplicity among bright stars. An abundance analysis \citep{erspamer&north2003} showed a mild $\lambda$\,Boo pattern. A new spectrum obtained for this work has the spectral type A9.5\,V\,kA7mA6. There is some mild metal weakness but $\lambda$4481 is normal -- this is not a $\lambda$\,Boo star.
\textbf{Recommendation:} non-member.

\subsection{HD\,216847}
Spectral type hF0mA3\,Vn $\lambda$\,Boo \citep{paunzenetal2001}, with a comment in the third paper of that series: ``The hydrogen lines agree very well with F0, although it is a very rapid rotator, it is clearly metal-weak'' \citep{paunzen2001}. Inconsistent UV flux and suspected binary \citep{gerbaldietal2003}. We indicate only probable membership until a spectrum can be classified against rapidly rotating standards to ensure it is of the $\lambda$\,Boo type.
\textbf{Recommendation:} probable member.

\subsection{HD\,217782 (2\,And)}
Identified by \citet{parenago1958} as a $\lambda$\,Boo star because it was an early A star below the main sequence, but no metal weakness was actually reported \citep{sargent1965}. Falls on the straight line representing normal unreddened stars in the $\Delta a$ photometry study of \citet{maitzen&pavlovski1989}. \citet{abt&morrell1995} gave the classification A1\,V. \citet{faraggianaetal1990} found none of the UV criteria for $\lambda$\,Boo membership are met, but the star does have the required negative $\Delta m_1$ index. There is disagreement in the literature over whether this is a $\lambda$\,Boo star, with at least two references arguing on each side. This is probably why its membership is uncertain in the \citet{rensonetal1990} catalogue. Described as `LB, shell' in the IR study of \citet[][see additional comments, end Sect.\,4, there]{andrillatetal1995}. Evidence of \ion{Na}{i} D lines and circumstellar \ion{Ca}{ii} K line absorption \citep{haucketal1998}. Appears in the list of \citet{faraggianaetal2004} of stars with composite spectra, following the comment `composite, Hipparcos' in \citet{gerbaldietal2003}\red{; the WDS catalogue \citep{masonetal2001} lists a companion at 0.2$''$ that is 2.5\,mag fainter}. We have obtained a new spectrum and classified it on the MK system as A2\,IIIn.
\textbf{Recommendation:} non-member.

\subsection{HD\,218396 (HR\,8799)}
Identified as a weak-lined star by \citet{abt&morrell1995}. One of the few stars \citet{gerbaldietal2003} did not call a composite, with a spectral type listed there as kA5hF0mA5\,V $\lambda$\,Boo. HR\,8799 is known to host four planets \citep[e.g.][]{soummeretal2011}. It is also a $\gamma$\,Dor star, and a whole paper linking the $\lambda$\,Boo and $\gamma$\,Dor classes focussed on this object (\citealt{gray&kaye1999}; see also \citealt{moyaetal2010a}, \citealt{wrightetal2011}, and \citealt{sodoretal2014} for asteroseimic studies of this star). The reader is referred to  \citet{gray&kaye1999} for more details.
\textbf{Recommendation:} member.

\subsection{HD\,220061}
Listed in the \citet{rensonetal1990} catalogue, and also called $\lambda$\,Boo by \citet{abt&morrell1995}. This was called a `normal star' of spectral type A5\,V in the IR study of \citet{andrillatetal1995}. The spectral type in \citet{gerbaldietal2003} is A8\,V\,kA5mA5 $\lambda$4481-weak (probably taken from \citealt{grayetal2001a}), and it is noted there as an RV variable, but \citet{griffinetal2012} found no line-profile variations to support binarity and played down those `RV variations'. They reported only mild metal deficiencies of $<0.5$\,dex, and described the $\lambda$\,Boo classification of this star as `questionable' and `tentative'. A cursory inspection of the UV spectrum (this work) was inconclusive on the $\lambda$\,Boo nature, possibly because of the late hydrogen line type. At $v\sin i=150$\,km\,s$^{-1}$ \citep{bernacca&perinotto1970}, an abundance analysis may be difficult. This star could be considered as an extremely mild $\lambda$\,Boo star, but the evidence is clearly discordant.
\textbf{Recommendation:} \blu{uncertain member}.

\subsection{HD\,220278}
Its spectral type in Vol.\,IV of the Michigan Catalogue \citep{houk&smith-moore1988} is A3\,V, and \citet{cowleyetal1969} gave A5\,Vn. Its spectrum is considered composite \citep[e.g.][]{faraggianaetal2004} because it is separated by less than 2$''$ from a star less than 2.2\,mag fainter. \red{Actual values of the separation and magnitude difference from the WDS catalogue are 0.2$''$ and 1.1\,mag \citep{masonetal2001}. HD\,220278} does appear in the catalogue of $\lambda$\,Boo stars by \citet{rensonetal1990}, and is given as `$\lambda$\,Boo, CP(Mg wk), close binary' in \citet{ohanesyan2008}, where the EW of the $\lambda\lambda2786-2810$ and 4481 features are described for many stars, but its status as a true $\lambda$\,Boo star is clearly doubtful.
\textbf{Recommendation:} uncertain member.

\subsection{HD\,221756}
Spectral type of `A1\,Va$^{+}$ ($\lambda$\,Boo) $\dots$ P/NHL' in \citet{gray1988}, with the note: `The K line and the metallic-line spectrum are similar to the A1 standards, except that $\lambda$4481 appears weak. The hydrogen lines show very broad wings with slightly weak cores. This star may be a transitional type between type NHL and type PHL. This star is a suspected photometric variable.' Re-observation for this work the star for this work yielded the spectral type `A3\,Va$^{+}$\,kA1mA0.5~($\lambda$\,Boo)'. LTE abundance analysis \citep{stuerenburg1993} is incomplete and shows only mild metal deficiency ($\sim0.5$\,dex), but it does have $\lambda$\,Boo features in the infrared \citep{andrillatetal1995}, and has a circumstellar \ion{Na}{i} D line. Key physical parameters are available \citep{iliev&barzova1995}. NLTE abundances from \citet{kampetal2001} show solar O and S; [N/H]$=-0.5$. \citet{paunzenetal1999b} included this star in their paper on `Accurate LTE abundances of seven well established $\lambda$\,Bootis stars', finding [C/H]$=0.0$ and metal abundances on the order $-0.5$\,dex. \citet{paunzenetal2003} reported both dust and circumstellar gas. The 60-$\upmu$m flux measurement is perhaps the most excessive they measured (their figure\:1). \citet{gerbaldietal2003} noted RV variability, though \citet{griffinetal2012} noted its RV to be constant to well within their obtained precision, and found no evidence for an SB2 spectrum -- a conclusion that they say was also reached by \citet{stutz&paunzen2006}.
\textbf{Recommendation:} member.

\subsection{HD\,222303}
\citet{maitzen&pavlovski1989} complained of unknown reddening for this star (whose $b-y$ value is 0.414!), but commented it would still be in the $\lambda$\,Boo regime in their $a$ vs.\ $b-y$ plot. It was perhaps the high $b-y$ that led \citet{paunzenetal1997a} to reject this star as a $\lambda$\,Boo member. It does appear in the \citet{rensonetal1990} catalogue, and its spectral type from \citet{abt1984} is `A6VmA1 $\lambda$\,Boo?'. The metal weakness is thus evident and a red colour is not quite sufficient ground for exclusion on its own. We obtained a new spectrum of this star, which is certainly peculiar but not typical for a $\lambda$\,Boo star. The spectrum does not match at any type. We tentatively give the rough spectral type A9\,III (met wk A3), but `metal weak, late-A giant' is as precise as it is possible to be. Because of the peculiar spectrum we cannot definitively rule out this star as a member of the $\lambda$\,Boo class. Further study is recommended.
\textbf{Recommendation:} uncertain member.

\subsection{HD\,223352}
\citet{abt&morrell1995} classified as A0\,Vp (Lam Boo)n. \blu{This is a 2.4-d eclipsing binary \citep{shobbrook2005} within the Blanco\,1 cluster. Inspection of the published light curve, at its somewhat limited resolution, indicates the secondary contributes on the order 10-20\:per\:cent of the combined light output, and its lines may be visible in a spectrum of good signal-to-noise.} Str\"omgren photometry ($\beta=2.893$, $b-y=-0.004$, $m_1= 0.153$, $c_1= 1.019$; \citealt{hauck&mermilliod1998}) would suggest a slightly metal weak star around A2. In this work, we re-examined the spectrum and gave the spectral type A0\,Van\,kB9 ((Lam Boo)), with the comment `$\lambda$4481 is weak. \ion{He}{i} and hydrogen-line profiles put spectral type firmly at A0, but K-line and metallic-line spectrum generally weak and closer to B9. A very mild (marginal) Lambda Boo.' \blu{Taking both the spectroscopic and binary information into account, we consider this a probable member of the $\lambda$\,Boo group.}
\textbf{Recommendation:} \blu{probable member.}

\subsection{HD\,225180}
A1\,IVp $\lambda$\,Boo according to \citet{abt&morrell1995}, but a few literature spectral types suggesting an early-A giant exist, including A1\,II-III by \citet{gray&garrison1987}. \citet{haucketal1998} recommended this star be excluded from the $\lambda$\,Boo group. Its \ion{Na}{i} doublet warranted special mention and a figure (their figure\:4), and they wrote that `these lines are strong and all the lines of the optical spectrum possess a shell component. The \ion{Na}{i} doublet presents a double structure in which a photospheric and a red-shift shell component may be identified. The shell component has an asymmetric background.'
\textbf{Recommendation:} non-member.

\subsection{HD\,225218}
Another \citet{abt&morrell1995} $\lambda$\,Boo star in a double or multiple system, having `duplicity induced variability' \citep{gerbaldietal2003}. The star is a slow rotator, probably as a result of the binarity. The B9\,III spectral type from \citet{cowleyetal1969} does not agree with the A3\,V classifications from \citet{barry1970} and \citet{abt&morrell1995}. No strong evidence for $\lambda$\,Boo membership. We obtained two new spectra of this star, and in neither does it look like a $\lambda$\,Boo star. The H lines could be placed at either A3\,IV or A5\,IV, but the stronger cores argue slightly for the former. Both the \ion{Ca}{ii} K line and \specline{Ca}{i}{4226} are weak compared to other metals, and lines of \ion{Sr}{ii} are slightly strong, so this could be a mild Am star. The compromise type that describes both spectra satisfactorily is A4\,IV. \red{The WDS catalogue \citep{masonetal2001} lists a separation of 0.1$''$ for this star, with a note that it is a spectroscopic binary, which might explain the difficulty in classifying the spectrum.}
\textbf{Recommendation:} non-member.

\subsection{HD\,228509}
The literature on this star is scarce. The $uvby\beta$ photometry from \citet{hauck&mermilliod1998} indicates a late-A temperature type with metal line strength near A0. We obtained a new spectrum of this star and classify it as A9\,V(n)\,kA5mA5, with a slight weakness in the $\lambda$4481 line. Whether or not the $\lambda$4481 line is weak enough to warrant $\lambda$\,Boo membership is uncertain, therefore a high-resolution spectrum and full abundance analysis, including volatile elements, will be the discriminant. The high $v\sin i$ may prove obstructive in this regard.
\textbf{Recommendation:} uncertain member.

\subsection{HD\,245185}
\citet{gray&corbally1998} observed this star but did not note any $\lambda$\,Boo peculiarities, instead they classified it as `A3\,Vae\,Bd$<$\,Nem1'. \citet{folsometal2012} wrote: ``We find He, C, N and O abundances that are consistent with solar, and iron peak abundances that are $\sim$0.8 dex below solar. While the uncertainties are relatively large for this star, the strong iron peak underabundances indicate $\lambda$\,Boo peculiarities.'' Their analysis showed that metals are $\sim$$2\sigma$ below solar. As such a young star, it is probably a spectrum variable.
\textbf{Recommendation:} uncertain member.

\subsection{HD\,261904 (NGC\,2264-138)}
\citet{gray&corbally1998} gave the spectral type A0\,Va$^{-}$ (($\lambda$\,Boo)) with the comment that the star has a slightly weak \specline{Mg}{ii}{4481} line and other slightly weak metal lines. \citeauthor{paunzen2001}'s (2001) list of `new and confirmed' $\lambda$\,Boo stars gives the spectral type as A0.5\,V $\lambda$\,Boo. \citet{andrievskyetal2002} determined the Mg and Fe abundances to be super-solar, and thus stated this star `can be definitely ruled out' as a member of the $\lambda$\,Boo group. Inspection of the classification resolution spectrum in \citeauthor{paunzen2000}'s (2000) PhD thesis indicates a star that is earlier than A0.5\,V, in that the hydrogen cores are too shallow and helium lines too strong for this spectral type.
\textbf{Recommendation:} uncertain member.

\subsection{HD\,278937}
\citet{gray&corbally1998} observed this star (IP\,Per) on two separate occasions, two years apart, and found the metal weakness to change in severity. They classified it as kA3hA7mA4\,III:er, where the uncertainty applies to the luminosity class and the emission features are redshifted compared to the photospheric H-line cores. They described it as ``slightly metal-weak, even though it does not show all of the characteristics of a $\lambda$\,Boo star.'' \citet{folsometal2012} carried out an abundance analysis, and wrote: ``We find solar abundances for C, N, O and S, while iron peak elements as well as Na, Mg and Si are between 0.5 and 0.7\,dex underabundant. This star shows clear $\lambda$\,Boo peculiarities.''
\textbf{Recommendation:} probable member.

\subsection{HD\,290492}
\citet{paunzen&gray1997} recorded this star as A0.5\,Vb ($\lambda$\,Boo) with the comment `Mild $\lambda$\,Bootis character, star belongs to the Orion OB1 belt \citep{guetter1981}, classified therein as A0\,V. Close binary system ($\Delta m = 0.9$, $d = 2''$) which was resolved without any contamination.' \citet{marchettietal2001}, on the other hand, said of HD\,290492 that it is a close binary `with values of the separation and of the magnitude difference such that only a composite spectrum can be observed.' and directly contested the \citet{paunzen&gray1997} result (see their section\:5). \red{According to the WDS catalogue \citep{masonetal2001}, the separation is 0.7$''$ and the magnitude difference is 0.6\,mag.} When we obtained a spectrum the seeing was poor, and we did not see this as a binary. Our classification is A1\,V(n)\,kA0.5mA0.5, and we note that the spectrum is generally slightly metal weak; \specline{Mg}{ii}{4481} is not extra weak. Clearly for this system it will be necessary to obtain \red{resolved} high-resolution spectra and carry out a full abundance analysis of each star, including for volatile elements, before a firm membership evaluation can take place.
\textbf{Recommendation:} uncertain member.

\subsection{HD\,290799}
Identified as a young $\lambda$\,Boo star by \citet{gray&corbally1993}, where it is listed as A2\,Vb\,$\lambda$\,Boo PHL. Described as one of the true ZAMS stars of that survey of OB associations for $\lambda$\,Boo stars. [paraphrasing:] The hydrogen line \textit{core} type is A6-A7, the metallic line type is A2 and the \ion{Ca}{ii} K line type is slightly earlier. \specline{Mg}{ii}{4481} is slightly earlier than the metallic-line spectrum. ``The [$m_1$] index confirms the metal weak character of the star. The $\beta$ index is too small for an early-A-type dwarf, another typical signature of a PHL $\lambda$\,Bootis star.'' Key physical parameters are available \citep{iliev&barzova1995}. An abundance analysis by \citet{andrievskyetal2002} confirmed this star's membership in the $\lambda$\,Boo class, with normal C and O, and [Z]$=-1.0$\,dex. The NLTE sodium abundance was determined there, from the $D_1$ and $D_2$ doublet, to be $+0.45$\,dex. Additionally, we obtained a new spectrum, which we classify as A7\,V\,kA2mA2~$\lambda$\,Boo, with the comment: PHL, cores match A7, but the wings are closer to A2. The metal line morphology matches the A7 type, but at a strength consistent with A2. The $\lambda4481$ line is substantially weakened.
\textbf{Recommendation:} member.

\subsection{HD\,294253}
Spectral type of B9.5\,Va ($\lambda$\,Boo) in \citet{paunzen&gray1997}, with the comment that this is one of the hottest as well as youngest $\lambda$\,Boo stars. We re-observed this star at 3.6-\AA\ resolution for this work, and assigned the spectral type `A0\,Va\,kB8.5~($\lambda$\,Boo).' \citet{andrievskyetal2002} conducted an abundance analysis, but the results were ambiguous as to the star's $\lambda$\,Boo nature, in that although O was solar, metals were only slightly subsolar, and only a few metals could have their abundances measured.
\textbf{Recommendation:} member.

\subsection{\blu{TYC\,3680-215-1 (BSD\,8-403)}}
A $\lambda$\,Boo member on Gray's website that was unpublished until now. We reobserved the star for this work. Its spectrum is consistent with an extreme $\lambda$\,Boo star with a hydrogen-line type of A0\,Va$^{-}$. No metallic lines are visible, with the exception of a narrow \ion{Ca}{ii} K line, which is probably interstellar. Diffuse interstellar band at $\lambda$4420 is visible, again suggesting significant interstellar absorption. This extreme metallic-line weakness makes it impossible to assign a metallic-line class. We thus admit BSD8-403 to the $\lambda$\,Boo class. Note that the B and V magnitudes on SIMBAD are incorrect; magnitudes in V, B-V and U-B are $11.10$, $0.29$ and $0.14$, respectively \citep{mermilliodetal1997}.
\textbf{Recommendation:} member.

\subsection{\blu{TYC\,4774-866-1 (T\,Ori)}}
\citet{gray&corbally1998} looked at T\,Ori and did not note any $\lambda$\,Boo peculiarity, even though they were looking for young $\lambda$\,Boo stars. They classified it as `A3\,IVeb\,Bd$>$\,Npc1', that is, an emission line star with a slightly blue-shifted component with respect to H-line cores, having a moderately strong Balmer decrement, and displaying a P\,Cygni profile. \citet{folsometal2012} concluded they could see $\lambda$\,Boo peculiarities, and wrote: ``The abundances for He, C, N and O are all consistent with solar, while the S abundance is almost 3$\sigma$ above solar$\dots$ The iron peak abundances are clearly $\sim$0.5 dex below solar. We conclude that T\,Ori has clear $\lambda$\,Boo peculiarities.'' We were therefore motivated to reobserve this target. From our spectrum, we assign the spectral type A8\,Ve\,kA2mA2~Bd$>$\,Npc1. The $\lambda4481$ line is normal for an A8 star. This is almost certainly a spectrum variable; in our particular spectrum it did not appear to be a $\lambda$\,Boo star, but at some times it may have a $\lambda$\,Boo nature.
\textbf{Recommendation:} uncertain member.

\section{Table and conclusions}
\label{sec:table}
The information in the extended table (Table\:\ref{tab:bigtable}) in this section is a compilation of literature values alongside values we provide ourselves. The SIMBAD identifier, parallax, $V$ and $B-V$ values are taken directly from the SIMBAD data base. Upon noticing some inaccuracies in the SIMBAD $V$ and $B-V$ values, we have checked a sample against The General Catalogue of Photometric Data (Vol. II; \citealt{mermilliodetal1997}). The agreement was satisfactory for almost all the test sample, so for simplicity we adopted the SIMBAD values for every star.

Membership recommendations for inclusion in the $\lambda$\,Boo class (fourth column) come from the discussions in the previous section and are given here as follows: $\checkmark$ denotes a member; $\circ$ a probable member; ? an uncertain member; and $\times$ a non-member. 

Transverse velocities have been calculated from SIMBAD proper motions and parallaxes. These are provided to help eliminate thick disc or Population II objects. One cannot use proper motions alone because nearby stars may have high proper motions yet only small space velocities (e.g.\ Vega). For stars without parallaxes, no transverse velocity could be calculated. We do, however, recognise the result of \citet{paunzenetal2014}, which was that intermediate Pop.\,II objects cannot be distinguished from $\lambda$\,Boo stars by velocities alone. Indeed, we make no attempt to do this, as is evident from the previous section. \blu{We do not provide an additional column for the uncertainties on the transverse velocities, but these are provided to be indicative only, and were used with very little weight in our membership evaluations.}

The spectral types column contains spectral types as provided in \S\ref{sec:evaluations}. When multiple are available, we have chosen which is the most appropriate, given the literary sources and the membership. Where possible, we provide our own spectral types based on spectra obtained for this work. Otherwise, we attempt to provide the most trustworthy spectral type from the literature for each object, favouring where possible, those in agreement with the membership recommendations in \S\ref{sec:evaluations}. This sometimes results in $\lambda$\,Boo classifications for stars we have rejected. We refer the reader to \S\ref{sec:evaluations} in those instances, and strongly recommend cross-referencing our evaluations before quoting any spectral types from Table\:\ref{tab:bigtable}.

Finally, we have provided $v\sin i$ for as many stars as possible, based only on values available in the literature, and we provide a reference for each value as a superscript. Those references can be found in Table\:\ref{tab:references}. For stars without $v\sin i$ values provided, we encourage the community to conduct abundance analyses and to determine $v\sin i$ for the continued investigation of stars belonging the $\lambda$\,Boo class.

The number of stars falling in the member, probable member, uncertain member and non-member groups are \blu{64, 19, 26 and 103}, respectively.

\begin{table*}
\centering
\caption{Summary table for the stars in \S\,\ref{sec:evaluations}. Col.\:3 indicates the membership recommendation given in \S\,\ref{sec:evaluations}, with symbols: $\checkmark =$ member; $\circ=$ probable member; ? = uncertain member; $\times=$ non-member. Col.\:4 the transverse velocity in km\,s$^{-1}$. Cols\:5\:\&\:6 are Johnson $V$ and $B-V$ magnitudes. Col.\:7 contains spectral types as specified in the text. Col.\:8 gives the projected stellar rotational velocity in km\,s$^{-1}$, for which references are given in Table\:\ref{tab:references}.}
\label{tab:bigtable}
\rowcolors{1}{}{lightgray}
\begin{tabular}{l l H c H r r r c c}
\toprule
HD number & SIMBAD identifier & Obj. type & Member? & $\pi$& T. Vel.& $V$&$B-V$&Sp.T.& $v\sin i$ \\
\midrule
HD	\phantom{11111}3	&	HR 1	&$	*i*	$&$	?	$&$	7.22	$&$	21.9	$&$	6.71	$&$	0.06	$&	A0\,Vn~($\lambda$\,Boo)	&$	228	^{a}	$	\\
HD	\phantom{111}319	&	HR 12	&$	*	$&$	\checkmark	$&$	12.76	$&$	28.9	$&$	5.94	$&$	0.13	$&	kA1hA7mA2\,Vb~$\lambda$\,Boo	&$	59	^{a}	$	\\
HD	\phantom{11}2904	&	HR 129	&$	*	$&$	?	$&$	6.59	$&$	25.6	$&$	6.38	$&$	-0.05	$&	A0\,Vnn~($\lambda$\,Boo)	&$	243	^{a}	$	\\
HD	\phantom{11}4158	&	HD 4158	&$	*	$&$	\checkmark	$&$		$&$		$&$	9.54	$&$	0.22	$&	F3\,V\,m-3	&$			$	\\
HD	\phantom{11}5789	&	HR 283	&$	*i*	$&$	\times	$&$	8.40	$&$	15.6	$&$	6.06	$&$	-0.03	$&	B9.5\,Vnn~($\lambda$\,Boo)	&$	300	^{b}	$	\\
HD	\phantom{11}6173	&	HD 6173	&$	*	$&$	?	$&$		$&$		$&$	8.52	$&$	0.13	$&	A0\,IIIn ?	&$			$	\\
HD	\phantom{11}6870	&	V* BS Tuc	&$	dS*	$&$	\checkmark	$&$	9.63	$&$	59.9	$&$	7.45	$&$	0.26	$&	kA2hA7mA2~$\lambda$\,Boo	&$	130	^{s}	$	\\
HD	\phantom{11}7908	&	HD 7908	&$	*	$&$	\checkmark	$&$	12.98	$&$	13.9	$&$	7.32	$&$	0.24	$&	hF0mA3~$\lambda$\,Boo	&$			$	\\
HD	\phantom{11}9100	&	V* VX Psc	&$	dS*	$&$	\times	$&$	6.84	$&$	40.5	$&$	6.02	$&$	0.14	$&	A3\,Vb~$\lambda$\,Boo	&$	123	^{a}	$	\\
HD	\phantom{1}11413	&	V* BD Phe	&$	dS*	$&$	\checkmark	$&$	12.96	$&$	17.7	$&$	5.94	$&$	0.14	$&	kA0.5hF0mA0.5\,Va~$\lambda$\,Boo	&$	139	^{a}	$	\\
HD	\phantom{1}11502	&	HR 545	&$	*i*	$&$	\times	$&$		$&$		$&$	4.70	$&$	-0.14	$&	A0\,IV-V(n)\,kB8~($\lambda$\,Boo)	&$	179	^{b}	$	\\
HD	\phantom{1}11503	&	V* gam Ari	&$	a2*	$&$	\times	$&$	21.00	$&$	29.6	$&$	4.52	$&$	-0.03	$&	knA0hA3\,(IV)\,SiSr	&$	201	^{c}	$	\\
HD	\phantom{1}11905	&	HR 562	&$	*	$&$	\times	$&$	3.27	$&$	11.0	$&$	6.77	$&$	-0.07	$&	B9\,HgMn	&$	68	^{d}	$	\\
HD	\phantom{1}13755	&	V* CV Phe	&$	dS*	$&$	\checkmark	$&$	5.80	$&$	10.6	$&$	7.83	$&$	0.28	$&	hF2mA5\,V~$\lambda$\,Boo	&$			$	\\
HD	\phantom{1}15164	&	HD 15164	&$	*	$&$	\checkmark	$&$	7.83	$&$	23.2	$&$	8.27	$&$	0.31	$&	F1 V kA7mA6 ($\lambda$\,Boo)?	&$	$	\\
HD	\phantom{1}15165\,A	&	V* VW Ari	&$	dS*	$&$	\checkmark	$&$	7.83	$&$	23.2	$&$	6.69	$&$	0.25	$&	F2 V kA2mA2 $\lambda$\,Boo?	&$	129	^{a}	$	\\
HD	\phantom{1}16811	&	* 34 Ari	&$	*	$&$	\times	$&$	8.56	$&$	30.5	$&$	5.74	$&$	-0.03	$&	A0\,IVn	&$	175	^{a}	$	\\
HD	\phantom{1}16955	&	HR 803	&$	*	$&$	\times	$&$	9.59	$&$	3.9	$&$	6.36	$&$	0.10	$&	A3\,V	&$	175	^{a}	$	\\
HD	\phantom{1}16964\,A/B 	&	HD 16964	&$	*i*	$&$	\times	$&$	2.83	$&$	7.8	$&$	8.86	$&$	0.08	$&	A0\,IV-V; A0.5\,IVn	&$			$	\\
HD	\phantom{1}17138	&	V* RZ Cas	&$	Al*	$&$	\circ	$&$	15.44	$&$	11.5	$&$	6.26	$&$	0.14	$&	A3\,V	&$	81	^{b}	$	\\
HD	\phantom{1}21335	&	HR 1036	&$	*	$&$	\times	$&$	8.18	$&$	21.3	$&$	6.57	$&$	0.15	$&	A3\,IVn	&$	217	^{a}	$	\\
HD	\phantom{1}22470	&	V* EG Eri	&$	a2*	$&$	\times	$&$	6.70	$&$	24.2	$&$	5.23	$&$	-0.13	$&	B9\,Si	&$	75	^{c}	$	\\
HD	\phantom{1}23258	&	HR 1137	&$	*	$&$	\checkmark	$&$	12.87	$&$	11.1	$&$	6.08	$&$	0.03	$&	kB9.5hA3mB9.5\,V~($\lambda$\,Boo)	&$	123	^{a}	$	\\
HD	\phantom{1}23392	&	HD 23392	&$	*	$&$	\checkmark	$&$	4.34	$&$	13.1	$&$	8.24	$&$	0.05	$&	A0\,Va$^{-}$~($\lambda$\,Boo)	&$			$	\\
HD	\phantom{1}24712	&	V* DO Eri	&$	a2*	$&$	\times	$&$	20.32	$&$	18.6	$&$	5.99	$&$	0.30	$&	A9\,Vp SrCrEu	&$	18	^{c}	$	\\
HD	\phantom{1}24472	&	HD 24472	&$	*	$&$	\circ	$&$	11.29	$&$	18.3	$&$	7.10	$&$	0.30	$&	*	&$			$	\\
HD	\phantom{1}26801	&	HD 26801	&$	*i*	$&$	\times	$&$	5.40	$&$	4.9	$&$	7.72	$&$	0.00	$&	A0\,III~$\lambda$\,Boo?	&$			$	\\
HD	\phantom{1}27404	&	V* V1140 Tau	&$	a2*	$&$	\times	$&$	4.40	$&$	38.1	$&$	7.92	$&$	0.26	$&	Ap\,Si	&$	37	^{u}	$	\\
HD	\phantom{1}30422	&	V* EX Eri	&$	dS*	$&$	\checkmark	$&$	17.8	$&$	4.8	$&$	6.18	$&$	0.17	$&	A7\,V\,kA3mA3~($\lambda$\,Boo)	&$	128	^{a}	$	\\
HD	\phantom{1}30739	&	* 2 Ori	&$	*	$&$	\times	$&$	14.53	$&$	9.8	$&$	4.35	$&$	0.02	$&	A0.5\,IVn	&$	218	^{b}	$	\\
HD	\phantom{1}31293	&	HD 31293	&$	Or*	$&$	\times	$&$	7.18	$&$	16.4	$&$	7.06	$&$	0.12	$&	A0\,Vaer\,Bd$<$\,Nem5	&$	116	^{n}	$	\\
HD	\phantom{1}31295	&	* 7 Ori	&$	*i*	$&$	\checkmark	$&$	28.04	$&$	22.9	$&$	4.66	$&$	0.08	$&	kA0hA3mA0\,Va$^{-}$~$\lambda$\,Boo	&$	108	^{b}	$	\\
HD	\phantom{1}34787	&	* 16 Cam	&$	*i*	$&$	\times	$&$	9.70	$&$	28.0	$&$	5.25	$&$	-0.02	$&	A0\,IIIn	&$	217	^{a}	$	\\
HD	\phantom{1}34797	&	V* TX Lep	&$	a2*	$&$	\times	$&$	4.09	$&$	12.0	$&$	6.54	$&$	-0.11	$&	B7\,Vp He-wk	&$	80	^{e}	$	\\
HD	\phantom{1}35242	&	V* V1649 Ori	&$	dS*	$&$	\checkmark	$&$	11.80	$&$	13.1	$&$	6.34	$&$	0.12	$&	kA0.5hA3mA1\,Va~$\lambda$\,Boo	&$	86	^{a}	$	\\
HD	\phantom{1}36496	&	HR 1853	&$	*	$&$	?	$&$	12.86	$&$	7.6	$&$	6.29	$&$	0.21	$&	A5\,Vn	&$	196	^{a}	$	\\
HD	\phantom{1}36726	&	HD 36726	&$	*i*	$&$	\checkmark	$&$		$&$		$&$	8.85	$&$	0.09	$&	kA0hA5mA0\,V~$\lambda$\,Boo	&$	80	^{g}	$	\\
HD	\phantom{1}37411	&	HD 37411	&$	Or*	$&$	\checkmark	$&$		$&$		$&$	9.86	$&$	0.11	$&	hA2\,Vae\,kB8mB8~$\lambda$\,Boo	&$			$	\\
HD	\phantom{1}37886	&	HD 37886	&$	*	$&$	\times	$&$		$&$		$&$	9.05	$&$	-0.06	$&	Ap\,HgMn	&$	19	^{t}	$	\\
HD	\phantom{1}38043	&	HD 38043	&$	*	$&$	\checkmark	$&$	2.72	$&$	5.1	$&$	9.45	$&$	0.22	$&	F1\,V\,kA5mA3~$\lambda$\,Boo?	&$			$	\\
HD	\phantom{1}38545	&	* 131 Tau	&$	*	$&$	\times	$&$	9.60	$&$	20.6	$&$	5.73	$&$	0.07	$&	A2\,Va$^{+}$~$\lambda$\,Boo	&$	191	^{a}	$	\\
HD	\phantom{1}39283	&	* ksi Aur	&$	*	$&$	\times	$&$	13.69	$&$	6.2	$&$	4.97	$&$	0.05	$&	A1\,Va	&$	68	^{b}	$	\\
HD	\phantom{1}39421	&	HR 2039	&$	*	$&$	?	$&$	10.11	$&$	15.1	$&$	5.96	$&$	0.08	$&	A1\,Vn	&$	227	^{a}	$	\\
HD	\phantom{1}40588	&	HR 2110	&$	*	$&$	\checkmark	$&$	13.9	$&$	6.9	$&$	6.19	$&$	0.09	$&	A3\,V\,kA0.5mA0~$\lambda$\,Boo	&$	123	^{a}	$	\\
HD	\phantom{1}41580	&	HD 41580	&$	*i*	$&$	\times	$&$	4.43	$&$	24.5	$&$	7.19	$&$	-0.06	$&	A1\,IIIp~Si 	&$			$	\\
HD	\phantom{1}42503	&	V* AU Col	&$	dS*	$&$	\checkmark	$&$	4.04	$&$	13.5	$&$	7.43	$&$	0.17	$&	hA9\,Vn\,kA2mA2~$\lambda$\,Boo	&$			$	\\
HD	\phantom{1}47152	&	* 53 Aur	&$	*	$&$	\times	$&$	9.41	$&$	11.6	$&$	5.75	$&$	0.01	$&	B9Mn+F0m	&$	33	^{c}	$	\\
HD	\phantom{1}54272	&	HD 54272	&$	RR*	$&$	?	$&$		$&$		$&$	8.77	$&$	0.25	$&	kA3hF2mA3\,V~$\lambda$\,Boo	&$			$	\\
HD	\phantom{1}56405	&	HR 2758	&$	*	$&$	\times	$&$	12.12	$&$	18.3	$&$	5.45	$&$	0.09	$&	A2\,Va	&$	159	^{a}	$	\\
HD	\phantom{1}64491	&	V* DD Lyn	&$	dS*	$&$	\checkmark	$&$	16.95	$&$	17.8	$&$	6.23	$&$	0.25	$&	F1\,Vs\,kA3mA3~$\lambda$\,Boo	&$	23	^{c}	$	\\
HD	\phantom{1}66684	&	CCDM J08056+2732AB	&$	**	$&$	\times	$&$	3.40	$&$	17.3	$&$	6.21	$&$	0.01	$&	B9\,Va+A1\,IVn	&$	75	^{a}	$	\\
HD	\phantom{1}66920	&	HR 3171	&$	*	$&$	\times	$&$	7.69	$&$	13.1	$&$	6.33	$&$	0.14	$&	A3\,V	&$			$	\\
HD	\phantom{1}68695	&	HD 68695	&$	*	$&$	?	$&$	7.69	$&$	6.6	$&$	9.87	$&$	0.06	$&	A0\,Ve	&$	44	^{n}	$	\\
HD	\phantom{1}68758	&	HR 3230	&$	*	$&$	\times	$&$	7.71	$&$	30.8	$&$	6.53	$&$	0.06	$&	A1\,IVp	&$	270	^{a}	$	\\
HD	\phantom{1}73210	&	HD 73210	&$	*	$&$	\times	$&$	4.83	$&$	37.1	$&$	6.75	$&$	0.16	$&	A5\,IIIs	&$	80	^{b}	$	\\
HD	\phantom{1}73345	&	V* CY Cnc	&$	dS*	$&$	\times	$&$	5.60	$&$	31.1	$&$	8.14	$&$	0.20	$&	A7\,V~kA8	&$	98	^{b}	$	\\
HD	\phantom{1}73872	&	HD 73872	&$	*iC	$&$	\times	$&$		$&$		$&$	8.34	$&$	0.21	$&	kA6hA8mA6\,V(n)	&$	180	^{b}	$	\\
HD	\phantom{1}74873	&	* 50 Cnc	&$	*	$&$	\checkmark	$&$	18.53	$&$	21.1	$&$	5.89	$&$	0.11	$&	kA0.5hA5mA0.5\,V~$\lambda$\,Boo	&$	130	^{i}	$	\\
HD	\phantom{1}74911	&	HD 74911	&$	*	$&$	\times	$&$	5.41	$&$	20.4	$&$	8.51	$&$	0.13	$&	A2\,IV~(4481-wk)	&$	190	^{g}	$	\\
\bottomrule
\end{tabular}
\label{LB1}
\end{table*}

\begin{table*}
\setcounter{table}{0}
\centering
\caption{continued from previous page.}
\rowcolors{1}{}{lightgray}
\begin{tabular}{l l H c H r r r c c}
\toprule
HD number & SIMBAD identifier & Obj. type & Member? & $\pi$& T. Vel.& $V$&$B-V$&Sp.T.& $v\sin i$ \\
\midrule
HD	\phantom{1}75654	&	V* HZ Vel	&$	dS*	$&$	\checkmark	$&$	13.15	$&$	27.1	$&$	6.38	$&$	0.22	$&	kA3hA7mA3\,V~$\lambda$\,Boo	&$	45	^{c}	$	\\
HD	\phantom{1}78316	&	V* kap Cnc	&$	a2*	$&$	\times	$&$	6.14	$&$	17.9	$&$	5.24	$&$	-0.09	$&	B8\,IIIp\,HgMnEu~4481-wk	&$	18	^{c}	$	\\
HD	\phantom{1}78661	&	HR 3635	&$	*	$&$	\times	$&$	25.69	$&$	14.1	$&$	6.48	$&$	0.33	$&	hF2mA8\,V	&$	82	^{c}	$	\\
HD	\phantom{1}79025	&	HR 3647	&$	*	$&$	\times	$&$	6.19	$&$	21.7	$&$	6.48	$&$	0.16	$&	A9\,Vn	&$			$	\\
HD	\phantom{1}79108	&	HR 3651	&$	*	$&$	\times	$&$	10.07	$&$	24.3	$&$	6.14	$&$	0.00	$&	A0\,V~$\lambda$\,Boo	&$	172	^{a}	$	\\
HD	\phantom{1}79469	&	LTT 12431	&$	PM*	$&$	\times	$&$	28.74	$&$	55.1	$&$	3.88	$&$	-0.06	$&	B9.5\,IV~(CII)	&$	100	^{b}	$	\\
HD	\phantom{1}80081	&	* 38 Lyn	&$	**	$&$	\times	$&$	26.13	$&$	23.5	$&$	3.82	$&$	0.06	$&	A2\,IV$^{-}$	&$	160	^{b}	$	\\
HD	\phantom{1}81104	&	* 21 UMa	&$	**	$&$	\times	$&$	5.10	$&$	40.9	$&$	7.66	$&$	0.13	$&	A3\,Van	&$			$	\\
HD	\phantom{1}81290	&	HD 81290	&$	*	$&$	\circ	$&$	4.22	$&$	18.3	$&$	8.89	$&$	0.26	$&	F2\,V\,kA3mA3~($\lambda$\,Boo?)	&$	56	^{h}	$	\\
HD	\phantom{1}82573	&	HR 3796	&$	*	$&$	\times	$&$	8.65	$&$	19.4	$&$	5.74	$&$	0.14	$&	A3\,V	&$	75	^{c}	$	\\
HD	\phantom{1}83041	&	V* AK Ant	&$	dS*	$&$	\circ	$&$	4.08	$&$	25.3	$&$	8.79	$&$	0.31	$&	F1\,V\,kA3mA2~$\lambda$\,Boo	&$	95	^{h}	$	\\
HD	\phantom{1}83277	&	HD 83277	&$	*	$&$	\circ	$&$	5.65	$&$	43.2	$&$	8.29	$&$	0.28	$&	F1.5\,V\,kA3mA3~$\lambda$\,Boo?	&$			$	\\
HD	\phantom{1}84123	&	HD 84123	&$	*	$&$	\checkmark	$&$	7.80	$&$	53.5	$&$	6.85	$&$	0.27	$&	hF2\,V\,kA6mA6~$\lambda$\,Boo &$	28	^{g}	$	\\
HD	\phantom{1}84948	&	HD 84948	&$	SB*	$&$	\circ	$&$	5.01	$&$	6.7	$&$	8.12	$&$	0.30	$&	F1.5\,Vs\,kA5mA5~$\lambda$\,Boo?	&$	45(A)/55(B)	^{h}	$	\\
HD	\phantom{1}87271	&	V* GM Leo	&$	dS*	$&$	\checkmark	$&$	6.13	$&$	4.3	$&$	7.13	$&$	0.18	$&	kA9hA9mA0\,V~$\lambda$\,Boo	&$			$	\\
HD	\phantom{1}87696	&	* 21 LMi	&$	dS*	$&$	\times	$&$	35.41	$&$	7.1	$&$	4.49	$&$	0.18	$&	A7\,V	&$	155	^{b}	$	\\
HD	\phantom{1}89239	&	HR 4041	&$	*	$&$	\times	$&$	6.98	$&$	18.6	$&$	6.53	$&$	-0.02	$&	B9.5\,V	&$	149	^{a}	$	\\
HD	\phantom{1}89353	&	V* AG Ant	&$	pA*	$&$	\times	$&$	1.23	$&$	74.7	$&$	5.53	$&$	0.23	$&	B9.5\,Ib-II	&$			$	\\
HD	\phantom{1}90821	&	HD 90821	&$	*	$&$	\times	$&$		$&$		$&$	9.45	$&$	0.09	$&	A3\,IV-V	&$	150	^{g}	$	\\
HD	\phantom{1}91130	&	* 33 LMi	&$	*i*	$&$	\checkmark	$&$	12.82	$&$	7.1	$&$	5.91	$&$	0.10	$&	kA0hA1mA0\,Va~($\lambda$\,Boo)	&$	207	^{a}	$	\\
HD	\phantom{1}97411	&	HR 4347	&$	*	$&$	\times	$&$	6.50	$&$	23.5	$&$	6.10	$&$	0.00	$&	A0\,V (4481-wk)	&$	33	^{a}	$	\\
HD	\phantom{1}97773	&	HD 97773	&$	*	$&$	?	$&$	4.73	$&$	23.3	$&$	7.55	$&$	0.25	$&	hA8mA3\,V~$\lambda$\,Boo?	&$			$	\\
HD	\phantom{1}97937	&	HR 4366	&$	*	$&$	?	$&$	15.01	$&$	20.3	$&$	6.66	$&$	0.27	$&	F1\,V\,kA9mA6	&$	133	^{a}	$	\\
HD	\phantom{1}98353	&	* 55 UMa	&$	*	$&$	\times	$&$	17.00	$&$	24.5	$&$	4.80	$&$	0.10	$&	A1\,Va; composite.	&$	60	^{b}	$	\\
HD	\phantom{1}98772	&	HR 4391	&$	*	$&$	\times	$&$	11.52	$&$	14.7	$&$	6.03	$&$	0.08	$&	A1\,Va	&$	249	^{a}	$	\\
HD	100546	&	HD 100546	&$	Be*	$&$	?	$&$	10.32	$&$	17.9	$&$	6.70	$&$	0.01	$&	A0\,Vae\,kB8	&$	55	^{p}	$	\\
HD	100740	&	HR 4464	&$	*	$&$	\times	$&$	9.35	$&$	19.6	$&$	6.57	$&$	0.13	$&	A3\,IVnn\,kA2.5mA2.5	&$	259	^{a}	$	\\
HD	101108	&	HD 101108	&$	*i*	$&$	\circ	$&$	4.91	$&$	8.1	$&$	8.89	$&$	0.15	$&	A3\,IV~(4481-wk)	&$	90	^{i}	$	\\
HD	101412	&	HD 101412	&$	Be*	$&$	\checkmark	$&$		$&$		$&$	9.29	$&$	0.18	$&	A3\,Va\,kA0mA0~$\lambda$\,Boo	&$	3	^{q}	$	\\
HD	102541	&	V* V1023 Cen	&$	dS*	$&$	\blu{\circ}	$&$	8.48	$&$	144.6	$&$	7.94	$&$	0.24	$&	A9\,V\,kA4mA4~($\lambda$\,Boo) &$			$	\\
HD	103483	&	V* DN UMa	&$	Al*	$&$	\times	$&$	4.72	$&$	10.7	$&$	6.54	$&$	0.10	$&	kA2hA5mA3\,V	&$	150	^{b}	$	\\
HD	105058	&	HD 105058	&$	V*	$&$	\checkmark	$&$	5.15	$&$	7.9	$&$	8.88	$&$	0.17	$&	hA8\,V\,kA0.5mA0.5~$\lambda$\,Boo	&$	130	^{e}	$	\\
HD	105199	&	HD 105199	&$	*	$&$	\times	$&$		$&$		$&$	9.83	$&$	0.14	$&	kA0.5hF0mA3\,V~$\lambda$\,Boo:	&$	20	^{b}	$	\\
HD	105260	&	HD 105260	&$	*	$&$	\blu{?}	$&$	2.68	$&$	78.5	$&$	9.21	$&$	0.22	$&	hF0mA5\,V	&$			$	\\
HD	105759	&	V* II Vir	&$	dS*	$&$	\checkmark	$&$	8.84	$&$	14.0	$&$	6.54	$&$	0.19	$&	A0	&$	120	^{h}	$	\\
HD	106223	&	HD 106223	&$	*	$&$	\checkmark	$&$	8.12	$&$	41.4	$&$	7.43	$&$	0.26	$&	F4\,V\,kA1.5mA1~$\lambda$\,Boo &$	80	^{g}	$	\\
HD	107223	&	HD 107223	&$	*	$&$	\times	$&$	4.40	$&$	35.8	$&$	8.19	$&$	0.10	$&	A1\,IVs	&$			$	\\
HD	107233	&	HD 107233	&$	*	$&$	\checkmark	$&$	11.27	$&$	22.7	$&$	7.36	$&$	0.24	$&	kA1hF0mA1\,V~$\lambda$\,Boo	&$	80	^{i}	$	\\
HD	108283	&	* 14 Com	&$	*iC	$&$	\times	$&$	11.82	$&$	7.0	$&$	4.92	$&$	0.26	$&	A9\,IVnp\,\ion{Sr}{ii}	&$	226	^{b}	$	\\
HD	108714	&	HD 108714	&$	*	$&$	\times	$&$	5.44	$&$	44.0	$&$	7.74	$&$	0.09	$&	A0	&$			$	\\
HD	108765	&	* 20 Com	&$	*	$&$	\times	$&$	11.90	$&$	16.3	$&$	5.68	$&$	0.09	$&	kA3hA3mA0\,V	&$	133	^{a}	$	\\
HD	109738	&	HD 109738	&$	*	$&$	\checkmark	$&$		$&$		$&$	8.30	$&$	0.20	$&	A9\,Vn\,kB9.5mA0~$\lambda$\,Boo	&$	166	^{h}	$	\\
HD	109980	&	* 9 CVn	&$	*	$&$	\times	$&$	14.74	$&$	9.2	$&$	6.35	$&$	0.19	$&	kA6hA8\,Vnn	&$	250	^{b}	$	\\
HD	110377	&	V* GG Vir	&$	dS*	$&$	\times	$&$	15.90	$&$	31.7	$&$	6.22	$&$	0.19	$&	A6\,Vp~($\lambda$\,Boo)	&$	175	^{a}	$	\\
HD	110411	&	V* rho Vir	&$	dS*	$&$	\checkmark	$&$	27.57	$&$	20.9	$&$	4.87	$&$	0.08	$&	kA0hA3mA0\,Va~$\lambda$\,Boo	&$	154	^{a}	$	\\
HD	111005	&	HD 111005	&$	*i*	$&$	\circ	$&$	6.24	$&$	32.4	$&$	7.97	$&$	0.33	$&	F2\,V\,kA5mA5~$\lambda$\,Boo?	&$	140	^{j}	$	\\
HD	111164	&	* 34 Vir	&$	*i*	$&$	\times	$&$	12.01	$&$	19.0	$&$	6.11	$&$	0.13	$&	A3\,IV-V	&$	191	^{a}	$	\\
HD	111604	&	V* DT CVn	&$	dS*	$&$	\checkmark	$&$	9.11	$&$	47.6	$&$	5.89	$&$	0.16	$&	kA1.5hA8mA1\,Vn~$\lambda$\,Boo	&$	200	^{b}	$	\\
HD	111786	&	V* MO Hya	&$	dS*	$&$	\checkmark	$&$	14.93	$&$	36.4	$&$	6.15	$&$	0.21	$&	F0\,V\,kA1mA1~$\lambda$\,Boo	&$	47	^{c}	$	\\
HD	111893	&	HR 4886	&$	*	$&$	\times	$&$	9.04	$&$	23.5	$&$	6.30	$&$	0.17	$&	A5\,IV-Vnn	&$	233	^{a}	$	\\
HD	112097	&	* 41 Vir	&$	*	$&$	\times	$&$	16.42	$&$	21.1	$&$	6.25	$&$	0.26	$&	kA7hF0mF0\,(V)	&$	71	^{c}	$	\\
HD	113848	&	* 39 Com	&$	*	$&$	\times	$&$	20.13	$&$	19.6	$&$	6.02	$&$	0.36	$&	kF0hF8mF3 p	&$	26	^{f}	$	\\
HD	114879	&	HD 114879	&$	*	$&$	\times	$&$	1.63	$&$	60.0	$&$	8.92	$&$	0.16	$&	A3\,V	&$			$	\\
HD	114930	&	HD 114930	&$	*	$&$	\times	$&$	7.37	$&$	28.5	$&$	9.01	$&$	0.29	$&	F1\,Vs	&$			$	\\
HD	118623	&	CCDM J13375+3617AB	&$	**	$&$	\times	$&$	16.42	$&$	28.4	$&$	4.82	$&$	0.23	$&	kA8hF0\,Vnn	&$	207	^{c}	$	\\
HD	119288	&	G 62-63	&$	PM*	$&$	\times	$&$	27.44	$&$	67.1	$&$	6.16	$&$	0.42	$&	F5\,V ((metal-weak))	&$	12	^{b}	$	\\
HD	120500	&	V* FQ Boo	&$	dS*	$&$	\checkmark	$&$	6.49	$&$	21.4	$&$	6.61	$&$	0.11	$&	kA1/5hA5mA1.5\,V~($\lambda$\,Boo)	&$	130	^{g}	$	\\
HD	120896	&	V* QT Vir	&$	dS*	$&$	\checkmark	$&$	2.66	$&$	34.4	$&$	8.51	$&$	0.25	$&	kA6hF0mA6\,V~$\lambda$\,Boo	&$	$	\\
HD	123299	&	THUBAN	&$	*	$&$	\times	$&$	10.76	$&$	26.0	$&$	3.68	$&$	-0.04	$&	A0\,IIIs	&$	15	^{b}	$	\\
HD	125162	&	LTT 14190~($=\lambda$\,Bo\"otis)	&$	PM*	$&$	\checkmark	$&$	32.94	$&$	35.4	$&$	4.18	$&$	0.08	$&	kB9hA3mB9\,Va~$\lambda$\,Boo	&$	123	^{c}	$	\\
HD	125489	&	HR 5368	&$	*	$&$	\times	$&$	15.81	$&$	12.1	$&$	6.19	$&$	0.19	$&	A7\,Vn	&$	159	^{a}	$	\\
HD	125889	&	HD 125889	&$	*	$&$	\circ	$&$		$&$		$&$	9.81	$&$	0.27	$&	F1\,V\,kA4mA4~($\lambda$\,Boo)	&$  $	\\
HD	128167	&	* sig Boo	&$	V*	$&$	\times	$&$	63.16	$&$	17.3	$&$	4.46	$&$	0.36	$&	kF2hF4mF1\,V	&$	5	^{b}	$	\\
HD	130158	&	* 55 Hya	&$	*	$&$	\times	$&$	4.39	$&$	16.8	$&$	5.61	$&$	-0.04	$&	A0\,II-IIIp~(Si)	&$	65	^{a}	$	\\
\bottomrule
\end{tabular}
\end{table*}

\begin{table*}
\setcounter{table}{0}
\centering
\caption{continued from previous page.}
\rowcolors{1}{}{lightgray}
\begin{tabular}{l l H c H r r r c c}
\toprule
HD number & SIMBAD identifier & Obj. type & Member? & $\pi$& T. Vel.& $V$&$B-V$&Sp.T.& $v\sin i$ \\
\midrule
HD	130767	&	HD 130767	&$	*	$&$	\checkmark	$&$	7.96	$&$	28.2	$&$	6.90	$&$	0.04	$&	A0\,Va~$\lambda$\,Boo	&$			$	\\
HD	138527	&	* 12 Ser	&$	*	$&$	\circ	$&$	6.95	$&$	4.8	$&$	6.22	$&$	-0.04	$&	B9.5\,Vp~($\lambda$\,Boo:)	&$	158	^{a}	$	\\
HD	139614	&	HD 139614	&$	pr*	$&$	\checkmark	$&$		$&$		$&$	8.24	$&$	0.23	$&	A7Ve 	&$	24	^{n}	$	\\
HD	141569	&	HD 141569	&$	pr*	$&$	\checkmark	$&$	8.61	$&$	14.7	$&$	7.12	$&$	0.08	$&	A2\,Ve\,kB9mB9~$\lambda$\,Boo	&$	228	^{n}	$	\\
HD	141851	&	* b Ser	&$	**	$&$	\times	$&$	20.10	$&$	22.5	$&$	5.10	$&$	0.13	$&	A2\,IVn	&$	229	^{a}	$	\\
HD	142666	&	HD 142666	&$	TT*	$&$	\times	$&$		$&$		$&$	8.82	$&$	0.55	$&	F0\,Vs\,shell? 	&$	65	^{n}	$	\\
HD	142703	&	V* HR Lib	&$	dS*	$&$	\checkmark	$&$	19.57	$&$	19.1	$&$	6.12	$&$	0.22	$&	kA1hF0mA1\,Va~$\lambda$\,Boo	&$	110	^{c}	$	\\
HD	142944	&	HD 142944	&$	*	$&$	\times	$&$		$&$		$&$	10.08	$&$	0.22	$&	A0\,V	&$			$	\\
HD	142994	&	V* IN Lup	&$	dS*	$&$	\checkmark	$&$		$&$		$&$	7.17	$&$	0.29	$&	F0\,V\,kA3mA3~$\lambda$\,Boo	&$			$	\\
HD	143148	&	HD 143148	&$	*	$&$	\times	$&$		$&$		$&$	7.39	$&$	0.28	$&	A7\,IVn	&$			$	\\
HD	144708	&	* 11 Sco	&$	*	$&$	\times	$&$	8.90	$&$	26.3	$&$	5.76	$&$	0.01	$&	B9\,Vp~($\lambda$\,Boo)nn	&$	275	^{a}	$	\\
HD	145782	&	HR 6040	&$	*	$&$	\times	$&$	7.71	$&$	31.4	$&$	5.63	$&$	0.12	$&	A3\,V	&$			$	\\
HD	148638	&	V* NP TrA	&$	dS*	$&$	?	$&$	5.33	$&$	23.4	$&$	7.90	$&$	0.19	$&	A2\,IV$^-$n~(4481-wk)	&$	160	^{j}	$	\\
HD	149130	&	HD 149130	&$	*	$&$	\circ	$&$	4.47	$&$	21.2	$&$	8.48	$&$	0.31	$&	kA7hF0mA7\,V~$\lambda$\,Boo	&$			$	\\
HD	149303	&	HR 6162	&$	*i*	$&$	?	$&$	14.43	$&$	12.8	$&$	5.66	$&$	0.11	$&	A3\,IV-V\,(4481-wk)~+~F9\,V	&$	275	^{j}	$	\\
HD	153747	&	V* V922 Sco	&$	dS*	$&$	\checkmark	$&$	5.54	$&$	13.6	$&$	7.42	$&$	0.12	$&	A7\,V\,kA0mA0~$\lambda$\,Boo	&$			$	\\
HD	153808	&	* eps Her	&$	SB*	$&$	\times	$&$	21.04	$&$	12.3	$&$	3.91	$&$	-0.01	$&	A0\,IV$^{+}$	&$	60	^{a}	$	\\
HD	154153	&	HR 6338	&$	*	$&$	\circ	$&$	14.35	$&$	9.3	$&$	6.20	$&$	0.25	$&	hF0mA5~($\lambda$\,Boo)	&$	125	^{c}	$	\\
HD	156954	&	HD 156954	&$	*	$&$	?	$&$	11.11	$&$	18.9	$&$	7.67	$&$	0.30	$&	F1\,Vs\,kA5mA4	&$	51	^{h}	$	\\
HD	159082	&	HR 6532	&$	*	$&$	\times	$&$	7.39	$&$	11.5	$&$	6.45	$&$	-0.01	$&	A0p\,HgMn	&$	22	^{a}	$	\\
HD	160928	&	HR 6597	&$	*	$&$	\times	$&$	13.87	$&$	1.9	$&$	5.88	$&$	0.14	$&	A2\,IV$^-$n	&$			$	\\
HD	161223	&	V* V2314 Oph	&$	dS*	$&$	\checkmark	$&$		$&$		$&$	7.43	$&$	0.33	$&	kA5hA9mA5\,V~($\lambda$\,Boo)	&$			$	\\
HD	161868	&	* gam Oph	&$	*	$&$	\times	$&$	31.73	$&$	11.7	$&$	3.75	$&$	0.04	$&	kA0hA1mA0\,V	&$	212	^{b}	$	\\
HD	168740	&	V* V346 Pav	&$	dS*	$&$	\checkmark	$&$	14.05	$&$	34.5	$&$	6.13	$&$	0.19	$&	A9\,V\,kA2mA2~$\lambda$\,Boo	&$	130	^{i}	$	\\
HD	168947	&	V* V704 CrA	&$	dS*	$&$	\checkmark	$&$		$&$		$&$	8.11	$&$	0.24	$&	F0\,V\,kA3mA3~$\lambda$\,Boo	&$			$	\\
HD	169009	&	HR 6878	&$	*	$&$	\times	$&$	9.75	$&$	9.8	$&$	6.34	$&$	0.12	$&	B9\,V He-wk	&$	44	^{a}	$	\\
HD	169022	&	KAUS AUSTRALIS	&$	*i*	$&$	\times	$&$	22.76	$&$	27.1	$&$	1.80	$&$	0.02	$&	A0\,II$^{-}$(n) (shell)	&$	236	^{c}	$	\\
HD	169142	&	HD 169142	&$	Em*	$&$	?	$&$		$&$		$&$	8.16	$&$	0.26	$&	F1\,V\,kA4mA5 var	&$	48	^{n}	$	\\
HD	170000	&	V* phi Dra	&$	a2*	$&$	\times	$&$	10.77	$&$	16.8	$&$	4.22	$&$	-0.10	$&	kB9hB9HeA0\,V (Si)	&$	75	^{c}	$	\\
HD	170680	&	HR 6944	&$	*i*	$&$	\checkmark	$&$	14.62	$&$	7.5	$&$	5.13	$&$	0.01	$&	A0\,Van\,kB9~($\lambda$\,Boo)	&$	222	^{a}	$	\\
HD	171948\,A	&	HD 171948\,A	&$	SB*	$&$	\checkmark	$&$	7.60	$&$	8.1	$&$	6.77	$&$	0.05	$&	A3\,Va$^{-}$\,kB8.5~$\lambda$\,Boo	&$	15	^{k}	$	\\
HD	171948\,B	&	HD 171948\,B	&$	SB*	$&$	\checkmark	$&$	$&$	$&$ 11.7	$&$	$&	&$	10	^{k}	$	\\
HD	172167	&	VEGA	&$	dS*	$&$	\circ	$&$	130.23	$&$	12.7	$&$	0.03	$&$	0.00	$&	A0\,Va	&$	5	^{b}	$	\\
HD	174005	&	HD 174005	&$	*i*	$&$	\checkmark	$&$	4.33	$&$	35.5	$&$	6.50	$&$	0.22	$&	A7\,V\,kA2\,mA2 $\lambda$\,Boo	&$	87	^{c}	$	\\
HD	175445	&	HD 175445	&$	*	$&$	\times	$&$	5.87	$&$	12.2	$&$	7.79	$&$	0.11	$&	A1.5\,Van	&$			$	\\
HD	177120	&	HD 177120	&$	*i*	$&$	\times	$&$	3.29	$&$	9.0	$&$	6.88	$&$	0.15	$&	A0.5\,IV (shell)	&$			$	\\
HD	177756	&	* lam Aql	&$	*	$&$	\times	$&$	26.37	$&$	16.7	$&$	3.43	$&$	-0.08	$&	B8.5\,Vn	&$	155	^{b}	$	\\
HD	179218	&	HD 179218	&$	Em*	$&$	\circ	$&$	3.94	$&$	25.5	$&$	7.39	$&$	0.09	$&	A0\,IVe	&$	69	^{n}	$	\\
HD	179791	&	HR 7288	&$	V*	$&$	\times	$&$	6.76	$&$	11.4	$&$	6.48	$&$	0.09	$&	A2\,IV	&$	196	^{a}	$	\\
HD	181470	&	HR 7338	&$	*	$&$	\times	$&$	5.24	$&$	13.7	$&$	6.26	$&$	0.00	$&	\red{B7.5\,V + G4\,III-IV}	&$	16	^{c}	$	\\
HD	183007	&	HR 7392	&$	*	$&$	?	$&$	16.03	$&$	43.5	$&$	5.71	$&$	0.19	$&	A8\,Vs\,kA2(p)mA3~$\lambda$\,Boo	&$			$	\\
HD	183324	&	V* V1431 Aql	&$	dS*	$&$	\checkmark	$&$	16.34	$&$	9.5	$&$	5.79	$&$	0.08	$&	kB9hA3mB9\,Va~$\lambda$\,Boo	&$	110	^{a}	$	\\
HD	184190	&	HD 184190	&$	*	$&$	\times	$&$		$&$		$&$	9.74	$&$	0.28	$&	A7/9s wl	&$			$	\\
HD	184779	&	HD 184779	&$	*	$&$	\checkmark	$&$		$&$		$&$	8.90	$&$	0.27	$&	F0.5\,V\,kA5mA4~($\lambda$\,Boo)	&$	$	\\
HD	187949	&	V* V505 Sgr	&$	Al*	$&$	\times	$&$	8.40	$&$	28.5	$&$	6.48	$&$	0.14	$&	A2\,IVn~+~F8\,IV	&$	101	^{b}	$	\\
HD	188164	&	HR 7588	&$	*	$&$	\checkmark	$&$	9.46	$&$	41.1	$&$	6.38	$&$	0.15	$&	A5\,IV-V\,kA2mA3~($\lambda$\,Boo)	&$			$	\\
HD	188728	&	* phi Aql	&$	*	$&$	\times	$&$	14.86	$&$	10.4	$&$	5.29	$&$	0.01	$&	A1m	&$	27	^{a}	$	\\
HD	191850	&	HD 191850	&$	V*	$&$	\checkmark	$&$		$&$		$&$	9.69	$&$	0.18	$&	F0\,V\,kA3mA3~$\lambda$\,Boo	&$	$	\\
HD	192424	&	HD 192424	&$	V*	$&$	\times	$&$	3.85	$&$	0.0	$&$	7.89	$&$	0.04	$&	A2\,Vp~$\lambda$\,Boo	&$			$	\\
HD	192640	&	V* V1644 Cyg	&$	dS*	$&$	\checkmark	$&$	23.42	$&$	19.9	$&$	4.96	$&$	0.14	$&	kA1.5hA7mA0.5~$\lambda$\,Boo	&$	25	^{b}	$	\\
HD	193063	&	IDS 20132+3923	&$	**	$&$	\times	$&$	0.37	$&$	52.3	$&$	7.73	$&$	-0.05	$&	B9\,III~+~B9\,IIIa	&$			$	\\
HD	193256	&	HD 193256	&$	*i*	$&$	\checkmark	$&$		$&$		$&$	7.64	$&$	0.19	$&	A9\,Vn\,kA2mA2~$\lambda$\,Boo	&$	240	^{m}	$	\\
HD	193281	&	CCDM J20205-2911AB	&$	**	$&$	?	$&$	7.35	$&$	0.3	$&$	6.30	$&$	0.17	$&	A2\,IVn	&$	95	^{a}	$	\\
HD	193322\,D&	HD 193322D	&$	*	$&$	\times	$&$		$&$		$&$	11.20	$&$		$&	B8\,V Si	&$			$	\\
HD	196821	&	HR 7903	&$	V*	$&$	\times	$&$	4.54	$&$	35.4	$&$	6.08	$&$	-0.04	$&	A0\,III$^{+}$p kB8mA1 (CP)	&$	22	^{c}	$	\\
HD	198160	&	HR 7959	&$	*i*	$&$	\checkmark	$&$		$&$		$&$	6.59	$&$	0.17	$&	A2\,Vann~$\lambda$\,Boo	&$	200	^{h}	$	\\
HD	198161	&	HR 7960	&$	*i*	$&$	\checkmark	$&$		$&$		$&$	6.59	$&$	0.17	$&	A3\,V	&$	180	^{h}	$	\\
HD	200841	&	HD 200841	&$	*	$&$	\times	$&$	1.32	$&$	7.4	$&$	8.28	$&$	0.05	$&	F2\,V\,kA7mA6~$\lambda$\,Boo?	&$			$	\\
HD	201019	&	HD 201019	&$	*	$&$	?	$&$	6.49	$&$	12.4	$&$	8.38	$&$	0.28	$&	F3 wk met	&$			$	\\
HD	201184	&	* chi Cap	&$	*i*	$&$	\times	$&$	18.14	$&$	16.5	$&$	5.28	$&$	0.01	$&	A0\,V	&$	212	^{a}	$	\\
HD	204041	&	HR 8203	&$	*	$&$	\checkmark	$&$	13.47	$&$	17.9	$&$	6.47	$&$	0.14	$&	kA1hA6mA1\,V~($\lambda$\,Boo)	&$	67	^{a}	$	\\
HD	204754	&	HR 8226	&$	V*	$&$	\times	$&$	2.56	$&$	38.1	$&$	6.15	$&$	0.09	$&	B5\,III-IVs	&$	19	^{c}	$	\\
HD	204965	&	HR 8237	&$	*	$&$	\times	$&$	6.31	$&$	20.0	$&$	6.02	$&$	0.08	$&	A2\,Vp~4481-wk	&$	96	^{a}	$	\\
HD	207978	&	* 15 Peg	&$	*	$&$	\times	$&$	36.43	$&$	11.2	$&$	5.54	$&$	0.39	$&	F2\,V	&$	3	^{b}	$	\\
HD	210111	&	HR 8437	&$	dS*	$&$	\checkmark	$&$	12.82	$&$	10.3	$&$	6.38	$&$	0.18	$&	kA2hA7mA2\,Vas~$\lambda$\,Boo	&$	54	^{c}	$	\\
\bottomrule
\end{tabular}
\end{table*}

\begin{table*}
\setcounter{table}{0}
\centering
\caption{continued from previous page.}
\rowcolors{1}{}{lightgray}
\begin{tabular}{l l H c H r r r c c}
\toprule
HD number & SIMBAD identifier & Obj. type & Member? & $\pi$& T. Vel.& $V$&$B-V$&Sp.T.& $v\sin i$ \\
\midrule
HD	210418	&	* tet Peg	&$	V*	$&$	\times	$&$	35.34	$&$	38.1	$&$	3.50	$&$	0.11	$&	hA2mA1\,IV-V	&$	122	^{b}	$	\\
HD	212061	&	* gam Aqr	&$	*i*	$&$	\times	$&$	19.92	$&$	30.9	$&$	3.85	$&$	-0.03	$&	B9.5\,III-IV	&$	75	^{b}	$	\\
HD	212150	&	HR 8525	&$	*	$&$	\times	$&$	3.99	$&$	17.5	$&$	6.63	$&$	0.01	$&	B9\,IIInp\,kA0	&$	196	^{a}	$	\\
HD	213669	&	V* DR Gru	&$	dS*	$&$	\circ	$&$	9.26	$&$	16.0	$&$	7.42	$&$	0.20	$&	F0.5\,V\,kA2.5mA2.5~$\lambda$\,Boo 	&$			$	\\
HD	214454	&	* 9 Lac	&$	*	$&$	\times	$&$	19.00	$&$	28.9	$&$	4.65	$&$	0.23	$&	F0\,Vp ($\lambda$\,Boo; met A6)	&$	90	^{b}	$	\\
HD	216847	&	HD 216847	&$	*	$&$	\circ	$&$	6.62	$&$	14.3	$&$	7.06	$&$	0.23	$&	hF0mA3\,Vn~$\lambda$\,Boo	&$	209	^{a}	$	\\
HD	217782	&	* 2 And	&$	*	$&$	\times	$&$	7.74	$&$	34.6	$&$	5.10	$&$	0.08	$&	A2\,IIIn	&$	212	^{a}	$	\\
HD	218396	&	V* V342 Peg	&$	gD*	$&$	\checkmark	$&$	25.38	$&$	22.2	$&$	5.95	$&$	0.26	$&	kA5hF0mA5\,V~$\lambda$\,Boo	&$	49	^{a}	$	\\
HD	220061	&	V* tau Peg	&$	dS*	$&$	\blu{?}	$&$	20.17	$&$	7.3	$&$	4.59	$&$	0.16	$&	kA5hA8mA5\,V 4481-wk	&$	150	^{b}	$	\\
HD	220278	&	* 97 Aqr	&$	*	$&$	?	$&$	15.30	$&$	36.6	$&$	5.22	$&$	0.21	$&	A5\,Vn	&$	175	^{a}	$	\\
HD	221756	&	V* V340 And	&$	dS*	$&$	\checkmark	$&$	12.45	$&$	19.0	$&$	5.56	$&$	0.09	$&	kA1hA3mA0.5\,Va$^{+}$~($\lambda$\,Boo)	&$	86	^{c}	$	\\
HD	222303	&	HD 222303	&$	*	$&$	?	$&$		$&$		$&$	9.16	$&$	0.57	$&	A9\,III: (met wk A3)	&$			$	\\
HD	223352	&	* del Scl	&$	**	$&$	\blu{\circ}	$&$	23.73	$&$	29.1	$&$	4.58	$&$	0.01	$&	kB9hA0mB9\,Van (($\lambda$\,Boo))	&$	299	^{a}	$	\\
HD	225180	&	* 9 Cas	&$	*i*	$&$	\times	$&$	0.84	$&$	7.5	$&$	5.90	$&$	0.25	$&	A3\,Vae	&$	33	^{c}	$	\\
HD	225218	&	ADS 30	&$	**	$&$	\times	$&$	2.59	$&$	31.1	$&$	6.12	$&$	0.15	$&	A4\,IV	&$	28	^{c}	$	\\
HD	228509	&	HD 228509	&$	*	$&$	?	$&$		$&$		$&$	9.24	$&$	0.21	$&	A9\,V(n)\,kA5mA5	&$			$	\\
HD	245185	&	HD 245185	&$	Or*	$&$	?	$&$		$&$		$&$	10.00	$&$	0.03	$&	A3\,Vae\,Bd$<$\,Nem1	&$	118	^{n}	$	\\
HD	261904	&	HD 261904	&$	*iC	$&$	?	$&$		$&$		$&$	10.30	$&$	0.08	$&	A0\,Va$^{-}$ (($\lambda$\,Boo))	&$	150	^{g}	$	\\
HD	278937	&	V* IP Per	&$	RI*	$&$	\circ	$&$		$&$		$&$	10.36	$&$	0.30	$&	kA3hA7mA4\,III:e	&$	80	{^n}	$	\\
HD	290492	&	CCDM J05313-0029AB	&$	**	$&$	?	$&$		$&$		$&$	9.39	$&$	0.08	$&	A0.5\,Vb~($\lambda$\,Boo)	&$			$	\\
HD	290799	&	V* V1790 Ori	&$	dS*	$&$	\checkmark	$&$		$&$		$&$	10.80	$&$	0.00	$&	A7\,V\,kA2mA2~$\lambda$\,Boo	&$	70	^{g}	$	\\
HD	294253	&	HD 294253	&$	*	$&$	\checkmark	$&$		$&$		$&$	9.67	$&$	0.01	$&	A0\,Va\,kB8.5~($\lambda$\,Boo) &$	70	^{g}	$	\\
\blu{TYC\,3680-215-1}	&	TYC 3680-215-1	&$	*	$&$	\checkmark	$&$		$&$		$&$	11.00	$&$	0.50	$&	A0\,Va$^{-}$~$\lambda$\,Boo	&$			$	\\
\blu{TYC\,4774-866-1}	&	V* T Ori	&$	Or*	$&$	?	$&$		$&$		$&$	11.25	$&$	0.39	$&	A3\,IVe	&$	163	^{r}	$	\\
\bottomrule
\end{tabular}
\end{table*}

\begin{table}
\centering
\caption{List of references for the $v\sin i$ values in the main table, ordered by most matching entries with Table\:\ref{tab:bigtable}.}
\label{tab:references}
\begin{tabular}{rl}
\toprule
Note & Reference\\
\midrule
$^{a}$ & \protect{\citet{royeretal2007}}\\
$^{b}$ & \protect{\citet{bernacca&perinotto1970}}\\
$^{c}$ & \protect{\citet{royeretal2002b}}\\
$^{d}$ & \protect{\citet{huangetal2010}}\\
$^{e}$ & \protect{\citet{uesugi&fukuda1982}}\\
$^{f}$ & \protect{\citet{schroederetal2009}}\\
$^{g}$ & \protect{\citet{andrievskyetal2002}}\\
$^{h}$ & \protect{\citet{heiteretal2002a}}\\
$^{i}$ & \protect{\citet{heiter2002}}\\
$^{j}$ & \protect{\citet{kampetal2001}}\\
$^{k}$ & \protect{\citet{paunzenetal1998a}}\\
$^{m}$ & \protect{\citet{stuerenburg1993}}\\
$^{n}$ & \protect{\citet{alecianetal2013a}}\\
$^{p}$ & \protect{\citet{acke&waelkens2004}}\\
$^{q}$ & \protect{\citet{cowleyetal2010}}\\
$^{r}$ & \protect{\citet{folsometal2012}}\\
$^{s}$ & \protect{\citet{rodgers1968}}\\
$^{t}$ & \protect{\citet{woolf&lambert1999}}\\ 
$^{u}$ & \protect{\citet{kudryavtsevetal2007}}\\
\bottomrule
\end{tabular}
\end{table}

\section*{acknowledgements}
\blu{We thank the referee, Ernst Paunzen, for his thorough reading of this long paper and for his useful comments,} \red{and thank Brian Skiff for his attention to detail in noticing some errors in an earlier version.} This research was supported by the Australian Research Council. Funding for the Stellar Astrophysics Centre is provided by the Danish National Research Foundation (grant agreement no.: DNRF106). The research is supported by the ASTERISK project (ASTERoseismic Investigations with SONG and Kepler) funded by the European Research Council (grant agreement no.: 267864). This program is supported by grants from the National Science Foundation to California State University Fullerton (AST-1211213), the College of Charleston (AST-1211221), and Appalachian State University (AST-1211215). JEN received additional support through NSFÕs Independent Research and Development program available to Program Officers.\\
This research has made extensive use of the SIMBAD data base and the VizieR catalogue access tool \citep{ochsenbeinetal2000}, operated at CDS, Strasbourg, France. Spectral types for some Ap stars were found in the \citet{skiff2013} Catalogue of Stellar Spectral Classifications, and checked in their original articles. IRAF is distributed by the National Optical Astronomy Observatory, which is operated by the Association of Universities for
Research in Astronomy, Inc. under cooperative agreement with the National Science Foundation.

\bibliography{2r_lamboo_notes} 

\begin{thebibliography}{235}
\expandafter\ifx\csname natexlab\endcsname\relax\def\natexlab#1{#1}\fi

\bibitem[{{Abt}(1984{\natexlab{a}})}]{abt1984b}
{Abt} H.~A., 1984{\natexlab{a}}, in The MK Process and Stellar Classification,
  {Garrison} R.~F., ed., p. 340

\bibitem[{{Abt}(1984{\natexlab{b}})}]{abt1984}
---, 1984{\natexlab{b}}, \apj, 285, 247

\bibitem[{{Abt}(1985)}]{abt1985}
---, 1985, \apjs, 59, 95

\bibitem[{{Abt}(1988)}]{abt1988}
---, 1988, \apj, 331, 922

\bibitem[{{Abt} \& {Cardona}(1983)}]{abt&cardona1983}
{Abt} H.~A., {Cardona} O., 1983, \apj, 272, 182

\bibitem[{{Abt} \& {Cardona}(1984)}]{abt&cardona1984}
---, 1984, \apj, 276, 266

\bibitem[{{Abt} \& {Morrell}(1995)}]{abt&morrell1995}
{Abt} H.~A., {Morrell} N.~I., 1995, \apjs, 99, 135

\bibitem[{{Acke}, {van den Ancker} \& {Dullemond}(2005){Acke}, {van den
  Ancker}, \& {Dullemond}}]{ackeetal2005}
{Acke} B., {van den Ancker} M.~E., {Dullemond} C.~P., 2005, \aap, 436, 209

\bibitem[{{Acke} \& {Waelkens}(2004)}]{acke&waelkens2004}
{Acke} B., {Waelkens} C., 2004, \aap, 427, 1009

\bibitem[{{Adelman} {et~al}\mbox{.}(1997){Adelman}, {Caliskan}, {Kocer}, \&
  {Bolcal}}]{adelmanetal1997}
{Adelman} S.~J., {Caliskan} H., {Kocer} D., {Bolcal} C., 1997, \mnras, 288, 470

\bibitem[{{Alecian} {et~al}\mbox{.}(2013){Alecian}, {Wade}, {Catala},
  {Grunhut}, {Landstreet}, {Bagnulo}, {B{\"o}hm}, {Folsom}, {Marsden}, \&
  {Waite}}]{alecianetal2013a}
{Alecian} E. {et~al.}, 2013, \mnras, 429, 1001

\bibitem[{{Andersen} \& {Nordstrom}(1977)}]{andersen&nordstrom1977}
{Andersen} J., {Nordstrom} B., 1977, \aaps, 29, 309

\bibitem[{{Andrievsky} {et~al}\mbox{.}(2002){Andrievsky}, {Chernyshova},
  {Paunzen}, {Weiss}, {Korotin}, {Beletsky}, {Handler}, {Heiter}, {Korotina},
  {St{\"u}tz}, \& {Weber}}]{andrievskyetal2002}
{Andrievsky} S.~M. {et~al.}, 2002, \aap, 396, 641

\bibitem[{{Andrievsky} {et~al}\mbox{.}(1995){Andrievsky}, {Chernyshova},
  {Usenko}, {Kovtyukh}, {Panchuk}, \& {Galazutdinov}}]{andrievskyetal1995}
{Andrievsky} S.~M., {Chernyshova} I.~V., {Usenko} I.~A., {Kovtyukh} V.~V.,
  {Panchuk} V.~E., {Galazutdinov} G.~A., 1995, \pasp, 107, 219

\bibitem[{{Andrillat}, {Jaschek} \& {Jaschek}(1995){Andrillat}, {Jaschek}, \&
  {Jaschek}}]{andrillatetal1995}
{Andrillat} Y., {Jaschek} C., {Jaschek} M., 1995, \aap, 299, 493

\bibitem[{{Appenzeller}(1967)}]{appenzeller1967}
{Appenzeller} I., 1967, \pasp, 79, 102

\bibitem[{{Balona}(1977)}]{balona1977}
{Balona} L.~A., 1977, \memras, 84, 101

\bibitem[{{Barry}(1970)}]{barry1970}
{Barry} D.~C., 1970, \apjs, 19, 281

\bibitem[{{Bartolini} {et~al}\mbox{.}(1980){Bartolini}, {Dapergolas},
  {Piccioni}, \& {Voli}}]{bartolinietal1980}
{Bartolini} C., {Dapergolas} A., {Piccioni} A., {Voli} M., 1980, Information
  Bulletin on Variable Stars, 1757, 1

\bibitem[{{Baschek} {et~al}\mbox{.}(1984){Baschek}, {Koeppen}, {Scholz},
  {Wehrse}, {Heck}, {Jaschek}, \& {Jaschek}}]{bascheketal1984}
{Baschek} B., {Koeppen} J., {Scholz} M., {Wehrse} R., {Heck} A., {Jaschek} C.,
  {Jaschek} M., 1984, \aap, 131, 378

\bibitem[{{Baschek} \& {Slettebak}(1988)}]{baschek&slettebak1988}
{Baschek} B., {Slettebak} A., 1988, \aap, 207, 112

\bibitem[{{Batten}, {Fletcher} \& {Mann}(1978){Batten}, {Fletcher}, \&
  {Mann}}]{battenetal1978}
{Batten} A.~H., {Fletcher} J.~M., {Mann} P.~J., 1978, Publications of the
  Dominion Astrophysical Observatory Victoria, 15, 121

\bibitem[{{Bernacca} \& {Perinotto}(1970)}]{bernacca&perinotto1970}
{Bernacca} P.~L., {Perinotto} M., 1970, Contributions dell'Osservatorio
  Astrofisica dell'Universita di Padova in Asiago, 239, 1

\bibitem[{{Bidelman}(1943)}]{bidelman1943}
{Bidelman} W.~P., 1943, \apj, 98, 61

\bibitem[{{Bidelman}(1988)}]{bidelman1988}
---, 1988, \pasp, 100, 1084

\bibitem[{{Bidelman}, {Ratcliff} \& {Svolopoulos}(1988){Bidelman}, {Ratcliff},
  \& {Svolopoulos}}]{bidelmanetal1988}
{Bidelman} W.~P., {Ratcliff} S.~J., {Svolopoulos} S., 1988, \pasp, 100, 828

\bibitem[{{Bohlender}, {Gonzalez} \& {Matthews}(1999){Bohlender}, {Gonzalez},
  \& {Matthews}}]{bohlenderetal1999}
{Bohlender} D.~A., {Gonzalez} J.-F., {Matthews} J.~M., 1999, \aap, 350, 553

\bibitem[{{Bohlender} \& {Walker}(1994)}]{bohlender&walker1994}
{Bohlender} D.~A., {Walker} G.~A.~H., 1994, \mnras, 266, 891

\bibitem[{{Booth} {et~al}\mbox{.}(2013){Booth}, {Kennedy}, {Sibthorpe},
  {Matthews}, {Wyatt}, {Duch{\^e}ne}, {Kavelaars}, {Rodriguez}, {Greaves},
  {Koning}, {Vican}, {Rieke}, {Su}, {Moro-Mart{\'{\i}}n}, \&
  {Kalas}}]{boothetal2013}
{Booth} M. {et~al.}, 2013, \mnras, 428, 1263

\bibitem[{{Breger}(1969)}]{breger1969a}
{Breger} M., 1969, \aj, 74, 166

\bibitem[{{Breger}(1979)}]{breger1979}
---, 1979, \pasp, 91, 5

\bibitem[{{Breger}, {Hutchins} \& {Kuhi}(1976){Breger}, {Hutchins}, \&
  {Kuhi}}]{bregeretal1976}
{Breger} M., {Hutchins} J., {Kuhi} L.~V., 1976, \apj, 210, 163

\bibitem[{{Budaj}(1997)}]{budaj1997}
{Budaj} J., 1997, \aap, 326, 655

\bibitem[{{Cayrel de Strobel} {et~al}\mbox{.}(1997){Cayrel de Strobel},
  {Soubiran}, {Friel}, {Ralite}, \& {Francois}}]{cayreldestrobeletal1997}
{Cayrel de Strobel} G., {Soubiran} C., {Friel} E.~D., {Ralite} N., {Francois}
  P., 1997, \aaps, 124, 299

\bibitem[{{Cenarro} {et~al}\mbox{.}(2007){Cenarro}, {Peletier},
  {S{\'a}nchez-Bl{\'a}zquez}, {Selam}, {Toloba}, {Cardiel},
  {Falc{\'o}n-Barroso}, {Gorgas}, {Jim{\'e}nez-Vicente}, \&
  {Vazdekis}}]{cenarroetal2007}
{Cenarro} A.~J. {et~al.}, 2007, \mnras, 374, 664

\bibitem[{{Cheng} {et~al}\mbox{.}(1992){Cheng}, {Bruhweiler}, {Kondo}, \&
  {Grady}}]{chengetal1992}
{Cheng} K.-P., {Bruhweiler} F.~C., {Kondo} Y., {Grady} C.~A., 1992, \apjl, 396,
  L83

\bibitem[{{Childress} {et~al}\mbox{.}(2014){Childress}, {Vogt}, {Nielsen}, \&
  {Sharp}}]{childressetal2014}
{Childress} M.~J., {Vogt} F.~P.~A., {Nielsen} J., {Sharp} R.~G., 2014, \apss,
  349, 617

\bibitem[{{Colomba}, {de Benedetto} \& {Ielo}(1991){Colomba}, {de Benedetto},
  \& {Ielo}}]{colombaetal1991}
{Colomba} A., {de Benedetto} G., {Ielo} A., 1991, Information Bulletin on
  Variable Stars, 3597, 1

\bibitem[{{Corbally}(1984)}]{corbally1984}
{Corbally} C.~J., 1984, \apjs, 55, 657

\bibitem[{{Corbally} \& {Garrison}(1980)}]{corbally&garrison1980}
{Corbally} C.~J., {Garrison} R.~F., 1980, \pasp, 92, 493

\bibitem[{{Corbally} \& {Gray}(1996)}]{corbally&gray1996}
{Corbally} C.~J., {Gray} R.~O., 1996, \aj, 112, 2286

\bibitem[{{Cowley} {et~al}\mbox{.}(1969){Cowley}, {Cowley}, {Jaschek}, \&
  {Jaschek}}]{cowleyetal1969}
{Cowley} A., {Cowley} C., {Jaschek} M., {Jaschek} C., 1969, \aj, 74, 375

\bibitem[{{Cowley} \& {Bidelman}(1979)}]{cowley&bidelman1979}
{Cowley} A.~P., {Bidelman} W.~P., 1979, \pasp, 91, 83

\bibitem[{{Cowley}(1991)}]{cowley1991}
{Cowley} C.~R., 1991, in IAU Symposium, Vol. 145, Evolution of Stars: the
  Photospheric Abundance Connection, {Michaud} G., {Tutukov} A.~V., eds., p.
  183

\bibitem[{{Cowley} {et~al}\mbox{.}(2010){Cowley}, {Hubrig}, {Gonz{\'a}lez}, \&
  {Savanov}}]{cowleyetal2010}
{Cowley} C.~R., {Hubrig} S., {Gonz{\'a}lez} J.~F., {Savanov} I., 2010, \aap,
  523, A65

\bibitem[{{Crawford}(1975)}]{crawford1975}
{Crawford} D.~L., 1975, \aj, 80, 955

\bibitem[{{Crawford}(1978)}]{crawford1978}
---, 1978, \aj, 83, 48

\bibitem[{{Crawford}(1979)}]{crawford1979}
---, 1979, \aj, 84, 1858

\bibitem[{{Cucchiaro} {et~al}\mbox{.}(1980){Cucchiaro}, {Jaschek}, {Jaschek},
  \& {Macau-Hercot}}]{cucchiaroetal1980}
{Cucchiaro} A., {Jaschek} M., {Jaschek} C., {Macau-Hercot} D., 1980, \aaps, 40,
  207

\bibitem[{{Cucchiaro} {et~al}\mbox{.}(1979){Cucchiaro}, {Macau-Hercot},
  {Jaschek}, \& {Jaschek}}]{cucchiaroetal1979}
{Cucchiaro} A., {Macau-Hercot} D., {Jaschek} M., {Jaschek} C., 1979, \aaps, 35,
  75

\bibitem[{{Desikachary} \& {McInally}(1979)}]{desikachary&mcinally1979}
{Desikachary} K., {McInally} C.~J., 1979, \mnras, 188, 67

\bibitem[{{Dommanget} \& {Nys}(2002)}]{dommagnet&nys2002}
{Dommanget} J., {Nys} O., 2002, VizieR Online Data Catalog, 1274, 0

\bibitem[{{Dopita} {et~al}\mbox{.}(2007){Dopita}, {Hart}, {McGregor}, {Oates},
  {Bloxham}, \& {Jones}}]{dopitaetal2007}
{Dopita} M., {Hart} J., {McGregor} P., {Oates} P., {Bloxham} G., {Jones} D.,
  2007, \apss, 310, 255

\bibitem[{{Dunkin}, {Barlow} \& {Ryan}(1997){Dunkin}, {Barlow}, \&
  {Ryan}}]{dunkinetal1997}
{Dunkin} S.~K., {Barlow} M.~J., {Ryan} S.~G., 1997, \mnras, 286, 604

\bibitem[{{Eggen}(1984)}]{eggen1984}
{Eggen} O.~J., 1984, \aj, 89, 1878

\bibitem[{{Eggleton} \& {Tokovinin}(2008)}]{eggleton&tokovinin2008}
{Eggleton} P.~P., {Tokovinin} A.~A., 2008, \mnras, 389, 869

\bibitem[{{Erspamer} \& {North}(2003)}]{erspamer&north2003}
{Erspamer} D., {North} P., 2003, \aap, 398, 1121

\bibitem[{{Faraggiana}(1990)}]{faraggiana1990}
{Faraggiana} R., 1990, in IUE Proposal, p. 3832

\bibitem[{{Faraggiana} \& {Bonifacio}(1999)}]{faraggiana&bonifacio1999}
{Faraggiana} R., {Bonifacio} P., 1999, \aap, 349, 521

\bibitem[{{Faraggiana} \& {Bonifacio}(2005)}]{faraggiana&bonifacio2005}
---, 2005, \aap, 436, 697

\bibitem[{{Faraggiana} {et~al}\mbox{.}(2004){Faraggiana}, {Bonifacio},
  {Caffau}, {Gerbaldi}, \& {Nonino}}]{faraggianaetal2004}
{Faraggiana} R., {Bonifacio} P., {Caffau} E., {Gerbaldi} M., {Nonino} M., 2004,
  \aap, 425, 615

\bibitem[{{Faraggiana} \& {Gerbaldi}(2003)}]{faraggiana&gerbaldi2003}
{Faraggiana} R., {Gerbaldi} M., 2003, \aap, 398, 697

\bibitem[{{Faraggiana}, {Gerbaldi} \& {Boehm}(1990){Faraggiana}, {Gerbaldi}, \&
  {Boehm}}]{faraggianaetal1990}
{Faraggiana} R., {Gerbaldi} M., {Boehm} C., 1990, \aap, 235, 311

\bibitem[{{Faraggiana}, {Gerbaldi} \& {Bonifacio}(2001){Faraggiana},
  {Gerbaldi}, \& {Bonifacio}}]{faraggianaetal2001b}
{Faraggiana} R., {Gerbaldi} M., {Bonifacio} P., 2001, \aap, 380, 286

\bibitem[{{Faraggiana} {et~al}\mbox{.}(2001){Faraggiana}, {Gerbaldi},
  {Bonifacio}, \& {Fran{\c c}ois}}]{faraggianaetal2001a}
{Faraggiana} R., {Gerbaldi} M., {Bonifacio} P., {Fran{\c c}ois} P., 2001, \aap,
  376, 586

\bibitem[{{Faraggiana}, {Gerbaldi} \& {Burnage}(1997){Faraggiana}, {Gerbaldi},
  \& {Burnage}}]{faraggianaetal1997}
{Faraggiana} R., {Gerbaldi} M., {Burnage} R., 1997, \aap, 318, L21

\bibitem[{{Feigelson}, {Lawson} \& {Garmire}(2003){Feigelson}, {Lawson}, \&
  {Garmire}}]{feigelsonetal2003}
{Feigelson} E.~D., {Lawson} W.~A., {Garmire} G.~P., 2003, \apj, 599, 1207

\bibitem[{{Folsom} {et~al}\mbox{.}(2012){Folsom}, {Bagnulo}, {Wade}, {Alecian},
  {Landstreet}, {Marsden}, \& {Waite}}]{folsometal2012}
{Folsom} C.~P., {Bagnulo} S., {Wade} G.~A., {Alecian} E., {Landstreet} J.~D.,
  {Marsden} S.~C., {Waite} I.~A., 2012, \mnras, 422, 2072

\bibitem[{{Garrison} \& {Gray}(1994)}]{garrison&gray1994}
{Garrison} R.~F., {Gray} R.~O., 1994, \aj, 107, 1556

\bibitem[{{Gerbaldi} \& {Faraggiana}(1993)}]{gerbaldi&faraggiana1993}
{Gerbaldi} M., {Faraggiana} R., 1993, in Astronomical Society of the Pacific
  Conference Series, Vol.~44, IAU Colloq. 138: Peculiar versus Normal Phenomena
  in A-type and Related Stars, {Dworetsky} M.~M., {Castelli} F., {Faraggiana}
  R., eds., p. 368

\bibitem[{{Gerbaldi} \& {Faraggiana}(2004)}]{gerbaldi&faraggiana2004}
---, 2004, in Astronomical Society of the Pacific Conference Series, Vol. 318,
  Spectroscopically and Spatially Resolving the Components of the Close Binary
  Stars, {Hilditch} R.~W., {Hensberge} H., {Pavlovski} K., eds., pp. 312--315

\bibitem[{{Gerbaldi}, {Faraggiana} \& {Lai}(2003){Gerbaldi}, {Faraggiana}, \&
  {Lai}}]{gerbaldietal2003}
{Gerbaldi} M., {Faraggiana} R., {Lai} O., 2003, \aap, 412, 447

\bibitem[{{Glagolevskij}(2011)}]{glagolevskij2011}
{Glagolevskij} Y.~V., 2011, Astrophysics, 54, 231

\bibitem[{{Gopka} {et~al}\mbox{.}(2007){Gopka}, {Yushchenko}, {Kim}, {Lambert},
  {Rostopchin}, {Kim}, {Jeon}, {Dorokhova}, {Tarasov}, \&
  {Chernyshova}}]{gopkaetal2007}
{Gopka} V. {et~al.}, 2007, in Astronomical Society of the Pacific Conference
  Series, Vol. 362, The Seventh Pacific Rim Conference on Stellar Astrophysics,
  {Kang} Y.~W., {Lee} H.-W., {Leung} K.-C., {Cheng} K.-S., eds., p. 249

\bibitem[{{Graham} \& {Slettebak}(1973)}]{graham&slettebak1973}
{Graham} J.~A., {Slettebak} A., 1973, \aj, 78, 295

\bibitem[{{Gray}(1986)}]{gray1986}
{Gray} R.~O., 1986, PhD thesis, UNIVERSITY OF TORONTO (CANADA).

\bibitem[{{Gray}(1988)}]{gray1988}
---, 1988, \aj, 95, 220

\bibitem[{{Gray}(1989)}]{gray1989}
---, 1989, \aj, 98, 1049

\bibitem[{{Gray}(2014)}]{gray2014}
---, 2014, in {GeoPlanet: Earth and Planetary Sciences}, Vol.~16,
  {Determination of Atmospheric Parameters of B-, A-, F- and G-Type Stars,
  Wroc{\l}aw, Poland}, Springer International Publishing, Switzerland, pp.
  75--84

\bibitem[{{Gray} \& {Corbally}(2009)}]{gray&corbally2009}
{Gray} R.~O., {Corbally}, J. C., 2009, {Stellar Spectral Classification}.
  {Princeton University Press}

\bibitem[{{Gray} \& {Corbally}(1993)}]{gray&corbally1993}
{Gray} R.~O., {Corbally} C.~J., 1993, \aj, 106, 632

\bibitem[{{Gray} \& {Corbally}(1998)}]{gray&corbally1998}
---, 1998, \aj, 116, 2530

\bibitem[{{Gray} \& {Corbally}(2002)}]{gray&corbally2002}
---, 2002, \aj, 124, 989

\bibitem[{{Gray} \& {Corbally}(2014)}]{gray&corbally2014}
---, 2014, \aj, 147, 80

\bibitem[{{Gray} {et~al}\mbox{.}(2003){Gray}, {Corbally}, {Garrison},
  {McFadden}, \& {Robinson}}]{grayetal2003}
{Gray} R.~O., {Corbally} C.~J., {Garrison} R.~F., {McFadden} M.~T., {Robinson}
  P.~E., 2003, \aj, 126, 2048

\bibitem[{{Gray}, {Corbally} \& {Philip}(1996){Gray}, {Corbally}, \&
  {Philip}}]{grayetal1996}
{Gray} R.~O., {Corbally} C.~J., {Philip} A.~G.~D., 1996, \aj, 112, 2291

\bibitem[{{Gray} \& {Garrison}(1987)}]{gray&garrison1987}
{Gray} R.~O., {Garrison} R.~F., 1987, \apjs, 65, 581

\bibitem[{{Gray} \& {Garrison}(1989{\natexlab{a}})}]{gray&garrison1989}
---, 1989{\natexlab{a}}, \apjs, 69, 301

\bibitem[{{Gray} \& {Garrison}(1989{\natexlab{b}})}]{gray&garrison1989b}
---, 1989{\natexlab{b}}, \apjs, 70, 623

\bibitem[{{Gray} \& {Kaye}(1999)}]{gray&kaye1999}
{Gray} R.~O., {Kaye} A.~B., 1999, \aj, 118, 2993

\bibitem[{{Gray}, {Napier} \& {Winkler}(2001){Gray}, {Napier}, \&
  {Winkler}}]{grayetal2001a}
{Gray} R.~O., {Napier} M.~G., {Winkler} L.~I., 2001, \aj, 121, 2148

\bibitem[{{Gray} \& {Olsen}(1991)}]{gray&olsen1991}
{Gray} R.~O., {Olsen} E.~H., 1991, \aaps, 87, 541

\bibitem[{{Grenier} {et~al}\mbox{.}(1999){Grenier}, {Burnage}, {Faraggiana},
  {Gerbaldi}, {Delmas}, {G{\'o}mez}, {Sabas}, \& {Sharif}}]{grenieretal1999}
{Grenier} S., {Burnage} R., {Faraggiana} R., {Gerbaldi} M., {Delmas} F.,
  {G{\'o}mez} A.~E., {Sabas} V., {Sharif} L., 1999, \aaps, 135, 503

\bibitem[{{Griffin}, {Gray} \& {Corbally}(2012){Griffin}, {Gray}, \&
  {Corbally}}]{griffinetal2012}
{Griffin} R.~E., {Gray} R.~O., {Corbally} C.~J., 2012, \aap, 547, A8

\bibitem[{{Guetter}(1981)}]{guetter1981}
{Guetter} H.~H., 1981, \aj, 86, 1057

\bibitem[{{Handler}(1999)}]{handler1999}
{Handler} G., 1999, Information Bulletin on Variable Stars, 4817, 1

\bibitem[{{Handler}, {Gray} \& {Shobbrook}(2000){Handler}, {Gray}, \&
  {Shobbrook}}]{handleretal2000}
{Handler} G., {Gray} R.~O., {Shobbrook} R.~R., 2000, Information Bulletin on
  Variable Stars, 4876, 1

\bibitem[{{Hashimoto} {et~al}\mbox{.}(2011){Hashimoto}, {Tamura}, {Muto},
  {Kudo}, {Fukagawa}, {Fukue}, {Goto}, {Grady}, {Henning}, {Hodapp}, {Honda},
  {Inutsuka}, {Kokubo}, {Knapp}, {McElwain}, {Momose}, {Ohashi}, {Okamoto},
  {Takami}, {Turner}, {Wisniewski}, {Janson}, {Abe}, {Brandner}, {Carson},
  {Egner}, {Feldt}, {Golota}, {Guyon}, {Hayano}, {Hayashi}, {Hayashi}, {Ishii},
  {Kandori}, {Kusakabe}, {Matsuo}, {Mayama}, {Miyama}, {Morino}, {Moro-Martin},
  {Nishimura}, {Pyo}, {Suto}, {Suzuki}, {Takato}, {Terada}, {Thalmann},
  {Tomono}, {Watanabe}, {Yamada}, {Takami}, \& {Usuda}}]{hashimotoetal2011}
{Hashimoto} J. {et~al.}, 2011, \apjl, 729, L17

\bibitem[{{Hauck}(1986)}]{hauck1986}
{Hauck} B., 1986, \aap, 154, 349

\bibitem[{{Hauck}, {Ballereau} \& {Chauville}(1998){Hauck}, {Ballereau}, \&
  {Chauville}}]{haucketal1998}
{Hauck} B., {Ballereau} D., {Chauville} J., 1998, \aaps, 128, 429

\bibitem[{{Hauck} \& {Mermilliod}(1998)}]{hauck&mermilliod1998}
{Hauck} B., {Mermilliod} M., 1998, \aaps, 129, 431

\bibitem[{{Hauck} \& {Slettebak}(1983)}]{hauck&slettebak1983}
{Hauck} B., {Slettebak} A., 1983, \aap, 127, 231

\bibitem[{{Heiter}(1998)}]{heiter1998}
{Heiter} U., 1998, Contributions of the Astronomical Observatory Skalnate
  Pleso, 27, 403

\bibitem[{{Heiter}(2002)}]{heiter2002}
---, 2002, \aap, 381, 959

\bibitem[{{Heiter} {et~al}\mbox{.}(1998){Heiter}, {Kupka}, {Paunzen}, {Weiss},
  \& {Gelbmann}}]{heiteretal1998}
{Heiter} U., {Kupka} F., {Paunzen} E., {Weiss} W.~W., {Gelbmann} M., 1998,
  \aap, 335, 1009

\bibitem[{{Heiter}, {Weiss} \& {Paunzen}(2002){Heiter}, {Weiss}, \&
  {Paunzen}}]{heiteretal2002a}
{Heiter} U., {Weiss} W.~W., {Paunzen} E., 2002, \aap, 381, 971

\bibitem[{{Hekker} {et~al}\mbox{.}(2009){Hekker}, {Fr{\'e}mat}, {Lampens}, {De
  Cat}, {Niemczura}, {Creevey}, \& {Zorec}}]{hekkeretal2009}
{Hekker} S., {Fr{\'e}mat} Y., {Lampens} P., {De Cat} P., {Niemczura} E.,
  {Creevey} O.~L., {Zorec} J., 2009, \mnras, 396, 1689

\bibitem[{{Heller} \& {Kramer}(1988)}]{heller&kramer1988}
{Heller} C.~H., {Kramer} K.~S., 1988, \pasp, 100, 583

\bibitem[{{Hoffleit} \& {Jaschek}(1982)}]{hoffleit&jaschek1982}
{Hoffleit} D., {Jaschek} C., 1982, {The Bright Star Catalogue. Fourth revised
  edition. (Containing data compiled through 1979).} {Yale University
  Observatory}

\bibitem[{{Holweger} \& {Rentzsch-Holm}(1995)}]{holweger&rentzsch-holm1995}
{Holweger} H., {Rentzsch-Holm} I., 1995, \aap, 303, 819

\bibitem[{{Holweger} \& {Stuerenburg}(1991)}]{holweger&stuerenburg1991}
{Holweger} H., {Stuerenburg} S., 1991, \aap, 252, 255

\bibitem[{{Horn} {et~al}\mbox{.}(1996){Horn}, {Kubat}, {Harmanec}, {Koubsky},
  {Hadrava}, {Simon}, {Stefl}, \& {Skoda}}]{hornetal1996}
{Horn} J., {Kubat} J., {Harmanec} P., {Koubsky} P., {Hadrava} P., {Simon} V.,
  {Stefl} S., {Skoda} P., 1996, \aap, 309, 521

\bibitem[{{Houk}(1978)}]{houk1978}
{Houk} N., 1978, {Michigan catalogue of two-dimensional spectral types for the
  HD stars}. {University of Michigan}

\bibitem[{{Houk}(1982)}]{houk1982}
---, 1982, {Michigan Catalogue of Two-dimensional Spectral Types for the HD
  stars. Volume\_3. Declinations -40$^{\circ}$ to -26$^{\circ}$.} {University
  of Michigan}

\bibitem[{{Houk} \& {Cowley}(1975)}]{houk&cowley1975}
{Houk} N., {Cowley} A.~P., 1975, {University of Michigan Catalogue of
  two-dimensional spectral types for the HD stars. Volume I. Declinations -90\_
  to -53$^{\circ}$.} {University of Michigan}

\bibitem[{{Houk} \& {Smith-Moore}(1988)}]{houk&smith-moore1988}
{Houk} N., {Smith-Moore} M., 1988, {Michigan Catalogue of Two-dimensional
  Spectral Types for the HD Stars. Volume 4, Declinations -26$^{\circ}$ to
  -12$^{\circ}$.} {University of Michigan}

\bibitem[{{Huang}, {Gies} \& {McSwain}(2010){Huang}, {Gies}, \&
  {McSwain}}]{huangetal2010}
{Huang} W., {Gies} D.~R., {McSwain} M.~V., 2010, \apj, 722, 605

\bibitem[{{Hube}(1970)}]{hube1970}
{Hube} D.~P., 1970, \memras, 72, 233

\bibitem[{{Ibano{\v g}lu} {et~al}\mbox{.}(2006){Ibano{\v g}lu}, {Soydugan},
  {Soydugan}, \& {Dervi{\c s}o{\v g}lu}}]{ibanogluetal2006}
{Ibano{\v g}lu} C., {Soydugan} F., {Soydugan} E., {Dervi{\c s}o{\v g}lu} A.,
  2006, \mnras, 373, 435

\bibitem[{{Iliev} \& {Barzova}(1993{\natexlab{a}})}]{iliev&barzova1993b}
{Iliev} I.~K., {Barzova} I.~S., 1993{\natexlab{a}}, \apss, 208, 277

\bibitem[{{Iliev} \& {Barzova}(1993{\natexlab{b}})}]{iliev&barzova1993a}
---, 1993{\natexlab{b}}, in Astronomical Society of the Pacific Conference
  Series, Vol.~44, IAU Colloq. 138: Peculiar versus Normal Phenomena in A-type
  and Related Stars, {Dworetsky} M.~M., {Castelli} F., {Faraggiana} R., eds.,
  p. 423

\bibitem[{{Iliev} \& {Barzova}(1995)}]{iliev&barzova1995}
---, 1995, \aap, 302, 735

\bibitem[{{Iliev} \& {Barzova}(1998)}]{iliev&barzova1998}
{Iliev} I.~K., {Barzova} S.~I., 1998, Contributions of the Astronomical
  Observatory Skalnate Pleso, 27, 441

\bibitem[{{Iliev} {et~al}\mbox{.}(2001){Iliev}, {Paunzen}, {Barzova},
  {Andrievsky}, {Chernyshova}, \& {Kamp}}]{ilievetal2001}
{Iliev} I.~K., {Paunzen} E., {Barzova} I., {Andrievsky} S.~M., {Chernyshova}
  I.~V., {Kamp} I., 2001, Information Bulletin on Variable Stars, 5178, 1

\bibitem[{{Iliev} {et~al}\mbox{.}(2002){Iliev}, {Paunzen}, {Barzova},
  {Griffin}, {Kamp}, {Claret}, \& {Koen}}]{ilievetal2002}
{Iliev} I.~K., {Paunzen} E., {Barzova} I.~S., {Griffin} R.~F., {Kamp} I.,
  {Claret} A., {Koen} C., 2002, \aap, 381, 914

\bibitem[{{Ilijic} {et~al}\mbox{.}(1998){Ilijic}, {Rosandic}, {Dominis},
  {Planinic}, \& {Pavlovski}}]{ilijicetal1998}
{Ilijic} S., {Rosandic} M., {Dominis} D., {Planinic} M., {Pavlovski} K., 1998,
  Contributions of the Astronomical Observatory Skalnate Pleso, 27, 467

\bibitem[{{Jackisch}(1972)}]{jackisch1972}
{Jackisch} G., 1972, Astronomische Nachrichten, 294, 1

\bibitem[{{Jaschek} \& {Jaschek}(1980)}]{jaschek&jaschek1980}
{Jaschek} M., {Jaschek} C., 1980, \aaps, 42, 115

\bibitem[{{Kamp} {et~al}\mbox{.}(2001){Kamp}, {Iliev}, {Paunzen}, {Pintado},
  {Solano}, \& {Barzova}}]{kampetal2001}
{Kamp} I., {Iliev} I.~K., {Paunzen} E., {Pintado} O.~I., {Solano} E., {Barzova}
  I.~S., 2001, \aap, 375, 899

\bibitem[{{Kamp} \& {Paunzen}(2002)}]{kamp&paunzen2002}
{Kamp} I., {Paunzen} E., 2002, \mnras, 335, L45

\bibitem[{{King}(1994)}]{king1994}
{King} J.~R., 1994, \mnras, 269, 209

\bibitem[{{Koen} {et~al}\mbox{.}(1995){Koen}, {Kilkenny}, {van Wyk}, {Roberts},
  \& {Marang}}]{koenetal1995}
{Koen} C., {Kilkenny} D., {van Wyk} F., {Roberts} G., {Marang} F., 1995,
  \mnras, 277, 217

\bibitem[{{Koen} {et~al}\mbox{.}(2003){Koen}, {Paunzen}, {van Wyk}, {Marang},
  {Chernyshova}, \& {Andrievsky}}]{koenetal2003}
{Koen} C., {Paunzen} E., {van Wyk} F., {Marang} F., {Chernyshova} I.~V.,
  {Andrievsky} S.~M., 2003, \mnras, 338, 931

\bibitem[{{Kohoutek}(2001)}]{kohoutek2001}
{Kohoutek} L., 2001, \aap, 378, 843

\bibitem[{{Kudryavtsev} {et~al}\mbox{.}(2006){Kudryavtsev}, {Romanyuk},
  {Elkin}, \& {Paunzen}}]{kudryavtsevetal2006}
{Kudryavtsev} D.~O., {Romanyuk} I.~I., {Elkin} V.~G., {Paunzen} E., 2006,
  \mnras, 372, 1804

\bibitem[{{Kudryavtsev} {et~al}\mbox{.}(2007){Kudryavtsev}, {Romanyuk},
  {Semenko}, \& {Solov'ev}}]{kudryavtsevetal2007}
{Kudryavtsev} D.~O., {Romanyuk} I.~I., {Semenko} E.~A., {Solov'ev} G.~A., 2007,
  Astrophysical Bulletin, 62, 147

\bibitem[{{Kuschnig} {et~al}\mbox{.}(1996){Kuschnig}, {Gelbmann}, {Paunzen}, \&
  {Weiss}}]{kuschnigetal1996a}
{Kuschnig} R., {Gelbmann} M., {Paunzen} E., {Weiss} W.~W., 1996, Information
  Bulletin on Variable Stars, 4310, 1

\bibitem[{{Kuschnig}, {Paunzen} \& {Weiss}(1994{\natexlab{a}}){Kuschnig},
  {Paunzen}, \& {Weiss}}]{kuschnigetal1994a}
{Kuschnig} R., {Paunzen} E., {Weiss} W.~W., 1994{\natexlab{a}}, Information
  Bulletin on Variable Stars, 4069, 1

\bibitem[{{Kuschnig}, {Paunzen} \& {Weiss}(1994{\natexlab{b}}){Kuschnig},
  {Paunzen}, \& {Weiss}}]{kuschnigetal1994b}
---, 1994{\natexlab{b}}, Information Bulletin on Variable Stars, 4070, 1

\bibitem[{{Kuschnig}, {Paunzen} \& {Weiss}(1997){Kuschnig}, {Paunzen}, \&
  {Weiss}}]{kuschnigetal1997}
---, 1997, Information Bulletin on Variable Stars, 4483, 1

\bibitem[{{Landstreet} {et~al}\mbox{.}(1975){Landstreet}, {Borra}, {Angel}, \&
  {Illing}}]{landstreetetal1975}
{Landstreet} J.~D., {Borra} E.~F., {Angel} J.~R.~P., {Illing} R.~M.~E., 1975,
  \apj, 201, 624

\bibitem[{{Lemke}(1989)}]{lemke1989}
{Lemke} M., 1989, \aap, 225, 125

\bibitem[{{Liu} {et~al}\mbox{.}(1997){Liu}, {Gies}, {Xiong}, {Riddle},
  {Bagnuolo}, {Barry}, {Ferrara}, {Hartkopf}, {Hooda}, {Mason}, {McAlister},
  {Roberts}, \& {Sowers}}]{liuetal1997}
{Liu} N. {et~al.}, 1997, \apj, 485, 350

\bibitem[{{Lloyd}(1981)}]{lloyd1981}
{Lloyd} C., 1981, \mnras, 195, 805

\bibitem[{{Maitzen} \& {Pavlovski}(1989{\natexlab{a}})}]{maitzen&pavlovski1989}
{Maitzen} H.~M., {Pavlovski} K., 1989{\natexlab{a}}, \aap, 219, 253

\bibitem[{{Maitzen} \&
  {Pavlovski}(1989{\natexlab{b}})}]{maitzen&pavlovski1989b}
---, 1989{\natexlab{b}}, \aaps, 81, 335

\bibitem[{{Marchetti}, {Faraggiana} \& {Bonifacio}(2001){Marchetti},
  {Faraggiana}, \& {Bonifacio}}]{marchettietal2001}
{Marchetti} E., {Faraggiana} R., {Bonifacio} P., 2001, \aap, 370, 524

\bibitem[{{Martinez} {et~al}\mbox{.}(1998){Martinez}, {Koen}, {Handler}, \&
  {Paunzen}}]{martinezetal1998}
{Martinez} P., {Koen} C., {Handler} G., {Paunzen} E., 1998, \mnras, 301, 1099

\bibitem[{{Mason} {et~al}\mbox{.}(2001){Mason}, {Wycoff}, {Hartkopf},
  {Douglass}, \& {Worley}}]{masonetal2001}
{Mason} B.~D., {Wycoff} G.~L., {Hartkopf} W.~I., {Douglass} G.~G., {Worley}
  C.~E., 2001, \aj, 122, 3466

\bibitem[{{Matthews} \& {Wehlau}(1985)}]{matthews&wehlau1985}
{Matthews} J.~M., {Wehlau} W.~H., 1985, Information Bulletin on Variable Stars,
  2725, 1

\bibitem[{{Mermilliod}, {Mermilliod} \& {Hauck}(1997){Mermilliod},
  {Mermilliod}, \& {Hauck}}]{mermilliodetal1997}
{Mermilliod} J.-C., {Mermilliod} M., {Hauck} B., 1997, \aaps, 124, 349

\bibitem[{{Mora} {et~al}\mbox{.}(2001){Mora}, {Mer{\'{\i}}n}, {Solano},
  {Montesinos}, {de Winter}, {Eiroa}, {Ferlet}, {Grady}, {Davies}, {Miranda},
  {Oudmaijer}, {Palacios}, {Quirrenbach}, {Harris}, {Rauer}, {Collier Cameron},
  {Deeg}, {Garz{\'o}n}, {Penny}, {Schneider}, {Tsapras}, \&
  {Wesselius}}]{moraetal2001}
{Mora} A. {et~al.}, 2001, \aap, 378, 116

\bibitem[{{Morgan}, {Keenan} \& {Kellman}(1943){Morgan}, {Keenan}, \&
  {Kellman}}]{morganetal1943}
{Morgan} W.~W., {Keenan} P.~C., {Kellman} E., 1943, {An atlas of stellar
  spectra, with an outline of spectral classification}. "The University of
  Chicago press"

\bibitem[{{Moya} {et~al}\mbox{.}(2010){Moya}, {Amado}, {Barrado},
  {Garc{\'{\i}}a Hern{\'a}ndez}, {Aberasturi}, {Montesinos}, \&
  {Aceituno}}]{moyaetal2010a}
{Moya} A., {Amado} P.~J., {Barrado} D., {Garc{\'{\i}}a Hern{\'a}ndez} A.,
  {Aberasturi} M., {Montesinos} B., {Aceituno} F., 2010, \mnras, 405, L81

\bibitem[{{Mulders} {et~al}\mbox{.}(2013){Mulders}, {Paardekooper},
  {Pani{\'c}}, {Dominik}, {van Boekel}, \& {Ratzka}}]{muldersetal2013}
{Mulders} G.~D., {Paardekooper} S.-J., {Pani{\'c}} O., {Dominik} C., {van
  Boekel} R., {Ratzka} T., 2013, \aap, 557, A68

\bibitem[{{Murphy}(2014)}]{murphy2014}
{Murphy} S.~J., 2014, PhD thesis, Jeremiah Horrocks Institute, University of
  Central Lancashire, Preston, UK

\bibitem[{{Murphy} {et~al}\mbox{.}(2013){Murphy}, {Pigulski}, {Kurtz},
  {Su{\'a}rez}, {Handler}, {Balona}, {Smalley}, {Uytterhoeven}, {Szab{\'o}},
  {Thygesen}, {Elkin}, {Breger}, {Grigahc{\`e}ne}, {Guzik}, {Nemec}, \&
  {Southworth}}]{murphyetal2013a}
{Murphy} S.~J. {et~al.}, 2013, \mnras, 432, 2284

\bibitem[{{Nakos}, {Sinachopoulos} \& {van Dessel}(1995){Nakos},
  {Sinachopoulos}, \& {van Dessel}}]{nakosetal1995}
{Nakos} T., {Sinachopoulos} D., {van Dessel} E., 1995, \aaps, 112, 453

\bibitem[{{Narusawa} {et~al}\mbox{.}(2006{\natexlab{a}}){Narusawa}, {Ozaki},
  {Kambe}, \& {Sadakane}}]{narusawaetal2006a}
{Narusawa} S.-y., {Ozaki} S., {Kambe} E., {Sadakane} K., 2006{\natexlab{a}},
  \pasj, 58, 617

\bibitem[{{Narusawa} {et~al}\mbox{.}(2006{\natexlab{b}}){Narusawa}, {Ozaki},
  {Okyudo}, {Takano}, \& {Nakamura}}]{narusawaetal2006b}
{Narusawa} S.-y., {Ozaki} S., {Okyudo} M., {Takano} R., {Nakamura} Y.,
  2006{\natexlab{b}}, \pasp, 118, 809

\bibitem[{{Nordstr{\"o}m} {et~al}\mbox{.}(2004){Nordstr{\"o}m}, {Mayor},
  {Andersen}, {Holmberg}, {Pont}, {J{\o}rgensen}, {Olsen}, {Udry}, \&
  {Mowlavi}}]{nordstrometal2004}
{Nordstr{\"o}m} B. {et~al.}, 2004, \aap, 418, 989

\bibitem[{{Oblak}(1978)}]{oblak1978}
{Oblak} E., 1978, \aaps, 34, 453

\bibitem[{{Ochsenbein}, {Bauer} \& {Marcout}(2000){Ochsenbein}, {Bauer}, \&
  {Marcout}}]{ochsenbeinetal2000}
{Ochsenbein} F., {Bauer} P., {Marcout} J., 2000, \aaps, 143, 23

\bibitem[{{Ohanesyan}(2008)}]{ohanesyan2008}
{Ohanesyan} J.~B., 2008, Astrophysics, 51, 490

\bibitem[{{Olsen}(1979)}]{olsen1979}
{Olsen} E.~H., 1979, \aaps, 37, 367

\bibitem[{{Olsen}(1980)}]{olsen1980}
---, 1980, \aaps, 39, 205

\bibitem[{{Olsen}(1983)}]{olsen1983}
---, 1983, \aaps, 54, 55

\bibitem[{{Oppenheimer} {et~al}\mbox{.}(2008){Oppenheimer}, {Brenner},
  {Hinkley}, {Zimmerman}, {Sivaramakrishnan}, {Soummer}, {Kuhn}, {Graham},
  {Perrin}, {Lloyd}, {Roberts}, \& {Harrington}}]{oppenheimeretal2008}
{Oppenheimer} B.~R. {et~al.}, 2008, \apj, 679, 1574

\bibitem[{{Osawa}(1965)}]{osawa1965}
{Osawa} K., 1965, Annals of the Tokyo Astronomical Observatory, 9, 121

\bibitem[{{Parenago}(1958)}]{parenago1958}
{Parenago} P.~P., 1958, \sovast, 2, 151

\bibitem[{{Paunzen}(2000)}]{paunzen2000}
{Paunzen} E., 2000, PhD thesis, Univ. Wien

\bibitem[{{Paunzen}(2001)}]{paunzen2001}
---, 2001, \aap, 373, 633

\bibitem[{{Paunzen} {et~al}\mbox{.}(1999{\natexlab{a}}){Paunzen}, {Andrievsky},
  {Chernyshova}, {Klochkova}, {Panchuk}, \& {Handler}}]{paunzenetal1999b}
{Paunzen} E., {Andrievsky} S.~M., {Chernyshova} I.~V., {Klochkova} V.~G.,
  {Panchuk} V.~E., {Handler} G., 1999{\natexlab{a}}, \aap, 351, 981

\bibitem[{{Paunzen} {et~al}\mbox{.}(2001){Paunzen}, {Duffee}, {Heiter},
  {Kuschnig}, \& {Weiss}}]{paunzenetal2001}
{Paunzen} E., {Duffee} B., {Heiter} U., {Kuschnig} R., {Weiss} W.~W., 2001,
  \aap, 373, 625

\bibitem[{{Paunzen} \& {Gray}(1997)}]{paunzen&gray1997}
{Paunzen} E., {Gray} R.~O., 1997, \aaps, 126, 407

\bibitem[{{Paunzen} \& {Handler}(1996)}]{paunzen&handler1996}
{Paunzen} E., {Handler} G., 1996, Information Bulletin on Variable Stars, 4318,
  1

\bibitem[{{Paunzen} {et~al}\mbox{.}(2002{\natexlab{a}}){Paunzen}, {Handler},
  {Weiss}, {Nesvacil}, {Hempel}, {Romero-Colmenero}, {Vuthela}, {Reegen},
  {Shobbrook}, \& {Kilkenny}}]{paunzenetal2002a}
{Paunzen} E. {et~al.}, 2002{\natexlab{a}}, \aap, 392, 515

\bibitem[{{Paunzen} {et~al}\mbox{.}(2012){Paunzen}, {Heiter}, {Fraga}, \&
  {Pintado}}]{paunzenetal2012}
{Paunzen} E., {Heiter} U., {Fraga} L., {Pintado} O., 2012, \mnras, 419, 3604

\bibitem[{{Paunzen} {et~al}\mbox{.}(1998{\natexlab{a}}){Paunzen}, {Heiter},
  {Handler}, {Garrido}, {Solano}, {Weiss}, \& {Gelbmann}}]{paunzenetal1998a}
{Paunzen} E., {Heiter} U., {Handler} G., {Garrido} R., {Solano} E., {Weiss}
  W.~W., {Gelbmann} M., 1998{\natexlab{a}}, \aap, 329, 155

\bibitem[{{Paunzen} {et~al}\mbox{.}(2014{\natexlab{a}}){Paunzen}, {Iliev},
  {Fossati}, {Heiter}, \& {Weiss}}]{paunzenetal2014}
{Paunzen} E., {Iliev} I.~K., {Fossati} L., {Heiter} U., {Weiss} W.~W.,
  2014{\natexlab{a}}, \aap, 567, A67

\bibitem[{{Paunzen} {et~al}\mbox{.}(2002{\natexlab{b}}){Paunzen}, {Iliev},
  {Kamp}, \& {Barzova}}]{paunzenetal2002b}
{Paunzen} E., {Iliev} I.~K., {Kamp} I., {Barzova} I.~S., 2002{\natexlab{b}},
  \mnras, 336, 1030

\bibitem[{{Paunzen} {et~al}\mbox{.}(1999{\natexlab{b}}){Paunzen}, {Kamp},
  {Iliev}, {Heiter}, {Hempel}, {Weiss}, {Barzova}, {Kerber}, \&
  {Mittermayer}}]{paunzenetal1999a}
{Paunzen} E. {et~al.}, 1999{\natexlab{b}}, \aap, 345, 597

\bibitem[{{Paunzen} {et~al}\mbox{.}(2003){Paunzen}, {Kamp}, {Weiss}, \&
  {Wiesemeyer}}]{paunzenetal2003}
{Paunzen} E., {Kamp} I., {Weiss} W.~W., {Wiesemeyer} H., 2003, \aap, 404, 579

\bibitem[{{Paunzen} {et~al}\mbox{.}(2014{\natexlab{b}}){Paunzen}, {Skarka},
  {Holdsworth}, {Smalley}, \& {West}}]{paunzenetal2014a}
{Paunzen} E., {Skarka} M., {Holdsworth} D.~L., {Smalley} B., {West} R.~G.,
  2014{\natexlab{b}}, \mnras, 440, 1020

\bibitem[{{Paunzen}, {St{\"u}tz} \& {Maitzen}(2005){Paunzen}, {St{\"u}tz}, \&
  {Maitzen}}]{paunzenetal2005}
{Paunzen} E., {St{\"u}tz} C., {Maitzen} H.~M., 2005, \aap, 441, 631

\bibitem[{{Paunzen} {et~al}\mbox{.}(1997){Paunzen}, {Weiss}, {Heiter}, \&
  {North}}]{paunzenetal1997a}
{Paunzen} E., {Weiss} W.~W., {Heiter} U., {North} P., 1997, \aaps, 123, 93

\bibitem[{{Paunzen}, {Weiss} \& {Kuschnig}(1996){Paunzen}, {Weiss}, \&
  {Kuschnig}}]{paunzenetal1996a}
{Paunzen} E., {Weiss} W.~W., {Kuschnig} R., 1996, Information Bulletin on
  Variable Stars, 4302, 1

\bibitem[{{Paunzen} {et~al}\mbox{.}(1998{\natexlab{b}}){Paunzen}, {Weiss},
  {Kuschnig}, {Handler}, {Strassmeier}, {North}, {Solano}, {Gelbmann},
  {Kuenzli}, \& {Garrido}}]{paunzenetal1998c}
{Paunzen} E. {et~al.}, 1998{\natexlab{b}}, \aap, 335, 533

\bibitem[{{Paunzen} {et~al}\mbox{.}(1998{\natexlab{c}}){Paunzen}, {Weiss},
  {Martinez}, {Matthews}, {Pamyatnykh}, \& {Kuschnig}}]{paunzenetal1998b}
{Paunzen} E., {Weiss} W.~W., {Martinez} P., {Matthews} J.~M., {Pamyatnykh}
  A.~A., {Kuschnig} R., 1998{\natexlab{c}}, \aap, 330, 605

\bibitem[{{Paunzen}, {Weiss} \& {North}(1994){Paunzen}, {Weiss}, \&
  {North}}]{paunzenetal1994}
{Paunzen} E., {Weiss} W.~W., {North} P., 1994, Information Bulletin on Variable
  Stars, 4068, 1

\bibitem[{{Pavlovski}, {Schnell} \& {Maitzen}(1993){Pavlovski}, {Schnell}, \&
  {Maitzen}}]{pavlovskietal1993}
{Pavlovski} K., {Schnell} A., {Maitzen} H.~M., 1993, in Astronomical Society of
  the Pacific Conference Series, Vol.~44, IAU Colloq. 138: Peculiar versus
  Normal Phenomena in A-type and Related Stars, {Dworetsky} M.~M., {Castelli}
  F., {Faraggiana} R., eds., p. 429

\bibitem[{{Philip} \& {Hayes}(1983)}]{philip&hayes1983}
{Philip} A.~G.~D., {Hayes} D.~S., 1983, \apjs, 53, 751

\bibitem[{{Renson}, {Faraggiana} \& {Boehm}(1990){Renson}, {Faraggiana}, \&
  {Boehm}}]{rensonetal1990}
{Renson} P., {Faraggiana} R., {Boehm} C., 1990, Bulletin d'Information du
  Centre de Donnees Stellaires, 38, 137

\bibitem[{{Renson} \& {Manfroid}(2009)}]{renson&manfroid2009}
{Renson} P., {Manfroid} J., 2009, \aap, 498, 961

\bibitem[{{Richards} \& {Albright}(1999)}]{richards&albright1999}
{Richards} M.~T., {Albright} G.~E., 1999, \apjs, 123, 537

\bibitem[{{Rodgers}(1968)}]{rodgers1968}
{Rodgers} A.~W., 1968, \apj, 152, 109

\bibitem[{{Rodr{\'{\i}}guez} \& {Breger}(2001)}]{rodriguez&breger2001}
{Rodr{\'{\i}}guez} E., {Breger} M., 2001, \aap, 366, 178

\bibitem[{{Rodr{\'{\i}}guez}, {L{\'o}pez-Gonz{\'a}lez} \& {L{\'o}pez de
  Coca}(2000){Rodr{\'{\i}}guez}, {L{\'o}pez-Gonz{\'a}lez}, \& {L{\'o}pez de
  Coca}}]{rodriguezetal2000}
{Rodr{\'{\i}}guez} E., {L{\'o}pez-Gonz{\'a}lez} M.~J., {L{\'o}pez de Coca} P.,
  2000, \aaps, 144, 469

\bibitem[{{Royer} {et~al}\mbox{.}(2002){Royer}, {Grenier}, {Baylac},
  {G{\'o}mez}, \& {Zorec}}]{royeretal2002b}
{Royer} F., {Grenier} S., {Baylac} M.-O., {G{\'o}mez} A.~E., {Zorec} J., 2002,
  \aap, 393, 897

\bibitem[{{Royer}, {Zorec} \& {G{\'o}mez}(2007){Royer}, {Zorec}, \&
  {G{\'o}mez}}]{royeretal2007}
{Royer} F., {Zorec} J., {G{\'o}mez} A.~E., 2007, \aap, 463, 671

\bibitem[{{Sadakane} \& {Nishida}(1986)}]{sadakane&nishida1986}
{Sadakane} K., {Nishida} M., 1986, \pasp, 98, 685

\bibitem[{{Saffe} {et~al}\mbox{.}(2008){Saffe}, {G{\'o}mez}, {Pintado}, \&
  {Gonz{\'a}lez}}]{saffeetal2008}
{Saffe} C., {G{\'o}mez} M., {Pintado} O., {Gonz{\'a}lez} E., 2008, \aap, 490,
  297

\bibitem[{{Sargent}(1965)}]{sargent1965}
{Sargent} W.~L.~W., 1965, \apj, 142, 787

\bibitem[{{Schr{\"o}der}, {Reiners} \& {Schmitt}(2009){Schr{\"o}der},
  {Reiners}, \& {Schmitt}}]{schroederetal2009}
{Schr{\"o}der} C., {Reiners} A., {Schmitt} J.~H.~M.~M., 2009, \aap, 493, 1099

\bibitem[{{Shobbrook}(2005)}]{shobbrook2005}
{Shobbrook} R.~R., 2005, Journal of Astronomical Data, 11, 7

\bibitem[{{Skarka}(2013)}]{skarka2013}
{Skarka} M., 2013, \aap, 549, A101

\bibitem[{{Skiff}(2013)}]{skiff2013}
{Skiff} B.~A., 2013, VizieR Online Data Catalog, 1, 2023

\bibitem[{{Slawson}, {Hill} \& {Landstreet}(1992){Slawson}, {Hill}, \&
  {Landstreet}}]{slawsonetal1992}
{Slawson} R.~W., {Hill} R.~J., {Landstreet} J.~D., 1992, \apjs, 82, 117

\bibitem[{{Slettebak}(1952)}]{slettebak1952}
{Slettebak} A., 1952, \apj, 115, 575

\bibitem[{{Slettebak}(1954)}]{slettebak1954}
---, 1954, \apj, 119, 146

\bibitem[{{Slettebak}(1963)}]{slettebak1963}
---, 1963, \apj, 138, 118

\bibitem[{{Slettebak}(1968)}]{slettebak1968}
---, 1968, \apj, 154, 933

\bibitem[{{Slettebak}(1975)}]{slettebak1975}
---, 1975, \apj, 197, 137

\bibitem[{{Slettebak}, {Wright} \& {Graham}(1968){Slettebak}, {Wright}, \&
  {Graham}}]{slettebaketal1968}
{Slettebak} A., {Wright} R.~R., {Graham} J.~A., 1968, \aj, 73, 152

\bibitem[{{S{\'o}dor} {et~al}\mbox{.}(2014){S{\'o}dor}, {Chen{\'e}}, {De Cat},
  {Bogn{\'a}r}, {Wright}, {Marois}, {Walker}, {Matthews}, {Kallinger}, {Rowe},
  {Kuschnig}, {Guenther}, {Moffat}, {Rucinski}, {Sasselov}, \&
  {Weiss}}]{sodoretal2014}
{S{\'o}dor} {\'A}. {et~al.}, 2014, ArXiv e-prints

\bibitem[{{Solano} \& {Paunzen}(1998)}]{solano&paunzen1998}
{Solano} E., {Paunzen} E., 1998, in ESA Special Publication, Vol. 413,
  Ultraviolet Astrophysics Beyond the IUE Final Archive, {Wamsteker} W.,
  {Gonzalez Riestra} R., {Harris} B., eds., p. 129

\bibitem[{{Solano} \& {Paunzen}(1999)}]{solano&paunzen1999}
---, 1999, \aap, 348, 825

\bibitem[{{Solano} {et~al}\mbox{.}(2001){Solano}, {Paunzen}, {Pintado},
  {C{\'o}rdoba}, \& {Varela}}]{solanoetal2001}
{Solano} E., {Paunzen} E., {Pintado} O.~I., {C{\'o}rdoba}, {Varela} J., 2001,
  \aap, 374, 957

\bibitem[{{Soubiran} {et~al}\mbox{.}(2010){Soubiran}, {Le Campion}, {Cayrel de
  Strobel}, \& {Caillo}}]{soubiranetal2010}
{Soubiran} C., {Le Campion} J.-F., {Cayrel de Strobel} G., {Caillo} A., 2010,
  \aap, 515, A111

\bibitem[{{Soummer} {et~al}\mbox{.}(2011){Soummer}, {Brendan Hagan}, {Pueyo},
  {Thormann}, {Rajan}, \& {Marois}}]{soummeretal2011}
{Soummer} R., {Brendan Hagan} J., {Pueyo} L., {Thormann} A., {Rajan} A.,
  {Marois} C., 2011, \apj, 741, 55

\bibitem[{{St\"urenburg}(1993)}]{stuerenburg1993}
{St\"urenburg} S., 1993, \aap, 277, 139

\bibitem[{{St{\"u}tz} \& {Paunzen}(2006)}]{stutz&paunzen2006}
{St{\"u}tz} C., {Paunzen} E., 2006, \aap, 458, L17

\bibitem[{{Szczygie{\l}} \& {Fabrycky}(2007)}]{szczygiel&fabrycky2007}
{Szczygie{\l}} D.~M., {Fabrycky} D.~C., 2007, \mnras, 377, 1263

\bibitem[{{Takeda} \& {Sadakane}(1997)}]{takeda&sadakane1997}
{Takeda} Y., {Sadakane} K., 1997, \pasj, 49, 571

\bibitem[{{Uesugi} \& {Fukuda}(1982)}]{uesugi&fukuda1982}
{Uesugi} A., {Fukuda} I., 1982, {Catalogue of stellar rotational velocities
  (revised)}. {Department Of Astronomy, Kyoto Univ}

\bibitem[{{Venn} \& {Lambert}(1990)}]{venn&lambert1990}
{Venn} K.~A., {Lambert} D.~L., 1990, \apj, 363, 234

\bibitem[{{Vican}(2012)}]{vican2012}
{Vican} L., 2012, \aj, 143, 135

\bibitem[{{Vogt} {et~al}\mbox{.}(1998){Vogt}, {Kerschbaum}, {Maitzen}, \&
  {Faundez-Abans}}]{vogtetal1998}
{Vogt} N., {Kerschbaum} F., {Maitzen} H.~M., {Faundez-Abans} M., 1998, \aaps,
  130, 455

\bibitem[{{Warren} \& {Hoffleit}(1987)}]{warren&hoffleit1987}
{Warren}, Jr. W.~H., {Hoffleit} D., 1987, in Bulletin of the American
  Astronomical Society, Vol.~19, Bulletin of the American Astronomical Society,
  p. 733

\bibitem[{{Weiss} {et~al}\mbox{.}(1994){Weiss}, {Paunzen}, {Kuschnig}, \&
  {Schneider}}]{weissetal1994}
{Weiss} W.~W., {Paunzen} E., {Kuschnig} R., {Schneider} H., 1994, \aap, 281,
  797

\bibitem[{{Woolf} \& {Lambert}(1999)}]{woolf&lambert1999}
{Woolf} V.~M., {Lambert} D.~L., 1999, \apjl, 520, L55

\bibitem[{{Wraight} {et~al}\mbox{.}(2012){Wraight}, {Fossati}, {Netopil},
  {Paunzen}, {Rode-Paunzen}, {Bewsher}, {Norton}, \& {White}}]{wraightetal2012}
{Wraight} K.~T., {Fossati} L., {Netopil} M., {Paunzen} E., {Rode-Paunzen} M.,
  {Bewsher} D., {Norton} A.~J., {White} G.~J., 2012, \mnras, 420, 757

\bibitem[{{Wright} {et~al}\mbox{.}(2011){Wright}, {Chen{\'e}}, {De Cat},
  {Marois}, {Mathias}, {Macintosh}, {Isaacs}, {Lehmann}, \&
  {Hartmann}}]{wrightetal2011}
{Wright} D.~J. {et~al.}, 2011, \apjl, 728, L20

\bibitem[{{Zasche} {et~al}\mbox{.}(2012){Zasche}, {Uhl{\'a}{\v r}}, {{\v
  S}lechta}, {Wolf}, {Harmanec}, {Nemravov{\'a}}, \& {Kor{\v
  c}{\'a}kov{\'a}}}]{zascheetal2012}
{Zasche} P., {Uhl{\'a}{\v r}} R., {{\v S}lechta} M., {Wolf} M., {Harmanec} P.,
  {Nemravov{\'a}} J.~A., {Kor{\v c}{\'a}kov{\'a}} D., 2012, \aap, 542, A78

\bibitem[{{Zverko} {et~al}\mbox{.}(2008){Zverko}, {{\v Z}i{\v z}{\v
  n}ovsk{\'y}}, {Mikul{\'a}{\v s}ek}, \& {Iliev}}]{zverkoetal2008}
{Zverko} J., {{\v Z}i{\v z}{\v n}ovsk{\'y}} J., {Mikul{\'a}{\v s}ek} Z.,
  {Iliev} I.~K., 2008, Contributions of the Astronomical Observatory Skalnate
  Pleso, 38, 467

\end{thebibliography}
\end{document}